\documentclass[aps,prb,reprint,amsfonts,amsmath,amssymb,longbibliography,nofootinbib,balancelastpage]{revtex4-2}
\usepackage{hyperref}
\usepackage{graphicx}
\usepackage{bm}
\usepackage{newtxtext}
\usepackage[varg]{newtxmath}
\usepackage[normalem]{ulem}
\usepackage{braket}
\usepackage{mathrsfs}
\usepackage{xcolor}
\hypersetup{colorlinks=true,linkcolor=blue,citecolor=blue,urlcolor=blue}
\usepackage{mathtools}
\usepackage{mleftright}
\mleftright
\usepackage{empheq}
\usepackage{orcidlink}
\usepackage{placeins}

\DeclareMathOperator*{\Tr}{Tr}

\DeclareMathOperator{\re}{Re}
\DeclareMathOperator{\im}{Im}

\newcommand{\llangle}{\langle\!\langle}
\newcommand{\rrangle}{\rangle\!\rangle}
\newcommand{\Lket}[1]{|#1\rangle\!\rangle}
\newcommand{\Lbra}[1]{\llangle #1|}
\newcommand{\Lbraket}[2]{\llangle #1|#2\rrangle}

\begin{document}
\title{Floquet Green's functions for lattice electrons driven by Gaussian quantum light}
\author{\mbox{Atsushi Ono\,\orcidlink{0000-0002-8813-4470}}}
\affiliation{\mbox{Department of Physics, Graduate School of Science, Tohoku University, Sendai 980-8578, Japan}}
\date{\today}

\begin{abstract}
We formulate Floquet Green's functions for noninteracting single-band lattice electrons driven by a reservoir-stabilized single-mode Gaussian quantum light source.
The source is prescribed externally and is not updated by the many-electron polarization, while an active electronic probe still conditions the source evolution through the Peierls coupling.
The two time arguments of a Green's function share one source history: the lesser and greater components are obtained by convolving a shared-history four-endpoint kernel with the continuous bath kernels before the final source trace, while the bath canonical anticommutation relation yields an equal-time covariance that seeds the retarded and advanced one-leg propagations.
The Peierls coupling is treated nonperturbatively within the prescribed-source model, and classical Floquet theory is recovered in the appropriate limit.
Numerical calculations on a minimal one-dimensional model show finite-coupling quantum-source corrections beyond a prescribed classical drive, together with spectral reconstruction and occupancy redistribution for squeezed vacuum and squeezed coherent sources.
The squeezing parameter and phase provide additional control knobs, beyond classical amplitude modulation, for both sideband structure and occupied weight.
This work provides a theoretical framework for quantum Floquet engineering of condensed matter with an externally prescribed quantum light source.
\end{abstract}

\maketitle

\tableofcontents

\section{Introduction}\label{sec:intro}

Floquet engineering has emerged as a powerful approach to tailoring band structure, transport, and correlated phases in periodically driven quantum matter \cite{Bukov2015, Oka2019, Rudner2020}.
In its standard formulation for solids, the electromagnetic field is treated as a prescribed classical background, entering the lattice Hamiltonian as a time-dependent parameter through the Peierls substitution.
For noninteracting electrons, two-time Green's functions in a periodic steady state inherit the discrete time-translation symmetry, and can be analyzed in Sambe (Floquet) space \cite{Shirley1965, Sambe1973}, where harmonic sidebands and replicas of the undriven band structure naturally emerge.
Green's functions constructed in this Sambe space were introduced early on for periodically driven and ac-biased transport problems \cite{Martinez2003, Kohler2005}.
For correlated systems, nonequilibrium dynamical mean-field theory under time-periodic driving was pioneered in Ref.~\cite{Schmidt2002}.
Floquet Green's functions of this type, combined with dissipative baths that stabilize the nonequilibrium steady state, were developed within the nonequilibrium dynamical mean-field theory \cite{Joura2008, Tsuji2008, Tsuji2009, Aoki2014}.
They have been applied to steady-state spectra and response functions of a wide variety of periodically driven systems, including photoinduced Hall responses and Floquet topological or correlated systems \cite{Oka2009, Kitagawa2011, Mikami2016, Lee2017, Qin2017, Peronaci2018}, driven Mott insulators with impact ionization and high-harmonic generation \cite{Sorantin2018, Murakami2018, Murakami2018b}, and cold-atom, quantum-dot, and light-controlled magnetic or superconducting platforms \cite{Ono2018, Qin2018, Ono2019, Asmar2020, Ke2020, Ke2022, Honeychurch2024, Herre2026, Okugawa2026}.
This classical external-field viewpoint has successfully described a broad range of phenomena such as light-induced band renormalization, dynamical localization \cite{Dunlap1986}, and driven topological phases \cite{Oka2009, Lindner2011}, with Floquet--Bloch sidebands and light-induced Hall responses observed experimentally \cite{Wang2013, McIver2020}.

While Floquet engineering and, more broadly, light-induced control of condensed matter \cite{Kirilyuk2010, delaTorre2021, Koshihara2022} have largely been developed within a classical-field approximation, cavity and circuit quantum electrodynamics platforms increasingly seek to harness rather than discard the quantum nature of light \cite{Kockum2019, Burkard2020, Haroche2020, ZareRameshti2022}.
Even in the absence of a coherent amplitude, the electromagnetic field exhibits vacuum fluctuations; coherent states carry photon number fluctuations; and squeezed Gaussian states feature anomalous correlations that have no classical counterpart \cite{WallsMilburn2008}.
More generally, quantum reservoir engineering provides a route to stabilizing nonclassical states \cite{Poyatos1996}; related dissipative stabilization has also been demonstrated for trapped-ion and microwave squeezed modes \cite{Kienzler2015, Dassonneville2021}.
This perspective underlies the growing effort to control quantum materials with cavity vacuum fields and few-photon states \cite{Kiffner2019, Hubener2021, Schlawin2022, Bloch2022, Lu2025, Bretscher2026}.
Hamiltonian-based quantum-Floquet approaches have retained the full quantized Peierls exponential in a photon-number basis to study few-photon control of effective interactions, light--matter-entangled order, and the quantum-to-classical Floquet crossover \cite{Sentef2020, Li2020, Sueiro2025}.
A useful reference point is an exactly solvable single-band cavity chain with a spatially uniform quantized Peierls field \cite{Eckhardt2022}, in which each electron momentum occupation number is separately conserved, so that a fixed electron configuration labels a conditional photon problem.
This block structure gives a controlled route in the single-band zero-temperature setting, but it does not directly scale to finite temperature or to multiband models with interband coherence.
A direct combination of the full quantized Peierls link with electronic Green's functions and lead self-energies has been developed for a cavity-embedded chain model \cite{Nguyen2024}; that setting addresses a static finite closed cavity coupled to leads via exact light--matter eigenstates.

Operator-valued formulations of quantum-light-driven systems have also been developed in complementary spectroscopic and transport settings.
In Liouville space, superoperator methods for quantum spectroscopy and nonequilibrium Green's functions evaluate electronic and optical correlators from open-system dynamics, including auxiliary master-equation mappings onto finite Lindblad systems and Floquet extensions of related strategies \cite{Harbola2008, Arrigoni2013, Dorda2014, Dorfman2016, Sorantin2018}; periodically driven Lindblad dynamics have further been formulated in an extended Liouville--Sambe space with steady-state correlators obtained from the corresponding resolvent \cite{Keliri2026}.
In quantum transport, nonclassical and quantized electromagnetic environments have been incorporated through phase-operator correlators \cite{Souquet2014, Bi2019} and through diagrammatic electron--photon Green's functions with mutual self-energies \cite{Gao2016, Agarwalla2016}.
More broadly, sharing environmental histories between the forward and backward branches is a standard structure in the Feynman--Vernon influence functional, closed-time-path Green's functions, and multitime process tensors \cite{Schwinger1961, FeynmanVernon1963, Keldysh1965, Pollock2018}.
These works provide precedents for operator-valued light and Liouville-space electronic Green's functions.

In parallel, the interface between strong-field physics and quantum optics has developed rapidly, encompassing fully quantized descriptions of intense laser--matter interactions and experimental observations of nonclassical high-harmonic emission \cite{Lewenstein2021, Stammer2023, Bhattacharya2023, Lange2024, Theidel2024, Theidel2025}.
A separate route treats driving by nonclassical light through phase-space decompositions over c-number field histories, expressing matter observables as averages over external-field realizations \cite{Gorlach2023, EvenTzur2023, Rasputnyi2024, Gothelf2025, Li2025, Li2026, Rivera-Dean2026, Imai2026a, Imai2026, Imai2026b}.
These constructions employ the Sudarshan--Glauber $P$ function or its generalized and positive-$P$ variants, whose weights need not be ordinary classical probabilities \cite{Sudarshan1963, Glauber1963, DrummondGardiner1980, WallsMilburn2008}, as well as diagonal and large-photon-number approximations that reduce to Husimi-$Q$-weighted averages.
It is useful to distinguish such exact or generalized decompositions from approximate incoherent-mixture prescriptions: in the diagonal, large-photon-number, and weak-coupling external-field approximations, the quantum nature of the drive enters chiefly through the trajectory weights.
Across this route, the photon source is not propagated as an operator-valued, dissipative degree of freedom during the electron evolution; this structure is well suited to bright fields with macroscopic photon numbers.

Against this background, we are not aware of a previous formulation that combines a reservoir-stabilized prescribed Gaussian source, nonperturbative propagation of the full operator-valued lattice Peierls link, a shared-history two-leg process, and continuous-bath lesser and greater Green's functions before the final source trace; this work establishes such a Floquet Green's-function framework for lattice electrons.
Within this approximation, we keep the full light--matter coupling operator without expanding it in the coupling strength, so that coherent drives, vacuum fluctuations, and squeezed fluctuations are treated nonperturbatively and on the same footing.
The two time arguments of a Green's function share one source history: the occupied components are defined by convolution of one shared-history kernel with the continuous bath kernels, while the retarded and advanced components are closed through the bath canonical anticommutation relation (CAR) identity.
The resulting kernels satisfy positivity, the Pauli bound, and the CAR, and admit a quasifree fermionic realization at the two-point level.
We also clarify the relation to classical Floquet theory and to weak-coupling perturbative approximations.

The remainder of this paper is organized as follows.
Section~\ref{sec:formalism} develops the prescribed-source shared-history formulation and its Sambe implementation.
Section~\ref{sec:benchmark} presents numerical results for the minimal single-band model.
Section~\ref{sec:discussion} discusses the physical picture and the scope of the approximations, and Sec.~\ref{sec:summary} summarizes this work.
Appendix~\ref{sec:app_born_scba} gives the Born and self-consistent Born approximation (SCBA) self-energies, Appendix~\ref{sec:app_implementation} details the frequency-domain implementation of the shared-history components, and Appendix~\ref{sec:app_cutoff} documents the numerical-convergence tests.

\section{Formalism}\label{sec:formalism}

\begin{figure*}[t]\centering
\includegraphics[scale=1]{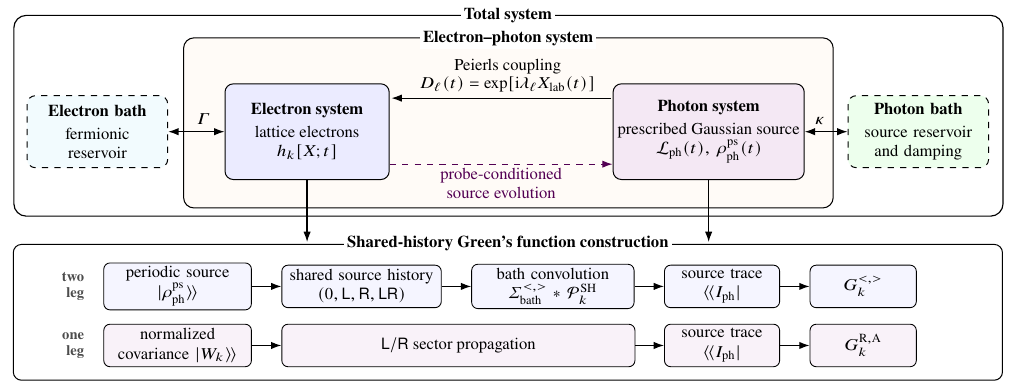}
\caption{Overview of the prescribed-source framework.
The total system consists of the electron--photon system coupled to a fermionic electron bath and a photon source bath.
The photon system is a reservoir-stabilized Gaussian source with steady state $\rho_{\mathrm{ph}}^{\mathrm{ss}}$, fixed by the parameters $\alpha_0$, $r$, and $\phi_0$; it acts on the lattice electrons through the gauge-covariant Peierls link operators $D_\ell(t)$ [Eq.~\eqref{eq:peierls_link_def}].
The lower panel summarizes the shared-history construction: four active-leg sectors and six endpoint orderings yield a two-leg response that is convolved with $\varSigma_{\mathrm{bath}}^{<,>}$ to obtain $G_k^{<,>}$, while the equal-time covariance $W_k$ supplies the one-leg evaluation of $G_k^{\mathrm{R,A}}$.
The source trace with $\Lbra{I_{\mathrm{ph}}}$ is applied only after the appropriate one-leg or two-leg propagation, and $G_k^{\mathrm{K}}$ is subsequently derived from $G_k^>+G_k^<$.
}
\label{fig:system_hierarchy}
\end{figure*}

\subsection{Physical setting and scope}

We use units with $\hbar=1$ in all dynamical equations.
Explicit factors of $\hbar$ are retained if necessary, such as in the Peierls coupling constant defined in Eq.~\eqref{eq:lambda_definition}.
We consider noninteracting single-band lattice electrons, whose Hamiltonian in real space is given by
\begin{align}
H_{\mathrm{el}} &= \sum_{ij} T_{ij} c_{i}^\dagger c_{j},
\label{eq:H0_realspace}
\end{align}
where $c_{i}^\dagger$ is the creation operator for an electron at site $i$, and $T_{ij}$ collects all bare hopping amplitudes.
We introduce a single-mode photon Hamiltonian with carrier frequency $\varOmega$,
\begin{align}
H_{\mathrm{ph}} = \varOmega a^\dagger a.
\label{eq:Hph_free}
\end{align}
Electromagnetic coupling is implemented in a gauge-covariant lattice form via Peierls link operators (see Sec.~\ref{sec:peierls_link}).

We focus on a regime where the photon state is stabilized by an external reservoir and can be represented as a prescribed quantum source with a periodic Markovian Liouvillian.
By ``prescribed'' we mean that the reservoir generator, its parameters, and the target periodic state are externally specified; this does not remove the multitime correlations carried by that source process.
An active electronic probe still conditions the source density operator through the Peierls coupling, so the forward and backward legs of the Schwinger--Keldysh contour share one source history.
The continuous fermionic bath is retained as an independent Gaussian input and is contracted after the shared source propagator has been constructed.
The object of interest is therefore a \emph{probe} Green's function for a single electronic addition or removal propagating in a reservoir-stabilized source process, rather than the exact Green's function of a reciprocal closed electron--photon system.

We study the time-periodic steady state (Floquet state) of electrons driven by the reservoir-stabilized source and formulate the corresponding Green's functions in Floquet--Sambe representation.
The electron bath supplies a local retarded linewidth and continuous lesser and greater kernels in the wide-band limit.
The retarded linewidth is determined by the retarded self-energy $\varSigma^{\mathrm{R}}=-\mathrm{i}\varGamma/2$, while the Fermi distribution remains in the lesser and greater self-energies $\varSigma_{\mathrm{bath}}^{<,>}(\omega)$.
We assume that the one-period source channel is relaxing, so its unit eigenvalue is simple and every nonstationary source mode decays.
The undamped limit generally lacks a unique periodic steady state and is outside the present formulation unless an initial state, a finite observation window, or an infinitesimal relaxation is prescribed separately.
Electron--electron interactions and self-consistent source backaction are outside the scope of this work.
The neglect of backaction is expected to be controlled when the photon linewidth $\kappa$ is large compared with the electron-induced photon self-energy, but establishing this criterion for concrete parameters is not attempted here; we study the electronic probe physics within the prescribed-source approximation.

The explicit source realizations developed below and used in the numerical calculations are pure Gaussian states, namely coherent, squeezed-vacuum, and squeezed-coherent states whose Bogoliubov-frame steady state is a vacuum.
Mixed Gaussian or squeezed-thermal reservoirs can be incorporated by modifying the photon Liouvillian, but they are not part of the main calculations.

The formalism, positivity and CAR proofs, and numerics below use the minimal one-dimensional nearest-neighbor model of Sec.~\ref{sec:peierls_link}, with one physical orbital per momentum.
Throughout, we assume that the photon field is spatially uniform and hence does not induce momentum transfer between electron Bloch states; in translation-invariant settings, the electron problem therefore decomposes into independent momentum sectors, and we work at fixed $k$.
Multiorbital mixing, superconducting anomalous sectors, and finite-band retarded bath memory require extensions of the sector structure and are left for future work.

\subsection{Gaussian quantum light source}

We denote the photon annihilation and creation operators by $a$ and $a^\dagger$ in the source frame, i.e., the frame rotating at the carrier frequency $\varOmega$, where the carrier rotation is made implicit.
The dimensionless quadrature is defined as
\begin{align}
X \equiv a + a^\dagger.
\label{eq:X_def}
\end{align}

\paragraph*{Coherent and squeezed Gaussian states.}
We introduce the displacement and squeezing operators in standard quantum-optics notation,
\begin{gather}
\mathcal{D}(\alpha)
=
\exp\left(\alpha a^\dagger-\alpha^* a\right),
\label{eq:displacement_op}\\
\mathcal{S}(r,\phi_0)
=
\exp\left[
\frac{r}{2}\left(
\mathrm{e}^{-\mathrm{i}\phi_0}a^2-\mathrm{e}^{+\mathrm{i}\phi_0}a^{\dagger 2}
\right)
\right],
\label{eq:squeeze_op}
\end{gather}
respectively.
For $\alpha\in\mathbb{C}$ and $r,\phi_0\in\mathbb{R}$, both $\mathcal{D}(\alpha)$ and $\mathcal{S}(r,\phi_0)$ are unitary, with $\mathcal{D}^\dagger(\alpha)=\mathcal{D}(-\alpha)$ and $\mathcal{S}^\dagger(r,\phi_0)=\mathcal{S}(-r,\phi_0)$.
With this convention, the conjugation identities used below are
\begin{gather}
\mathcal{D}(\alpha)a\mathcal{D}^\dagger(\alpha) = a-\alpha,
\label{eq:D_inverse_transform}\\
\mathcal{S}(r,\phi_0)a\mathcal{S}^\dagger(r,\phi_0)
= \cosh(r)\, a + \mathrm{e}^{+\mathrm{i}\phi_0} \sinh(r)\, a^\dagger.
\label{eq:S_inverse_transform}
\end{gather}

In the present setting, we assume that the photon mode is stabilized to a squeezed coherent Gaussian steady state of the form
\begin{align}
\rho_{\mathrm{ph}}^{\mathrm{ss}}
&=
\mathcal{D}(\alpha_0) \mathcal{S}(r,\phi_0) |0\rangle\langle 0| \mathcal{S}^\dagger(r,\phi_0) \mathcal{D}^\dagger(\alpha_0),
\label{eq:gaussian_state_def}
\end{align}
where $\alpha_0$ is the coherent amplitude, $r$ is the squeezing parameter, $\phi_0$ is the squeezing phase, and $|0\rangle$ is the vacuum of $a$.
The density matrix $\rho_{\mathrm{ph}}^{\mathrm{ss}}$ is understood in the source-frame interaction picture with respect to the free carrier rotation.
Expectation values in $\rho_{\mathrm{ph}}^{\mathrm{ss}}$ follow from standard Gaussian identities.
Denoting $\langle \bullet \rangle_{\mathrm{ss}}\equiv \Tr(\rho_{\mathrm{ph}}^{\mathrm{ss}} \bullet)$, one has
\begin{gather}
\langle a\rangle_{\mathrm{ss}}
= \alpha_0,
\label{eq:ss_mean_b}\\
\langle a^\dagger a\rangle_{\mathrm{ss}}
= |\alpha_0|^2+\sinh^2(r),
\label{eq:ss_mean_n}\\
\langle a^2 \rangle_{\mathrm{ss}}
= \alpha_0^2-\mathrm{e}^{+\mathrm{i}\phi_0} \sinh(r) \cosh(r).
\label{eq:ss_mean_bb}
\end{gather}
Here, $\sinh^2(r)$ is the normal covariance of the squeezed vacuum, while the last line gives the anomalous moment that carries the squeezing phase.

\paragraph*{Laboratory and Bogoliubov frames.}
The Bogoliubov mode $b$ whose vacuum is the squeezed coherent state is defined by
\begin{align}
b &\equiv
\mathcal{D}(\alpha_0)\mathcal{S}(r,\phi_0) a \mathcal{S}^\dagger(r,\phi_0)\mathcal{D}^\dagger(\alpha_0) \notag \\
&= \cosh(r) (a-\alpha_0) + \mathrm{e}^{+\mathrm{i}\phi_0}\sinh(r) (a^\dagger-\alpha_0^*).
\label{eq:bogoliubov_mode_def}
\end{align}
Solving for the source-frame mode gives
\begin{align}
a &= \alpha_0 + \cosh(r)\,b - \mathrm{e}^{+\mathrm{i}\phi_0}\sinh(r)\,b^\dagger.
\label{eq:a_in_terms_of_b}
\end{align}
In particular, $b$ annihilates the squeezed coherent state, so the steady photon state is the $b$-vacuum, $\rho_{\mathrm{ph}}^{\mathrm{ss}}=|0_b\rangle\langle 0_b|$ with $b|0_b\rangle=0$, which represents the same physical Gaussian state as Eq.~\eqref{eq:gaussian_state_def} in the $a$-basis.
By construction, $\langle b\rangle_{\mathrm{ss}} = \langle b^\dagger b\rangle_{\mathrm{ss}} = \langle b^2 \rangle_{\mathrm{ss}} = 0$.

The laboratory counterpart of the source-frame quadrature $X$ in Eq.~\eqref{eq:X_def} is the interaction-picture operator generated by the free carrier rotation in Eq.~\eqref{eq:Hph_free},
\begin{gather}
a(t)
= \mathrm{e}^{+\mathrm{i}\varOmega a^\dagger a t} a \mathrm{e}^{-\mathrm{i}\varOmega a^\dagger a t}
= a \mathrm{e}^{-\mathrm{i}\varOmega t},
\label{eq:at_interaction}\\
X_{\mathrm{lab}}(t)
= a(t)+a^\dagger(t)
= a\mathrm{e}^{-\mathrm{i}\varOmega t}+a^\dagger\mathrm{e}^{+\mathrm{i}\varOmega t},
\label{eq:Xlab_def}
\end{gather}
where $t$ denotes laboratory time.
In this convention, the carrier-frequency rotation is kept explicitly in $X_{\mathrm{lab}}(t)$; the Liouvillian $\mathcal{L}_{\mathrm{ph}}$ below describes the engineered source dynamics in the rotating Bogoliubov frame and does not double-count the free carrier rotation, so $\rho_{\mathrm{ph}}^{\mathrm{ss}}$ is time independent whereas $X_{\mathrm{lab}}(t)$ contains $\mathrm{e}^{\mp\mathrm{i}\varOmega t}$.

Substituting Eq.~\eqref{eq:a_in_terms_of_b} for $a$ into Eq.~\eqref{eq:Xlab_def} yields the decomposition
\begin{align}
X_{\mathrm{lab}}(t) = X_{\mathrm{cl}}(t) + f(t)\,b + f^*(t)\,b^\dagger,
\label{eq:Xlab_bogoliubov}
\end{align}
where $b$ is a bosonic mode in the Bogoliubov frame, and
\begin{gather}
X_{\mathrm{cl}}(t) = \alpha_0 \mathrm{e}^{-\mathrm{i}\varOmega t} + \alpha_0^* \mathrm{e}^{+\mathrm{i}\varOmega t},
\label{eq:Xcl_def}\\
f(t) = \cosh(r)\,\mathrm{e}^{-\mathrm{i}\varOmega t} - \sinh(r)\,\mathrm{e}^{-\mathrm{i}\phi_0}\mathrm{e}^{+\mathrm{i}\varOmega t}.
\label{eq:f_def}
\end{gather}
The fluctuation about the classical coordinate is therefore
\begin{align}
\delta X_{\mathrm{lab}}(t)
\equiv
X_{\mathrm{lab}}(t)-X_{\mathrm{cl}}(t)
=
f(t)\,b + f^*(t)\,b^\dagger.
\label{eq:deltaX_def}
\end{align}
For a coherent source, writing $\alpha_0=|\alpha_0|\mathrm{e}^{\mathrm{i}\varphi_\alpha}$ gives
\begin{align}
X_{\mathrm{cl}}(t)
&= 2|\alpha_0|\cos(\varOmega t-\varphi_\alpha).
\label{eq:Xcl_coherent}
\end{align}
The key point is that the light--matter coupling always contains the laboratory coordinate $X_{\mathrm{lab}}(t)$ [Eq.~\eqref{eq:peierls_link_def}], even though the steady state is represented compactly as the $b$-vacuum.
Gaussianity is exploited here mainly in two ways: one-time expectation values of Peierls exponentials admit closed forms via Gaussian characteristic functions, as in the mean-link formulas of Sec.~\ref{sec:benchmark}, and the pure Gaussian steady states used below are represented as the $b$-vacuum of a Bogoliubov mode stabilized by the single-mode damping channel of Eq.~\eqref{eq:lindblad_bogoliubov_main}.
The shared-history Green's functions then propagate the operator-valued link in a finite Fock representation of that source; that Fock propagation itself does not require a Wick closure of photon correlators.

\paragraph*{Source Liouvillian.}
The prescribed Gaussian source used in the Green's-function calculation is specified by a Markovian photon Liouvillian of Gorini--Kossakowski--Sudarshan--Lindblad form \cite{Gorini1976, Lindblad1976}, $\mathcal{L}_{\mathrm{ph}}:\mathcal{B}(\mathcal{H}_{\mathrm{ph}})\to\mathcal{B}(\mathcal{H}_{\mathrm{ph}})$ acting on photon density operators, where $\mathcal{B}(\mathcal{H}_{\mathrm{ph}})$ denotes the space of linear operators on the photon Hilbert space $\mathcal{H}_{\mathrm{ph}}$ (at a finite Fock cutoff, the complex $N_b\times N_b$ matrices).
In the rotating Bogoliubov frame introduced above, the pure Gaussian sources considered in the main calculations are represented by the damping generator
\begin{align}
\mathcal{L}_{\mathrm{ph}}\rho
&= \kappa \mathcal{L}_{\mathrm{diss}}[b]\rho,
\label{eq:lindblad_bogoliubov_main}
\end{align}
with $\mathcal{L}_{\mathrm{diss}}[b]\rho = b\rho b^\dagger-\frac{1}{2}\{b^\dagger b,\rho\}$.
The rate $\kappa$ is the relaxation rate of the engineered photon source.
The unique stationary state is $\rho_{\mathrm{ph}}^{\mathrm{ss}}=|0_b\rangle\langle 0_b|$, which is a standard property of a zero-temperature single-mode damping channel \cite{WallsMilburn2008}.
In the notation of the general sector construction below, this stationary state is identified with the periodic source state through $\rho_{\mathrm{ph}}^{\mathrm{ps}}(t)=\rho_{\mathrm{ph}}^{\mathrm{ss}}$, which is time independent in the rotating Bogoliubov frame.
The generator $\mathcal{L}_{\mathrm{ph}}$ specifies the autonomous source channel, while the conditional probe action enters separately through the sector generators of Sec.~\ref{sec:shared_history_gf}.

\subsection{Gauge-covariant lattice Hamiltonian}\label{sec:peierls_link}

Minimal coupling to an electromagnetic field can be implemented in a gauge-covariant lattice form by dressing the hopping amplitudes with Peierls links and including a scalar potential,
\begin{align}
H_{\mathrm{el}}[\{A_{ij}\},\{\varPhi_i\}](t)
= \sum_{ij}
T_{ij} U_{ij}(t) c_{i}^\dagger c_{j}
+ q\sum_i \varPhi_i(t) c_{i}^\dagger c_{i},
\label{eq:He_with_APhi}
\end{align}
with the Peierls link on an oriented bond $j\to i$,
\begin{align}
U_{ij}(t)
&= \exp\left[
\frac{\mathrm{i} q}{\hbar}
A_{ij}(t)
\right].
\label{eq:peierls_link_Uij}
\end{align}
Here, $q$ denotes the carrier charge, $\varPhi_i(t)$ is the scalar potential on site $i$, and $A_{ij}(t)$ is the bond potential along $j\to i$.
In this paper, we work in the temporal gauge $\varPhi_i=0$, so the coupling enters only through the Peierls links $U_{ij}(t)$.
Specializing to a long-wavelength single mode that is spatially uniform on the scale of a lattice bond, we write the laboratory vector potential as
\begin{align}
\bm{A}_{\mathrm{lab}}(t)
= \bm{e} A_{\mathrm{zpf}} X_{\mathrm{lab}}(t),
\label{eq:A_single_mode}
\end{align}
where $\bm{e}$ is the polarization unit vector and $A_{\mathrm{zpf}}$ is the zero-point fluctuation amplitude (set by the mode volume and frequency).
We take
\begin{align}
A_{ij}(t) = \bm{A}_{\mathrm{lab}}(t)\cdot\bm{d}_{ij},
\quad
\bm{d}_{ij} = \bm{r}_i-\bm{r}_j,
\label{eq:Aij_uniform}
\end{align}
where $\bm{r}_i$ is the real-space position of lattice site $i$.
The corresponding Peierls link on a bond $\ell:j\to i$ is then
\begin{align}
D_\ell(t) = \exp\bigl[\mathrm{i}\lambda_\ell X_{\mathrm{lab}}(t)\bigr],
\label{eq:peierls_link_def}
\end{align}
with the real bond-dependent Peierls coefficient
\begin{align}
\lambda_\ell
= \frac{q}{\hbar} A_{\mathrm{zpf}} \bm{e}\cdot \bm{d}_\ell,
\label{eq:lambda_definition}
\end{align}
where $\bm{d}_\ell=\bm{d}_{ij}$ for that bond.
The central feature of the present approach is that we keep the full exponential operator in Eq.~\eqref{eq:peierls_link_def}, rather than expanding it in powers of $\lambda_\ell$.
Equation~\eqref{eq:peierls_link_def} is the single-mode specialization of Eq.~\eqref{eq:peierls_link_Uij} in temporal gauge; the classical Floquet link is recovered by replacing $X_{\mathrm{lab}}(t)$ with the c-number mean $X_{\mathrm{cl}}(t)$ of Eq.~\eqref{eq:Xcl_def}.

For a single photon mode, Eq.~\eqref{eq:peierls_link_def} is a displacement operator in the photon Hilbert space.
In the bare source-frame mode, one can write
\begin{gather}
D_\ell(t)
= \exp\left[\beta_\ell(t)\,a^\dagger-\beta_\ell^*(t)\,a\right]
= \mathcal{D}(\beta_\ell(t)),
\label{eq:peierls_as_displacement}\\
\beta_\ell(t) = \mathrm{i} \lambda_\ell \mathrm{e}^{+\mathrm{i}\varOmega t}.
\label{eq:peierls_displacement_amplitude}
\end{gather}
For the prescribed Gaussian source, a more useful representation follows from the Bogoliubov decomposition in Eq.~\eqref{eq:Xlab_bogoliubov}: the link factorizes into a classical phase and a displacement of the $b$ mode,
\begin{align}
D_\ell(t)
= \exp\bigl[\mathrm{i}\lambda_\ell X_{\mathrm{cl}}(t)\bigr]
\exp\bigl[\mathrm{i}\lambda_\ell\bigl(f(t)b + f^*(t) b^\dagger\bigr)\bigr],
\label{eq:D_factorization_bogoliubov}
\end{align}
where the second factor is a displacement operator of the $b$ mode,
\begin{gather}
\exp\bigl[\mathrm{i}\lambda_\ell\bigl(f(t)b + f^*(t) b^\dagger\bigr)\bigr]
=
\mathcal{D}_b \left(\beta_{\ell}^{(b)}(t)\right),
\label{eq:bogoliubov_displacement_op}
\\
\beta_{\ell}^{(b)}(t)=\mathrm{i}\lambda_\ell f^*(t),
\label{eq:bogoliubov_displacement_amplitude}
\end{gather}
with $\mathcal{D}_b(\beta)\equiv\exp(\beta b^\dagger-\beta^* b)$ and $f(t)$ the Bogoliubov-frame quadrature coefficient in Eq.~\eqref{eq:f_def}.
The coherent part and the quantum fluctuation part of the link are thereby separated explicitly, which also provides an efficient route to numerical evaluation with controlled photon cutoffs (Sec.~\ref{sec:numerical_notes}).

In translation-invariant settings, the bare hopping and the Peierls coefficients depend only on the relative bond vector, so that $T_{ij}$ and $\lambda_\ell$ are both indexed by the same bond $\ell:j\to i$.
Fourier transforming the electron operators, $c_{j}=N_k^{-1/2}\sum_k \mathrm{e}^{\mathrm{i} \bm{k}\cdot\bm{r}_j}c_{k}$ with $N_k$ the number of unit cells, block-diagonalizes the Hamiltonian in momentum,
\begin{align}
H_{\mathrm{el}}(t)
&= \sum_k
c_{k}^\dagger h_{k}(t) c_{k},
\label{eq:He_from_hk}
\end{align}
where $h_{k}(t)$ is the operator-valued coefficient (acting on the photon Hilbert space) multiplying $c_{k}^\dagger c_{k}$; this step is generic and does not yet use the single-mode Peierls link of Eq.~\eqref{eq:peierls_link_def}.
Carrying out the transform explicitly gives $h_{k}(t)$ as a sum over bonds (or hopping channels) weighted by vertex factors and the corresponding link operators,
\begin{align}
h_{k}(t) = \sum_\ell \gamma_{\ell}(k) D_\ell(t),
\label{eq:hk_peierls_general}
\end{align}
where the sum runs over every bond $\ell$ of Eq.~\eqref{eq:peierls_link_def}, and $\gamma_{\ell}(k)$ is the momentum-space matrix element for the hopping channel carried by bond $\ell$.

For the remainder of this work, including the numerical calculations of Sec.~\ref{sec:benchmark}, we specialize Eq.~\eqref{eq:hk_peierls_general} to a one-dimensional single-band chain with nearest-neighbor hopping only, setting the lattice spacing to unity so that the site coordinate is $r_j=j$.
In this case, Eq.~\eqref{eq:hk_peierls_general} reduces to
\begin{align}
h_k(t) = -t_{\mathrm{h}}\,\mathrm{e}^{-\mathrm{i} k} D(t) + \mathrm{H.c.},
\label{eq:hk_1d}
\end{align}
where
\begin{align}
D(t) \equiv \exp[\mathrm{i}\lambda X_{\mathrm{lab}}(t)]
\end{align}
is the link operator for one directed nearest-neighbor bond, with the corresponding Peierls coefficient denoted by $\lambda=|\lambda_\ell|$.
For a coherent source, combining this coupling strength with the classical coordinate of Eq.~\eqref{eq:Xcl_coherent} defines the classical Peierls phase amplitude
\begin{align}
A_{\mathrm{cl}} \equiv 2\lambda|\alpha_0|,
\label{eq:Acl_def}
\end{align}
used for coherent-limit comparisons in the single-band model of Sec.~\ref{sec:benchmark}.

\subsection{Shared-history formulation of Green's functions}\label{sec:shared_history_gf}

Although the source generator is prescribed, the source density operator retains the conditional update generated by the active single-particle probe throughout each electronic propagation interval, so the two time arguments of a fermionic Green's function define two propagation legs that share one source history.
We call this the shared-history construction: both legs are driven by one prescribed source through one shared Kraus sequence, in contrast to a product of independently source-averaged retarded and advanced propagators.
The logical order below follows from this construction: the wide-band Langevin equation yields a variation-of-constants (Duhamel) formula whose occupied correlators require one shared two-leg source average; that average is evaluated by the four-sector process in Liouville space; and only afterwards is the average convolved with $\varSigma_{\mathrm{bath}}^{<,>}$ to obtain $G_k^{<,>}$.

\subsubsection{Two-time Green's functions}

We begin with the standard real-time components of the contour-ordered Green's function \cite{Schwinger1961, Keldysh1965, Aoki2014}.
For electron operators in a fixed momentum sector, the equal-time CAR reads
\begin{align}
\{c_{k}(t),c_{k}^\dagger(t)\} = 1,
\quad
\{c_{k}(t),c_{k}(t)\} = 0.
\label{eq:electron_car}
\end{align}
The contour-ordered Green's function is
\begin{align}
G_k(z,z')
&=
-\mathrm{i} \langle \mathcal{T}_{C} c_k(z)c_k^\dagger(z') \rangle,
\label{eq:contour_gf_def}
\end{align}
where $z,z'$ lie on the Schwinger--Keldysh contour $C$ and $\mathcal{T}_{C}$ denotes contour ordering along $C$.
The greater and lesser components are
\begin{align}
G_k^>(t,t')
&=
-\mathrm{i}\langle c_k(t)c_k^\dagger(t')\rangle,
\label{eq:greater_def}
\\
G_k^<(t,t')
&=
+\mathrm{i}\langle c_k^\dagger(t')c_k(t)\rangle,
\label{eq:lesser_def}
\end{align}
respectively.
The retarded, advanced, and Keldysh components are
\begin{align}
G_k^{\mathrm{R}}(t,t')
&=
\varTheta(t-t')
\left[
G_k^>(t,t')-G_k^<(t,t')
\right],
\label{eq:retarded_def}
\\
G_k^{\mathrm{A}}(t,t')
&=
-\varTheta(t'-t)
\left[
G_k^>(t,t')-G_k^<(t,t')
\right],
\label{eq:advanced_def}
\\
G_k^{\mathrm{K}}(t,t')
&=
G_k^>(t,t')+G_k^<(t,t').
\label{eq:keldysh_def}
\end{align}
Here, $\varTheta$ is the Heaviside step function with the convention $\varTheta(0)=1/2$, so that the equal-time CAR, $G_k^>(t,t)-G_k^<(t,t)=-\mathrm{i}$, implies $G_k^{\mathrm{R}}(t,t)=-\mathrm{i}/2$ and $G_k^{\mathrm{A}}(t,t)=+\mathrm{i}/2$.
Hermiticity and the component definitions imply
\begin{gather}
G_k^{\mathrm{A}}(t,t')
= \left[G_k^{\mathrm{R}}(t',t)\right]^*,
\label{eq:RA_hermiticity}
\\
G_k^{\mathrm{R}}(t,t')-G_k^{\mathrm{A}}(t,t')
= G_k^>(t,t')-G_k^<(t,t'),
\label{eq:RA_difference}
\\
G_k^{\gtrless}(t,t')
= \frac{1}{2} \left[
G_k^{\mathrm{K}}(t,t')
\pm
\left(G_k^{\mathrm{R}}(t,t')-G_k^{\mathrm{A}}(t,t')\right)
\right].
\label{eq:keldysh_inversion}
\end{gather}
Equation~\eqref{eq:RA_difference} is the two-time component identity associated with the fermionic anticommutator; its equal-time limit is fixed by the CAR [Eq.~\eqref{eq:electron_car}].
These identities are used below as consistency relations, not as identities that define the lesser and greater functions.

\subsubsection{Electron bath and Langevin representation}
\label{sec:branch_wideband}

The electron bath is kept as a continuous Gaussian bath.
For
\begin{align}
H_{\mathrm{bath}}
=
\sum_p\varepsilon_pd_p^\dagger d_p,
\quad
H_{\mathrm{hyb}}
=
\sum_p
\left(
c_k^\dagger V_pd_p+d_p^\dagger V_p^*c_k
\right),
\end{align}
the hybridization function is
\begin{align}
\varGamma(\varepsilon)
= 2\pi\sum_p|V_p|^2\delta(\varepsilon-\varepsilon_p),
\end{align}
where $\delta$ denotes the Dirac delta.
In the wide-band limit $\varGamma(\varepsilon)\to\varGamma>0$, the Lamb shift is absorbed into $h_k$, and the retarded, advanced, lesser, and greater self-energies are
\begin{gather}
\varSigma_{\mathrm{bath}}^{\mathrm{R}}(\omega)
= \varSigma_{\mathrm{bath}}^{\mathrm{A}}(\omega)^*
= -\frac{\mathrm{i}\varGamma}{2},
\label{eq:wbl_sigma_RA}
\\
\varSigma_{\mathrm{bath}}^<(\omega)
= \mathrm{i}\varGamma f_{\mathrm{bath}}(\omega),
\label{eq:wbl_sigma_lesser_freq}
\\
\varSigma_{\mathrm{bath}}^>(\omega)
= -\mathrm{i}\varGamma[1-f_{\mathrm{bath}}(\omega)],
\label{eq:wbl_sigma_enlarged}
\end{gather}
where
\begin{align}
f_{\mathrm{bath}}(\omega)
= \frac{1}{\mathrm{e}^{\beta_{\mathrm{bath}}(\omega-\mu_{\mathrm{bath}})}+1}.
\end{align}
Here, $\beta_{\mathrm{bath}}$ and $\mu_{\mathrm{bath}}$ denote the inverse temperature and chemical potential of the electron bath, respectively.
With the Fourier convention used here, the individual wide-band kernels are understood as distributions,
\begin{align}
\varSigma_{\mathrm{bath}}^<(\tau)
&=
+\frac{\mathrm{i}\varGamma}{2}\delta(\tau)
+
\bar\varSigma_{\mathrm{bath}}(\tau),
\label{eq:wbl_sigma_lesser_time}
\\
\varSigma_{\mathrm{bath}}^>(\tau)
&=
-\frac{\mathrm{i}\varGamma}{2}\delta(\tau)
+
\bar\varSigma_{\mathrm{bath}}(\tau),
\label{eq:wbl_sigma_greater_time}
\end{align}
where the $\delta(\tau)$ terms are the contact contributions and the noncontact part
\begin{align}
\bar\varSigma_{\mathrm{bath}}(\tau)
&=
\int\frac{\mathrm{d}\omega}{2\pi}\,
\mathrm{e}^{-\mathrm{i}\omega\tau}
\left[
-\frac{\mathrm{i}\varGamma}{2}
\tanh\frac{\beta_{\mathrm{bath}}(\omega-\mu_{\mathrm{bath}})}{2}
\right]
\label{eq:wbl_sigma_noncontact}
\end{align}
is common to the two components.
The difference of the two noise kernels is
\begin{gather}
\varSigma_{\mathrm{bath}}^>(\omega)
-
\varSigma_{\mathrm{bath}}^<(\omega)
=
-\mathrm{i}\varGamma,
\label{eq:bath_car_identity_freq}
\\
\varSigma_{\mathrm{bath}}^>(t-t')
-
\varSigma_{\mathrm{bath}}^<(t-t')
=
-\mathrm{i}\varGamma\delta(t-t').
\label{eq:bath_car_identity}
\end{gather}
Equations~\eqref{eq:bath_car_identity_freq} and \eqref{eq:bath_car_identity} are the bath-side consequences of the fermionic CAR for the reservoir modes.

The Heisenberg equations generated by $H_{\mathrm{bath}}+H_{\mathrm{hyb}}$ together with the Peierls-dressed probe Hamiltonian $h_k[X;t]$ read
\begin{align}
\partial_t c_k(t)
&=
-\mathrm{i} h_k[X;t]\,c_k(t)
-\mathrm{i}\sum_p V_p\,d_p(t),
\\
\partial_t d_p(t)
&=
-\mathrm{i}\varepsilon_p d_p(t)
-\mathrm{i} V_p^*\,c_k(t).
\end{align}
Integrating the bath equation from $t = -\infty$ gives
\begin{align}
d_p(t)
&=
d_p^{(0)}(t)
-
\mathrm{i}\int_{-\infty}^{t}\mathrm{d} s\,
\mathrm{e}^{-\mathrm{i}\varepsilon_p(t-s)}V_p^*\,c_k(s),
\label{eq:bath_formal_solution}
\end{align}
where $d_p^{(0)}(t)$ is the freely evolving incoming bath operator, and substitution into the probe equation produces a retarded hybridization memory acting on $c_k$ together with the additive noise
\begin{align}
\xi_k(t)
&=
-\mathrm{i}\sum_p V_p\,d_p^{(0)}(t)
\label{eq:bath_noise_def}
\end{align}
built from that incoming field.
In the wide-band limit of Eq.~\eqref{eq:wbl_sigma_RA}, the memory reduces to the local damping $-\varGamma/2$, so
\begin{align}
\partial_t c_k(t)
&=
\left[
-\mathrm{i} h_k[X;t]-\frac{\varGamma}{2}
\right]c_k(t)
+
\xi_k(t),
\label{eq:langevin_probe_equation}
\end{align}
with the electron bath statistically independent of the prescribed photon source.
Thermal averages over the initial bath state reproduce the kernels already introduced,
\begin{align}
\langle\xi_k^\dagger(s')\xi_k(s)\rangle
&=
-\mathrm{i}\varSigma_{\mathrm{bath}}^<(s-s'),
\label{eq:bath_noise_lesser}
\\
\langle\xi_k(s)\xi_k^\dagger(s')\rangle
&=
+\mathrm{i}\varSigma_{\mathrm{bath}}^>(s-s'),
\label{eq:bath_noise_greater}
\end{align}
and the bath CAR identity in Eq.~\eqref{eq:bath_car_identity} implies
\begin{align}
\{\xi_k(s),\xi_k^\dagger(s')\}
&=
\varGamma\delta(s-s').
\end{align}

\subsubsection{Shared two-leg source average}
\label{sec:duhamel_twoleg}

The two time arguments of $G_k$ define a left and a right electronic leg, which we label by the upright sans-serif indices $\mathsf{L}$ and $\mathsf{R}$.
The homogeneous left-leg propagator generated by Eq.~\eqref{eq:langevin_probe_equation} is the time-ordered source operator $U_{\mathsf{L}}(t,s)$ obeying
\begin{align}
\partial_t U_{\mathsf{L}}(t,s)
=
\left[
-\mathrm{i} h_k[X;t]-\frac{\varGamma}{2}
\right]
U_{\mathsf{L}}(t,s),
\label{eq:UL_homogeneous}
\end{align}
with $U_{\mathsf{L}}(s,s)=I_{\mathrm{ph}}$, where $I_{\mathrm{ph}}$ is the identity on the photon Hilbert space; $U_{\mathsf{R}}(t',s')$ denotes the corresponding homogeneous propagator for $c_k^\dagger$.
Because $\varGamma>0$ and the prescribed source channel is relaxing, the homogeneous contribution from the remote initial time vanishes in the periodic steady state, and the variation-of-constants (Duhamel) formula yields
\begin{align}
c_k(t)
=
\int_{-\infty}^{t}\mathrm{d} s\,
U_{\mathsf{L}}(t,s)\xi_k(s),
\label{eq:duhamel_probe}
\end{align}
together with
\begin{align}
c_k^\dagger(t')
=
\int_{-\infty}^{t'}\mathrm{d} s'\,
\xi_k^\dagger(s')\,U_{\mathsf{R}}(t',s').
\label{eq:duhamel_probe_right}
\end{align}
Each integrand therefore carries a single bath-injection time and a single observation time per electronic leg.

Substituting Eqs.~\eqref{eq:duhamel_probe} and \eqref{eq:duhamel_probe_right} into the component definitions in Eqs.~\eqref{eq:greater_def}--\eqref{eq:lesser_def} and contracting the bath with Eqs.~\eqref{eq:bath_noise_lesser}--\eqref{eq:bath_noise_greater} yields the double convolution
\begin{align}
G_k^{\gtrless}(t,t')
&=
\int_{-\infty}^{t}\mathrm{d} s
\int_{-\infty}^{t'}\mathrm{d} s'\,
\mathcal{M}_k(t,t';s,s')
\varSigma_{\mathrm{bath}}^{\gtrless}(s-s')
\label{eq:duhamel_bath_skeleton}
\end{align}
where the source factor $\mathcal{M}_k$ is a joint moment of the homogeneous propagators $U_{\mathsf{L}}(t,s)$ and $U_{\mathsf{R}}(t',s')$.
For a deterministic c-number drive, the legs factorize after the source average, $\mathcal{M}_k=\langle U_{\mathsf{L}}\rangle_{\mathrm{ph}}\langle U_{\mathsf{R}}\rangle_{\mathrm{ph}}$.
For an operator-valued Peierls link, the Langevin equation alone does not determine the shared source average: Hilbert-space products such as $\Tr_{\mathrm{ph}}(\rho\,U_{\mathsf{R}}^\dagger U_{\mathsf{L}})$ and $\Tr_{\mathrm{ph}}(\rho\,U_{\mathsf{L}}U_{\mathsf{R}}^\dagger)$ are inequivalent when the drive is noncommutative, and neither encodes the chronological overlap and idle intervals of a shared open-source history.
We therefore define $\mathcal{M}_k$ by the shared-history four-endpoint kernel $\mathcal{P}_k^{\mathrm{SH}}$ of Sec.~\ref{sec:four_endpoint_kernel}, built from the sector propagators $\mathcal{U}_q$ on the six chronological orderings, rather than by a Hilbert-space product of one-leg operators or by a product of independently averaged one-leg propagators.
The resulting formulas for $G_k^{<,>}$ are stated in Sec.~\ref{sec:bath_car}.

The causal support $s\leq t$ and $s'\leq t'$ implies that every contribution to the convolution is labeled by four endpoints $(s,t,s',t')$.
On each chronological interval, the left leg is either active (between its injection and observation) or idle, and likewise for the right leg, yielding the four active-leg sectors $q\in\{0,\mathsf{L},\mathsf{R},\mathsf{LR}\}$ indexed by the same leg labels (with $\mathsf{LR}$ for simultaneous activation of both).
Activation and deactivation of a leg occur once per integrand contribution because Eqs.~\eqref{eq:duhamel_probe} and \eqref{eq:duhamel_probe_right} already resum the wide-band hybridization into $U_{\mathsf{L}/\mathsf{R}}$ and leave a single bath-noise insertion on each leg.
The remainder of Sec.~\ref{sec:shared_history_gf} constructs that shared two-leg process: Sec.~\ref{sec:source_liouville} represents the source, Sec.~\ref{sec:shared_kraus_sectors} realizes $\mathcal{P}_k^{\mathrm{SH}}$ through the sector propagators, and Sec.~\ref{sec:bath_car} performs the bath convolution.

\subsubsection{Source Liouville space}
\label{sec:source_liouville}

Let $\mathcal{H}_{\mathrm{ph}}$ denote the source Hilbert space and
\begin{align}
\mathcal{V}_{\mathrm{ph}}
\equiv
\mathcal{B}(\mathcal{H}_{\mathrm{ph}})
\label{eq:Vph_def}
\end{align}
its operator space.
We use the subscript ``ph'' throughout to reflect the Gaussian photon realization, while the sector construction itself only requires a finite-dimensional completely positive source process, so the constructions below apply verbatim to any such source.
For an orthonormal source basis $\{|n\rangle\}$ and any $A\in\mathcal{V}_{\mathrm{ph}}$, we define the row-stacked vectorization by
\begin{align}
\Lket{A}
&\equiv
\operatorname{vec}(A)
=
\sum_{mn}A_{mn}|m\rangle\otimes|n\rangle,
\label{eq:liouville_vec_basis}
\end{align}
or, in components,
\begin{align}
\operatorname{vec}(A)=(A_{11},A_{12},\ldots,A_{1N},A_{21},\ldots,A_{NN})^{\mathsf{T}},
\end{align}
with the Hilbert--Schmidt inner product on $\mathcal{V}_{\mathrm{ph}}$
\begin{align}
\Lbraket{A}{B}
&=
\Tr_{\mathrm{ph}}(A^\dagger B).
\label{eq:liouville_trace_rule}
\end{align}
The identity vector
\begin{align}
\Lket{I_{\mathrm{ph}}}
&=
\sum_n|n\rangle\otimes|n\rangle
\label{eq:trace_vector_basis}
\end{align}
defines the trace bra,
\begin{align}
\Lbraket{I_{\mathrm{ph}}}{A} = \Tr_{\mathrm{ph}} A.
\label{eq:liouville_trace_ph}
\end{align}
Trace preservation gives $\Lbra{I_{\mathrm{ph}}}\mathbb{L}_{\mathrm{ph}}(t)=0$.
Here and below, $\mathbb{L}_{\mathrm{ph}}$ denotes the row-vectorized matrix representation of the source superoperator $\mathcal{L}_{\mathrm{ph}}$.

For an operator $O$ on $\mathcal{H}_{\mathrm{ph}}$, left and right multiplication are represented by
\begin{align}
\hat{\mathsf{L}}[O]
= O\otimes I_{\mathrm{ph}},
\quad
\hat{\mathsf{R}}[O]
= I_{\mathrm{ph}}\otimes O^{\mathsf{T}},
\label{eq:left_right_superop_def}
\end{align}
so that
\begin{align}
\operatorname{vec}(OA)=\hat{\mathsf{L}}[O]\operatorname{vec}(A), \quad
\operatorname{vec}(AO)=\hat{\mathsf{R}}[O]\operatorname{vec}(A).
\end{align}
For the nearest-neighbor model of Eq.~\eqref{eq:hk_1d},
\begin{align}
\hat{\mathsf{L}}[h_k(t)]
&=
\gamma_+(k)\hat{\mathsf{L}}[D(t)]
+
\gamma_-(k)\hat{\mathsf{L}}[D^\dagger(t)],
\label{eq:hL_singleband}
\end{align}
with $\gamma_+(k)=-t_{\mathrm{h}}\mathrm{e}^{-\mathrm{i} k}$ and $\gamma_-(k)=-t_{\mathrm{h}}\mathrm{e}^{+\mathrm{i} k}$.

\subsubsection{Shared Kraus sequence and four active-leg sectors}
\label{sec:shared_kraus_sectors}

We now construct the shared-history sector process that realizes $\mathcal{P}_k^{\mathrm{SH}}$.
Let $\mathcal{E}_j=\mathcal{U}_{\mathrm{ph}}(t_{j+1},t_j)$ be the completely positive and trace-preserving (CPTP) source propagator on a short interval $\Delta t=t_{j+1}-t_j$.
For a finite-dimensional source, it has an operator-sum representation~\cite{Kraus1971, Choi1975}
\begin{align}
\mathcal{E}_j(A)
= \sum_{\alpha_j} K_{j\alpha_j}A K_{j\alpha_j}^\dagger,
\quad
\sum_{\alpha_j} K_{j\alpha_j}^\dagger K_{j\alpha_j}
&= I_{\mathrm{ph}}.
\label{eq:source_kraus_ops}
\end{align}
Define the probe step
\begin{align}
V_j
&=
\exp[-\mathrm{i} h_k(t_j)\Delta t]
\label{eq:probe_step}
\end{align}
and active-leg occupation indicators $n_{\mathsf{L}},n_{\mathsf{R}}\in\{0,1\}$, equal to one when the corresponding left or right probe leg is active on that step.
The elementary shared-history map is
\begin{align}
\mathcal{E}_j^{n_{\mathsf{L}}n_{\mathsf{R}}}(A)
&=
\mathrm{e}^{-(n_{\mathsf{L}}+n_{\mathsf{R}})\varGamma \Delta t/2}
\sum_{\alpha_j}
K_{j\alpha_j}
V_j^{n_{\mathsf{L}}} A V_j^{n_{\mathsf{R}}\dagger}
K_{j\alpha_j}^\dagger.
\label{eq:common_kraus_map}
\end{align}
The same Kraus label $\alpha_j$ is used on the left and right.
The resulting shared Kraus sequence $\{\alpha_j\}$ is the discrete realization of the shared source history: both legs are driven by one prescribed source through the same sequence of Kraus events.
Choosing independent Kraus labels would define a different model and would destroy the shared-history multipoint correlations.

Equation~\eqref{eq:common_kraus_map} uses a common Kraus realization on the left and right probe-coherence blocks, so the four sectors $(0,\mathsf{L},\mathsf{R},\mathsf{LR})$ are shared-history influence-kernel blocks of a single probe-controlled source propagator whose full block action is completely positive, rather than independent CPTP channels.
These sectors organize the two-point covariance construction of Sec.~\ref{sec:bath_car} and do not by themselves define quantum-regression correlators of a separately specified joint electron--source Lindblad dynamics.
The fermionic reservoir remains an independent Gaussian input and enters only afterwards, through the contraction with $\varSigma_{\mathrm{bath}}^{<,>}$.

Expanding Eq.~\eqref{eq:common_kraus_map} for small $\Delta t$, with $\mathcal{E}_j=I+\Delta t\,\mathcal{L}_{\mathrm{ph}}+\mathcal{O}(\Delta t^2)$ and $V_j=I-\mathrm{i} h_k(t_j)\Delta t+\mathcal{O}(\Delta t^2)$, gives $\mathcal{E}_j^{n_{\mathsf{L}}n_{\mathsf{R}}}=I+\Delta t\,\mathcal{K}_q+\mathcal{O}(\Delta t^2)$ for the active-leg configuration $q$ associated with $(n_{\mathsf{L}},n_{\mathsf{R}})$.
The resulting generators are
\begin{align}
\mathcal{K}_0(t)
&=
\mathbb{L}_{\mathrm{ph}}(t),
\label{eq:sector_generator_zero}
\\
\mathcal{K}_{\mathsf{L}}(t)
&=
\mathbb{L}_{\mathrm{ph}}(t)
-\mathrm{i}\hat{\mathsf{L}}[h_k(t)]
-\frac{\varGamma}{2},
\label{eq:branch_generator_left}
\\
\mathcal{K}_{\mathsf{R}}(t)
&=
\mathbb{L}_{\mathrm{ph}}(t)
+\mathrm{i}\hat{\mathsf{R}}[h_k(t)]
-\frac{\varGamma}{2},
\label{eq:branch_generator_right}
\\
\mathcal{K}_{\mathsf{LR}}(t)
&=
\mathbb{L}_{\mathrm{ph}}(t)
-\mathrm{i}\hat{\mathsf{L}}[h_k(t)]
+\mathrm{i}\hat{\mathsf{R}}[h_k(t)]
-\varGamma.
\label{eq:branch_generator_overlap}
\end{align}
The source Liouvillian occurs once, rather than twice, in the overlap generator, so the $\mathsf{LR}$ sector propagates both legs on one source history.
For $q\in\{0,\mathsf{L},\mathsf{R},\mathsf{LR}\}$, the causal sector propagator is
\begin{align}
\mathcal{U}_q(t_2,t_1)
&=
\mathcal{T}
\exp\left[
\int_{t_1}^{t_2}\mathrm{d} u\,\mathcal{K}_q(u)
\right],
\label{eq:sector_propagator}
\end{align}
where $\mathcal{T}$ denotes chronological time ordering and $t_2\ge t_1$.
Hermiticity of $h_k$ implies
\begin{align}
\left[
\mathcal{U}_{\mathsf{L}}(t,s)A
\right]^\dagger
&=
\mathcal{U}_{\mathsf{R}}(t,s)A^\dagger.
\label{eq:sector_adjoint}
\end{align}
Since $\mathcal{K}_{\mathsf{LR}}+\varGamma$ is a CPTP generator, $\mathcal{U}_{\mathsf{LR}}(t_2,t_1)$ factors as $e^{-\varGamma(t_2-t_1)}$ times a CPTP map and is therefore positive and trace-decreasing.

Activation and deactivation maps are identity embeddings on the source operator that only change the sector label; they implement the once-per-leg endpoint switches of Sec.~\ref{sec:duhamel_twoleg}.
For example, $\mathcal{A}_{\mathsf{L}}$ maps $0\to\mathsf{L}$ or $\mathsf{R}\to\mathsf{LR}$, while $\mathcal{D}_{\mathsf{L}}$ maps $\mathsf{L}\to0$ or $\mathsf{LR}\to\mathsf{R}$.
The right-leg maps act analogously.
For coincident left- and right-bath injections, we define the simultaneous activation
\begin{align}
\mathcal{A}_{\mathsf{LR}}
\equiv \mathcal{A}_{\mathsf{R}}\mathcal{A}_{\mathsf{L}}
= \mathcal{A}_{\mathsf{L}}\mathcal{A}_{\mathsf{R}},
\label{eq:simultaneous_activation}
\end{align}
which maps the source operator directly from the $0$ sector to the $\mathsf{LR}$ sector.
Since the activation maps are identity embeddings and $\mathcal{U}_q(s,s)=I$, this prescription agrees with either one-sided limit $s'\to s^\pm$.

\subsubsection{Four-endpoint kernel and six chronological orderings}
\label{sec:four_endpoint_kernel}

\begin{figure*}[t]\centering
\includegraphics[scale=1]{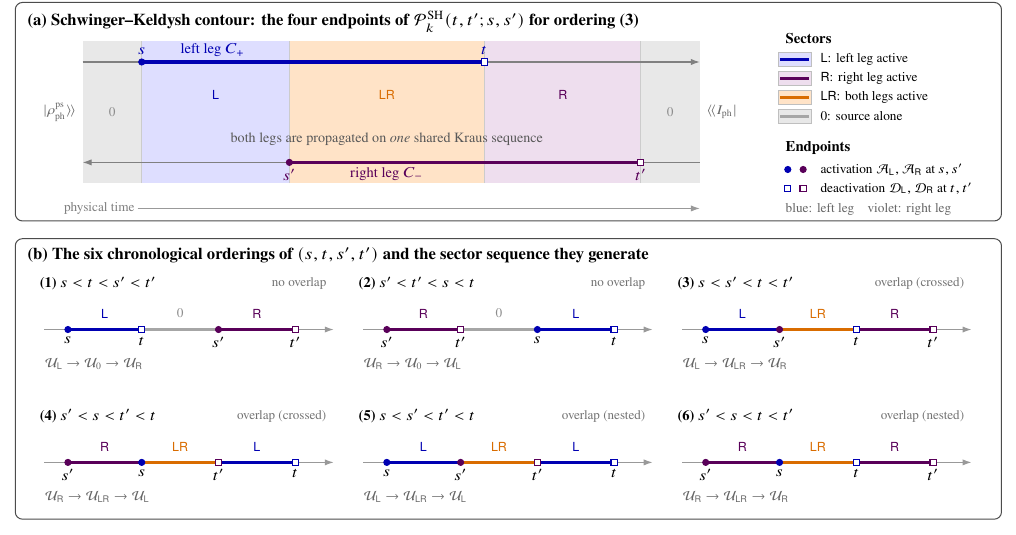}
\caption{Schwinger--Keldysh representation of the shared-history two-leg kernel.
(a)~Contour for the four-endpoint kernel $\mathcal{P}_k^{\mathrm{SH}}(t,t';s,s')$ of Eq.~\eqref{eq:four_endpoint_kernel}, drawn for ordering~(3), $s<s'<t<t'$.
The left ($C_+$) and right ($C_-$) legs are switched on at $s$ and $s'$ by $\mathcal{A}_{\mathsf{L}}$ and $\mathcal{A}_{\mathsf{R}}$ (filled circles) and switched off at $t$ and $t'$ by $\mathcal{D}_{\mathsf{L}}$ and $\mathcal{D}_{\mathsf{R}}$ (open squares).
The kernel starts from $\Lket{\rho_{\mathrm{ph}}^{\mathrm{ps}}(t_<)}$ at $t_<=\min(s,s')$ and is closed by $\Lbra{I_{\mathrm{ph}}}$ at $\max(t,t')$; both legs are propagated on one shared source history, and the shading marks the active sector $q\in\{0,\mathsf{L},\mathsf{R},\mathsf{LR}\}$.
(b)~The six chronological domains of $(s,t,s',t')$ with the corresponding sector sequences.
Orderings~(1) and (2) are nonoverlap contributions with an intervening source-only interval, while orderings~(3)--(6) contain a genuine $\mathsf{LR}$ overlap, either crossed or nested.
The occupied Green's functions follow by convolving $\mathcal{P}_k^{\mathrm{SH}}$ with $\varSigma_{\mathrm{bath}}^{<,>}$.}
\label{fig:common_source_sk}
\end{figure*}

Following the injection and observation endpoints of Sec.~\ref{sec:duhamel_twoleg}, let $s,s'$ be the bath injection times associated with the left and right legs, respectively, and let $t,t'$ be the corresponding observation times.
The causal constraints are
\begin{align}
s\leq t,
\quad
s'\leq t'.
\end{align}
Starting from the periodic source state at $t_<\equiv\min(s,s')$, the ordered product of sector propagators and endpoint maps is denoted by $\mathfrak{P}_k(t,t';s,s')$.
Closing with the source trace yields the scalar shared-history kernel
\begin{align}
\mathcal{P}_k^{\mathrm{SH}}(t,t';s,s')
&=
\Lbra{I_{\mathrm{ph}}}
\mathfrak{P}_k(t,t';s,s')
\Lket{\rho_{\mathrm{ph}}^{\mathrm{ps}}(t_<)},
\label{eq:four_endpoint_kernel}
\end{align}
which realizes $\mathcal{M}_k$ in Eq.~\eqref{eq:duhamel_bath_skeleton}.

The two causal intervals permit exactly six strict chronological orderings.
For $s<t<s'<t'$,
\begin{align}
\mathfrak{P}_k^{(1)}
&=
\mathcal{D}_{\mathsf{R}}
\mathcal{U}_{\mathsf{R}}(t',s')
\mathcal{A}_{\mathsf{R}}
\mathcal{U}_0(s',t)
\mathcal{D}_{\mathsf{L}}
\mathcal{U}_{\mathsf{L}}(t,s)
\mathcal{A}_{\mathsf{L}}.
\label{eq:ordering_one}
\end{align}
For $s'<t'<s<t$,
\begin{align}
\mathfrak{P}_k^{(2)}
&=
\mathcal{D}_{\mathsf{L}}
\mathcal{U}_{\mathsf{L}}(t,s)
\mathcal{A}_{\mathsf{L}}
\mathcal{U}_0(s,t')
\mathcal{D}_{\mathsf{R}}
\mathcal{U}_{\mathsf{R}}(t',s')
\mathcal{A}_{\mathsf{R}}.
\label{eq:ordering_two}
\end{align}
These are the two nonoverlap orderings and contain an intervening source-only propagator $\mathcal{U}_0$.
On that idle interval the source relaxes under $\mathcal{U}_0$ itself; no projector onto $\rho_{\mathrm{ph}}^{\mathrm{ps}}$ is inserted inside any sector interval, and $\rho_{\mathrm{ph}}^{\mathrm{ps}}$ enters the kernel only as the initial condition at $t_<$ in Eq.~\eqref{eq:four_endpoint_kernel}.
The remaining four orderings contain a genuine $\mathsf{LR}$ overlap interval, either crossed or nested.
For $s<s'<t<t'$,
\begin{align}
\mathfrak{P}_k^{(3)}
&=
\mathcal{D}_{\mathsf{R}}
\mathcal{U}_{\mathsf{R}}(t',t)
\mathcal{D}_{\mathsf{L}}
\mathcal{U}_{\mathsf{LR}}(t,s')
\mathcal{A}_{\mathsf{R}}
\mathcal{U}_{\mathsf{L}}(s',s)
\mathcal{A}_{\mathsf{L}}.
\label{eq:ordering_three}
\end{align}
For $s'<s<t'<t$,
\begin{align}
\mathfrak{P}_k^{(4)}
&=
\mathcal{D}_{\mathsf{L}}
\mathcal{U}_{\mathsf{L}}(t,t')
\mathcal{D}_{\mathsf{R}}
\mathcal{U}_{\mathsf{LR}}(t',s)
\mathcal{A}_{\mathsf{L}}
\mathcal{U}_{\mathsf{R}}(s,s')
\mathcal{A}_{\mathsf{R}}.
\label{eq:ordering_four}
\end{align}
For $s<s'<t'<t$,
\begin{align}
\mathfrak{P}_k^{(5)}
&=
\mathcal{D}_{\mathsf{L}}
\mathcal{U}_{\mathsf{L}}(t,t')
\mathcal{D}_{\mathsf{R}}
\mathcal{U}_{\mathsf{LR}}(t',s')
\mathcal{A}_{\mathsf{R}}
\mathcal{U}_{\mathsf{L}}(s',s)
\mathcal{A}_{\mathsf{L}}.
\label{eq:ordering_five}
\end{align}
For $s'<s<t<t'$,
\begin{align}
\mathfrak{P}_k^{(6)}
&=
\mathcal{D}_{\mathsf{R}}
\mathcal{U}_{\mathsf{R}}(t',t)
\mathcal{D}_{\mathsf{L}}
\mathcal{U}_{\mathsf{LR}}(t,s)
\mathcal{A}_{\mathsf{L}}
\mathcal{U}_{\mathsf{R}}(s,s')
\mathcal{A}_{\mathsf{R}}.
\label{eq:ordering_six}
\end{align}
Figure~\ref{fig:common_source_sk} summarizes the Schwinger--Keldysh contour with the shared source history and the six chronological domains.

Let $\mathcal{P}_k^{\mathrm{SH},(j)}=\Lbra{I_{\mathrm{ph}}}\mathfrak{P}_k^{(j)}\Lket{\rho_{\mathrm{ph}}^{\mathrm{ps}}(t_<)}$ be the scalar kernel built from the ordered product $\mathfrak{P}_k^{(j)}$ of ordering $j$.
Denote by $\bigl(u_1^{(j)},u_2^{(j)},u_3^{(j)},u_4^{(j)}\bigr)$ the permutation of $(s,s',t,t')$ in the sequence in which ordering $j$ lists them in Eqs.~\eqref{eq:ordering_one}--\eqref{eq:ordering_six}.
For instance, ordering $j=1$ lists $(s,t,s',t')$ and ordering $j=3$ lists $(s,s',t,t')$.
Ordering $j$ is then the condition $u_1^{(j)}<u_2^{(j)}<u_3^{(j)}<u_4^{(j)}$, and its indicator function is
\begin{align}
\chi_j(t,t';s,s')
&\equiv
\mathbf{1}_{\{
u_1^{(j)}<u_2^{(j)}<u_3^{(j)}<u_4^{(j)}
\}},
\label{eq:ordering_indicator}
\end{align}
which equals unity on that open chronological domain and vanishes otherwise; for example $\chi_1=\mathbf{1}_{\{s<t<s'<t'\}}$.
The six indicators partition the causal region away from the endpoint-coincidence hyperplanes; they label strict chronological domains and do not assign the coincidence set.
The kernel then decomposes as
\begin{align}
\mathcal{P}_k^{\mathrm{SH}}(t,t';s,s')
&=
\sum_{j=1}^{6}
\chi_j(t,t';s,s')
\mathcal{P}_k^{\mathrm{SH},(j)}(t,t';s,s').
\label{eq:six_ordering_decomposition}
\end{align}
On the bath-time diagonal $s=s'$, the kernel is defined by the simultaneous activation of Eq.~\eqref{eq:simultaneous_activation}, with zero-duration intervals represented by $\mathcal{U}_q(s,s)=I$; the contact weight $\varGamma$ enters through the bath CAR identity below.

\subsubsection{Bath convolution and the CAR identity}
\label{sec:bath_car}

\paragraph*{Occupied components.}
With the four-endpoint kernel in hand, we return to the occupied components anticipated in Sec.~\ref{sec:duhamel_twoleg}.
Closing the source factor in Eq.~\eqref{eq:duhamel_bath_skeleton} with the shared-history kernel,
\begin{align}
\mathcal{M}_k(t,t';s,s')
=
\mathcal{P}_k^{\mathrm{SH}}(t,t';s,s'),
\label{eq:shared_identified_with_P}
\end{align}
gives
\begin{align}
G_k^{\gtrless}(t,t')
&=
\int_{-\infty}^{t}\mathrm{d} s
\int_{-\infty}^{t'}\mathrm{d} s'\,
\mathcal{P}_k^{\mathrm{SH}}(t,t';s,s')
\varSigma_{\mathrm{bath}}^{\gtrless}(s-s').
\label{eq:keldysh_equation}
\end{align}
The fermionic-bath average is a Gaussian Wick contraction; the source factor is the shared Kraus kernel constructed above.

\paragraph*{Equal-time CAR covariance.}
The lesser and greater components are already fixed by the shared-history convolution in Eq.~\eqref{eq:keldysh_equation}.
Their difference is therefore fixed by the bath CAR identity in Eq.~\eqref{eq:bath_car_identity}.
The $\delta$-supported parts of Eqs.~\eqref{eq:wbl_sigma_lesser_time} and \eqref{eq:wbl_sigma_greater_time} collapse the convolution to the bath-time diagonal $s=s'$.
For $t\geq t' \geq s$, they give
\begin{align}
G_{k,\mathrm{ct}}^<(t,t')
&=
+\frac{\mathrm{i}\varGamma}{2}
\int_{-\infty}^{t'}\mathrm{d} s\,
\mathcal{P}_k^{\mathrm{SH}}(t,t';s,s),
\\
G_{k,\mathrm{ct}}^>(t,t')
&=
-\frac{\mathrm{i}\varGamma}{2}
\int_{-\infty}^{t'}\mathrm{d} s\,
\mathcal{P}_k^{\mathrm{SH}}(t,t';s,s),
\end{align}
in which the shared-history kernel $\mathcal{P}_k^{\mathrm{SH}}$ reduces to
\begin{align}
\mathcal{P}_k^{\Delta}(t,t';s)
&= \Lbra{I_{\mathrm{ph}}}
\mathcal{U}_{\mathsf{L}}(t,t')
\mathcal{U}_{\mathsf{LR}}(t',s)
\Lket{\rho_{\mathrm{ph}}^{\mathrm{ps}}(s)}.
\label{eq:P_delta}
\end{align}
The common noncontact contribution in Eq.~\eqref{eq:wbl_sigma_noncontact} cancels in their difference, yielding
\begin{align}
G_k^>(t,t')-G_k^<(t,t')
&=
-\mathrm{i}\varGamma
\int_{-\infty}^{t'}\mathrm{d} s\,
\mathcal{P}_k^{\Delta}(t,t';s),
\end{align}
and hence
\begin{align}
&G_k^>(t,t')-G_k^<(t,t')
\notag\\
&\quad=
-\mathrm{i}\varGamma
\int_{-\infty}^{t'}\mathrm{d} s\,
\Lbra{I_{\mathrm{ph}}}
\mathcal{U}_{\mathsf{L}}(t,t')
\mathcal{U}_{\mathsf{LR}}(t',s)
\Lket{\rho_{\mathrm{ph}}^{\mathrm{ps}}(s)}
\notag\\
&\quad=
-\mathrm{i}
\Lbra{I_{\mathrm{ph}}}
\mathcal{U}_{\mathsf{L}}(t,t')
\Lket{W_k(t')},
\label{eq:component_difference_from_Wk}
\end{align}
where the periodic equal-time CAR covariance
\begin{align}
W_k(t)
&=
\varGamma
\int_{-\infty}^{t}\mathrm{d} s\,
\mathcal{U}_{\mathsf{LR}}(t,s)
\rho_{\mathrm{ph}}^{\mathrm{ps}}(s)
\label{eq:equal_time_covariance}
\end{align}
is the unique $T$-periodic solution of
\begin{gather}
\partial_t W_k(t)
= \mathcal{K}_{\mathsf{LR}}(t)W_k(t) + \varGamma\rho_{\mathrm{ph}}^{\mathrm{ps}}(t),
\label{eq:covariance_equation}
\\
W_k(t+T) = W_k(t).
\label{eq:covariance_periodicity}
\end{gather}
Since $\mathcal{K}_{\mathsf{LR}}+\varGamma$ generates a CPTP evolution and $\varGamma>0$, the homogeneous propagator is exponentially stable, so the periodic solution is unique.
Positivity of $\mathcal{U}_{\mathsf{LR}}$ gives $W_k(t)\succeq0$, Hermiticity preservation gives $W_k^\dagger(t)=W_k(t)$, and the source trace gives
\begin{align}
\Tr_{\mathrm{ph}} W_k(t)
= \Lbraket{I_{\mathrm{ph}}}{W_k(t)}
= 1.
\label{eq:covariance_normalization}
\end{align}
Using $\partial_t\rho_{\mathrm{ph}}^{\mathrm{ps}}(t)=\mathcal{L}_{\mathrm{ph}}(t)\rho_{\mathrm{ph}}^{\mathrm{ps}}(t)$, one obtains
\begin{align}
W_k(t)-\rho_{\mathrm{ph}}^{\mathrm{ps}}(t)
&=
-\mathrm{i}\int_{-\infty}^{t}\mathrm{d} s\,
\mathcal{U}_{\mathsf{LR}}(t,s)
\bigl[h_k(s),\rho_{\mathrm{ph}}^{\mathrm{ps}}(s)\bigr],
\label{eq:Wk_minus_rhops}
\end{align}
so $W_k=\rho_{\mathrm{ph}}^{\mathrm{ps}}$ whenever $[h_k,\rho_{\mathrm{ph}}^{\mathrm{ps}}]=0$, including the classical or commutative limit.
In general, $W_k(t)\ne\rho_{\mathrm{ph}}^{\mathrm{ps}}(t)$: $W_k$ is the equal-time electronic CAR covariance generated by the bath anticommutator on the shared source history, whereas $\rho_{\mathrm{ph}}^{\mathrm{ps}}$ remains the prescribed photon state.

\paragraph*{Retarded and advanced components.}
The component identity in Eq.~\eqref{eq:retarded_def} then forces the retarded one-leg formula
\begin{align}
G_k^{\mathrm{R}}(t,t')
&=
-\mathrm{i}\varTheta(t-t')
\Lbra{I_{\mathrm{ph}}}
\mathcal{U}_{\mathsf{L}}(t,t')
\Lket{W_k(t')},
\label{eq:retarded_from_X}
\end{align}
with the advanced partner
\begin{align}
G_k^{\mathrm{A}}(t,t')
&=
+\mathrm{i}\varTheta(t'-t)
\Lbra{I_{\mathrm{ph}}}
\mathcal{U}_{\mathsf{R}}(t',t)
\Lket{W_k(t)}.
\label{eq:advanced_from_X}
\end{align}
For $f_{\mathrm{bath}}=0$, one has $\varSigma_{\mathrm{bath}}^<=0$ and hence $G_k^<=0$.
Equation~\eqref{eq:component_difference_from_Wk} then gives $G_k^{\mathrm{R}}=G_k^>$ for $t>t'$, as required by the component definitions.
The retarded and advanced components above are fixed by the shared-history kernels and the bath CAR identity, with seed $W_k$.
They are not identified with the quantum-regression correlators of a joint electron--photon Lindblad dynamics in which the fermionic bath is replaced by jump operators alone; in the empty-bath Markovian case, that dynamics yields the seed $\rho_{\mathrm{ph}}^{\mathrm{ps}}$ instead, and the two two-point processes differ whenever $W_k\neq\rho_{\mathrm{ph}}^{\mathrm{ps}}$.

For numerical work, it is convenient to denote
\begin{align}
\Lket{\mathcal{X}^{\mathrm{R}}(t,t')}
&=
\mathcal{U}_{\mathsf{L}}(t,t')
\Lket{W_k(t')},
\label{eq:left_branch_propagator}
\end{align}
so that
\begin{gather}
\partial_t\Lket{\mathcal{X}^{\mathrm{R}}(t,t')}
= \mathcal{K}_{\mathsf{L}}(t)
\Lket{\mathcal{X}^{\mathrm{R}}(t,t')},
\label{eq:retarded_aux_eom_dyn}
\\
\Lket{\mathcal{X}^{\mathrm{R}}(t',t')}
= \Lket{W_k(t')},
\label{eq:retarded_aux_eom}
\end{gather}
and
\begin{align}
G_k^{\mathrm{R}}(t,t')
&=
-\mathrm{i}\varTheta(t-t')
\Lbraket{I_{\mathrm{ph}}}{\mathcal{X}^{\mathrm{R}}(t,t')}.
\label{eq:photon_trace_readout}
\end{align}
The equal-time limit is fixed by $\Tr_{\mathrm{ph}}W_k=1$.

\subsubsection{Positivity, fermionic bounds, and quasifree reconstruction}
\label{sec:operator_reconstruction}

The Green's functions constructed above arise from shared-history averages and bath convolutions, not from Heisenberg correlators of a joint electron--photon Lindblad dynamics.
We therefore prove three two-point properties that are not self-evident for this construction: positivity of the lesser and greater kernels, the Pauli bound $0\leq n_k(t)\le1$, and a gauge-invariant quasifree (fermionic Gaussian) CAR realization.
Positivity and exact sum rules for retarded Floquet spectral densities have been established for periodically driven closed systems with a classical drive \cite{Uhrig2019}; the argument below addresses the lesser and greater kernels, the Pauli bound, and the CAR in the present setting.

\paragraph*{Shared-Kraus positivity and Pauli bound.}
The same shared Kraus sequence that defines the four sectors also supplies a direct positivity proof.
For binary left- and right-leg histories $\bm{a}=(a_0,\ldots,a_{N-1})$ and $\bm{b}=(b_0,\ldots,b_{N-1})$ with $a_j,b_j\in\{0,1\}$, define
\begin{align}
M_{\bm{\alpha}}[\bm{a}]
&=
K_{N-1,\alpha_{N-1}}V_{N-1}^{a_{N-1}}
\cdots
K_{0,\alpha_0}V_0^{a_0}.
\end{align}
Here, $K_{j\alpha_j}$ are the source Kraus operators of Eq.~\eqref{eq:source_kraus_ops} and $V_j$ is the probe step of Eq.~\eqref{eq:probe_step}, as in the elementary shared-history map of Eq.~\eqref{eq:common_kraus_map}.
Likewise, define $M_{\bm{\alpha}}[\bm{b}]$ by the replacement $a_j\to b_j$.
After removing the scalar electronic damping, the discrete source influence functional is
\begin{align}
\varPhi_\Delta[\bm{a},\bm{b}]
&=
\sum_{\bm{\alpha}}
\Tr_{\mathrm{ph}}
\left[
M_{\bm{\alpha}}[\bm{a}]
\rho_{\mathrm{ph}}
M_{\bm{\alpha}}[\bm{b}]^\dagger
\right].
\end{align}
For arbitrary histories $\bm{a}^{(r)}$ and coefficients $z_r\in\mathbb{C}$,
\begin{align}
\sum_{rs}z_r^*z_s
\varPhi_\Delta[\bm{a}^{(r)},\bm{a}^{(s)}]
&=
\sum_{\bm{\alpha}}
\Tr_{\mathrm{ph}}
\left[
Y_{\bm{\alpha}}\rho_{\mathrm{ph}} Y_{\bm{\alpha}}^\dagger
\right]
\ge0,
\end{align}
where $Y_{\bm{\alpha}}=\sum_r z_r^*M_{\bm{\alpha}}[\bm{a}^{(r)}]$.
The equality follows by substituting the definition of $\varPhi_\Delta$, interchanging the finite sums over $r,s$ and $\bm{\alpha}$, and collecting the path operators into $Y_{\bm{\alpha}}$.
Each summand is nonnegative whenever $\rho_{\mathrm{ph}}\ge0$.
Thus $\varPhi_\Delta$ is a Gram kernel on arbitrary binary left- and right-leg histories, including those with multiple activations and deactivations on either leg.
The physical four-endpoint kernel $\mathcal{P}_k^{\mathrm{SH}}$ corresponds to the restriction to a single contiguous activation window on each leg, labeled by $(s,t)$ and $(s',t')$; any such restriction of a Gram kernel remains a Gram kernel, and the continuum limit inherits that property.
The diagonal identity $\varPhi_\Delta[\bm{a},\bm{a}]=1$ follows from trace preservation.

The bath kernels
\begin{gather}
C_{\mathrm{bath}}^<(s,s')
= -\mathrm{i}\varSigma_{\mathrm{bath}}^<(s-s'), \\
C_{\mathrm{bath}}^>(s,s')
= +\mathrm{i}\varSigma_{\mathrm{bath}}^>(s-s')
\end{gather}
are positive definite as kernels in the time arguments $(s,s')$, because their Fourier densities are $\varGamma f_{\mathrm{bath}}\ge0$ and $\varGamma(1-f_{\mathrm{bath}})\ge0$.
By Eq.~\eqref{eq:keldysh_equation}, the electronic covariance kernels are the contractions
\begin{align}
-\mathrm{i} G_k^<(t,t')
&=
\int_{-\infty}^{t}\mathrm{d} s
\int_{-\infty}^{t'}\mathrm{d} s'\,
\mathcal{P}_k^{\mathrm{SH}}(t,t';s,s')
C_{\mathrm{bath}}^<(s,s'),
\\
+\mathrm{i} G_k^>(t,t')
&=
\int_{-\infty}^{t}\mathrm{d} s
\int_{-\infty}^{t'}\mathrm{d} s'\,
\mathcal{P}_k^{\mathrm{SH}}(t,t';s,s')
C_{\mathrm{bath}}^>(s,s').
\end{align}
Given the Gram property of $\mathcal{P}_k^{\mathrm{SH}}$, both $\mathcal{P}_k^{\mathrm{SH}}$ in the compound labels $(t,s)$ and $C_{\mathrm{bath}}^{<,>}$ in $(s,s')$ are positive-semidefinite kernels.
The Schur product theorem states that the entrywise product of two positive-semidefinite Gram matrices is again positive semidefinite, and therefore yields
\begin{align}
\int\mathrm{d} t\,\mathrm{d} t'\,
g^*(t)
\left[-\mathrm{i} G_k^<(t,t')\right]
g(t')
&\ge0,
\\
\int\mathrm{d} t\,\mathrm{d} t'\,
g^*(t)
\left[+\mathrm{i} G_k^>(t,t')\right]
g(t')
&\ge0
\end{align}
for every compactly supported test function $g$.
At equal time,
\begin{gather}
n_k(t)
= -\mathrm{i} G_k^<(t,t)
\ge0,
\\
1-n_k(t)
= +\mathrm{i} G_k^>(t,t)
\ge0,
\end{gather}
and hence
\begin{align}
0
\le
n_k(t)
\le
1.
\label{eq:pauli_bound}
\end{align}
The empty-bath limit gives $G^<=0$, and the full-bath limit gives $G^>=0$, independently of the source state and coupling.

\paragraph*{Operator reconstruction.}
We now show that the positive lesser and greater kernels obtained above admit an explicit realization as two-point correlation functions of fermionic operators on a CAR algebra, in the sense of a gauge-invariant quasifree (fermionic Gaussian) reconstruction \cite{ArakiWyss1964, Araki1970}.
Write the two-time covariance kernels as
\begin{align}
\mathcal{C}_k^<(t,t')&\equiv-\mathrm{i} G_k^<(t,t'),
\quad
\mathcal{C}_k^>(t,t')\equiv+\mathrm{i} G_k^>(t,t').
\label{eq:two_time_covariance_kernels}
\end{align}
For compactly supported test functions $f,g\in C_c^\infty(\mathbb{R})$, define the sesquilinear forms
\begin{align}
\mathcal{C}_{k}^{<}(f,g)
&\equiv
\int\mathrm{d} t\,\mathrm{d} t'\,
f^*(t)\left[-\mathrm{i} G_k^<(t,t')\right]g(t'),
\label{eq:C_less_smeared}
\\
\mathcal{C}_{k}^{>}(f,g)
&\equiv
\int\mathrm{d} t\,\mathrm{d} t'\,
f^*(t)\left[+\mathrm{i} G_k^>(t,t')\right]g(t').
\label{eq:C_greater_smeared}
\end{align}
The shared-Kraus Gram argument and positivity of the bath kernels imply
\begin{align}
\mathcal{C}_{k}^{<}(f,f)\geq0,
\quad
\mathcal{C}_{k}^{>}(f,f)\geq0.
\label{eq:C_positive_forms}
\end{align}
Their sum,
\begin{align}
\mathcal{A}_{k}(f,g)
&\equiv
\mathcal{C}_{k}^{<}(f,g)
+
\mathcal{C}_{k}^{>}(f,g)
\notag\\
&=
\int\mathrm{d} t\,\mathrm{d} t'\,
f^*(t)\,
\mathrm{i}\left[
G_k^>(t,t')-G_k^<(t,t')
\right]g(t')
\notag\\
&=
\int\mathrm{d} t\,\mathrm{d} t'\,
f^*(t)\,
\mathrm{i}\left[
G_k^{\mathrm{R}}(t,t')-G_k^{\mathrm{A}}(t,t')
\right]g(t'),
\label{eq:A_car_form}
\end{align}
is therefore positive semidefinite.
Moreover,
\begin{align}
0\leq \mathcal{C}_{k}^{<}\leq\mathcal{A}_{k}
\label{eq:C_less_bounded_by_A}
\end{align}
as quadratic forms, because $\mathcal{A}_{k}-\mathcal{C}_{k}^{<}=\mathcal{C}_{k}^{>}\geq0$.

Let
\begin{align}
\mathcal{N}_k
&=
\left\{
f\in C_c^\infty(\mathbb{R})
\ \middle|\
\mathcal{A}_k(f,f)=0
\right\},
\end{align}
and let $\mathcal{H}_k$ be the Hilbert-space completion of $C_c^\infty(\mathbb{R})/\mathcal{N}_k$ with respect to the inner product
\begin{align}
\langle[f],[g]\rangle_{\mathcal{H}_k}
=
\mathcal{A}_k(f,g),
\label{eq:Hk_inner_product}
\end{align}
where $[f]$ denotes the equivalence class of $f$ in the quotient.
The Cauchy--Schwarz inequality for the positive form $\mathcal{C}_k^<$ and Eq.~\eqref{eq:C_less_bounded_by_A} show that $\mathcal{C}_k^<$ vanishes on $\mathcal{N}_k$ and hence descends to $\mathcal{H}_k$.
The Riesz representation theorem then gives a unique bounded operator $S_k$ on $\mathcal{H}_k$ such that
\begin{align}
\mathcal{C}_k^<(f,g)
&=
\langle[f],S_k[g]\rangle_{\mathcal{H}_k},
\quad
0\leq S_k\leq I_{\mathcal{H}_k},
\label{eq:Sk_covariance_operator}
\end{align}
where $I_{\mathcal{H}_k}$ denotes the identity operator on $\mathcal{H}_k$.

An explicit CAR representation can now be given.
Let $J:\mathcal{H}_k\to\overline{\mathcal{H}_k}$ be the canonical antiunitary, and let $\psi(\bullet)$ be the canonical annihilation operator on the fermionic Fock space
\begin{align}
\mathcal{F}_-\left(
\mathcal{H}_k\oplus\overline{\mathcal{H}_k}
\right),
\end{align}
with Fock vacuum $|\Omega\rangle$.
The operator $\psi(\bullet)$ is antilinear in its argument.
For $\xi,\eta\in\mathcal{H}_k\oplus\overline{\mathcal{H}_k}$ it satisfies
\begin{align}
\{\psi(\xi),\psi^\dagger(\eta)\}
&=
\langle\xi,\eta\rangle_{\mathcal{H}_k\oplus\overline{\mathcal{H}_k}},
\quad
\{\psi(\xi),\psi(\eta)\}=0,
\label{eq:fock_car_psi}
\end{align}
and $\psi(\xi)|\Omega\rangle=0$.
Define the smeared fermionic operator
\begin{align}
c_k[f]
&\equiv
\psi\left(
(I_{\mathcal{H}_k}-S_k)^{1/2}[f]\oplus0
\right)
+
\psi^\dagger\left(
0\oplus J S_k^{1/2}[f]
\right).
\label{eq:explicit_car_reconstruction}
\end{align}
Equation~\eqref{eq:fock_car_psi} then gives
\begin{gather}
\{ c_k[f],c_k^\dagger[g] \}
= \mathcal{A}_k(f,g),
\label{eq:reconstructed_car}
\\
\{c_k[f],c_k[g]\} =0.
\end{gather}
The vacuum two-point functions are
\begin{align}
\langle\Omega|
c_k^\dagger[g]c_k[f]
|\Omega\rangle
&=
\mathcal{C}_k^<(f,g),
\label{eq:reconstructed_lesser}
\\
\langle\Omega|
c_k[f]c_k^\dagger[g]
|\Omega\rangle
&=
\mathcal{C}_k^>(f,g).
\label{eq:reconstructed_greater}
\end{align}
Equations~\eqref{eq:C_less_smeared}--\eqref{eq:C_greater_smeared} therefore become the ordinary operator definitions
\begin{align}
G_k^<(f,g)
&= +\mathrm{i} \langle\Omega | c_k^\dagger[g]c_k[f] | \Omega\rangle,
\\
G_k^>(f,g)
&= -\mathrm{i} \langle\Omega | c_k[f]c_k^\dagger[g] | \Omega\rangle.
\end{align}
For a finite time discretization, these are ordinary finite-dimensional kernel relations.
In the continuum, they are understood first for smeared fields; when the two-time kernels are regular, the notation $c_k(t)$ follows by the usual distributional limit.
The equal-time normalization $\Tr_{\mathrm{ph}}W_k(t)=1$ gives
\begin{align}
\mathcal{A}_k(t,t)=1,
\end{align}
and hence the pointwise equal-time CAR whenever the point evaluation is well defined.

This construction supplies one gauge-invariant quasifree realization of the two-point data, in the sense that the kernels are the two-point functions of a particle-number-conserving fermionic Gaussian state on a CAR algebra; it does not determine higher-order correlators of an independently specified joint dynamics.

\subsubsection{Sambe-space sector resolvents and readout}

For a periodic steady state, $G^x(t+T,t'+T)=G^x(t,t')$.
With $t_{\mathrm{av}}=(t+t')/2$ and $\tau=t-t'$, we use the Wigner convention
\begin{align}
F(t_{\mathrm{av}},\tau)
&=
\sum_p
\int\frac{\mathrm{d}\nu}{2\pi}\,
\mathrm{e}^{-\mathrm{i} p\varOmega t_{\mathrm{av}}}
\mathrm{e}^{-\mathrm{i}\nu\tau}
F_p(\nu),
\label{eq:wigner_conv}
\end{align}
in which $\nu$ is the relative frequency conjugate to $\tau$, and the reduced-zone Floquet convention
\begin{align}
F(t,t')
&=
\sum_{mn}
\int_{-\varOmega/2}^{\varOmega/2} \frac{\mathrm{d}\omega}{2\pi}\,
\mathrm{e}^{-\mathrm{i}(\omega+m\varOmega)t}
\mathrm{e}^{+\mathrm{i}(\omega+n\varOmega)t'}
F_{mn}(\omega).
\label{eq:sambe_conv}
\end{align}
The two representations are related by
\begin{align}
F_{mn}(\omega)
&=
F_{m-n}
\left(
\omega+\frac{m+n}{2}\varOmega
\right).
\label{eq:wigner_to_sambe}
\end{align}

Expand each sector generator of Eqs.~\eqref{eq:sector_generator_zero}--\eqref{eq:branch_generator_overlap} as
\begin{align}
\mathcal{K}_q(t)
&= \sum_l \mathrm{e}^{-\mathrm{i} l\varOmega t} \mathcal{K}_{q,l}.
\end{align}
The sector Floquet matrix is
\begin{align}
\left[\mathbb{K}_q\right]_{mn}
&=
\mathcal{K}_{q,m-n}
+
\mathrm{i} m\varOmega\delta_{mn}I_q,
\label{eq:branch_sambe_matrix}
\end{align}
where $I_q$ denotes the identity on the source Liouville space of sector $q$ (equivalently on $\mathcal{V}_{\mathrm{ph}}$ for every $q$).
The positive-time sector resolvent is
\begin{align}
\mathbb{R}_q(z)
&=
\left[
-\mathrm{i} zI_q-\mathbb{K}_q
\right]^{-1}.
\label{eq:branch_resolvents}
\end{align}
Equivalently,
\begin{align}
\sum_l
\left[
-\mathrm{i}(z+m\varOmega)\delta_{ml}I_q
-
\mathcal{K}_{q,m-l}
\right]
\left[\mathbb{R}_q(z)\right]_{ln}
&=
\delta_{mn}I_q.
\end{align}

When the four-endpoint convolution of Eq.~\eqref{eq:keldysh_equation} is transformed to the Sambe representation of Eq.~\eqref{eq:sambe_conv}, each chronological sector interval is replaced by a positive-time resolvent $\mathbb{R}_q(z)$ and therefore carries a definite frequency argument $z$.
That argument is fixed by the Fourier phases at the four endpoints $(t,t';s,s')$.
Extracting the observation component at the reduced-zone frequency $\omega$ of Eq.~\eqref{eq:sambe_conv} multiplies by $\mathrm{e}^{+\mathrm{i}\omega t}\mathrm{e}^{-\mathrm{i}\omega t'}$.
The bath kernel contributes $\varSigma_{\mathrm{bath}}^{\gtrless}(s-s')=\int(\mathrm{d}\nu/2\pi)\,\mathrm{e}^{-\mathrm{i}\nu(s-s')}\varSigma_{\mathrm{bath}}^{\gtrless}(\nu)$, with $\nu$ the bath injection frequency.
The resulting external phase factor is $\mathrm{e}^{+\mathrm{i}(\omega t-\omega t'-\nu s+\nu s')}$, which defines the endpoint charges
\begin{align}
q_t=+\omega,
\quad
q_{t'}=-\omega,
\quad
q_s=-\nu,
\quad
q_{s'}=+\nu.
\label{eq:endpoint_charges}
\end{align}
Writing the ordered times as $x_0<x_1<x_2<x_3$ with interval lengths $\Delta_j=x_j-x_{j-1}$ rewrites the phase as $\mathrm{e}^{+\mathrm{i}\sum_{j=1}^{3}z_j\Delta_j}$, where the resolvent frequency on interval $j$ is
\begin{align}
z_j
&=
\sum_{r=j}^{3}q_{x_r}.
\label{eq:frequency_assignment}
\end{align}
The resolvent on interval $j$ is therefore $\mathbb{R}_{q_j}(z_j)$.
The six chronological orderings of Fig.~\ref{fig:common_source_sk}(b) carry the resolvent frequencies $(z_1,z_2,z_3)$ as follows\footnote{For ordering~(1), $s<t<s'<t'$, the ordered endpoints are $(x_0,x_1,x_2,x_3)=(s,t,s',t')$, so Eq.~\eqref{eq:frequency_assignment} gives $z_1=q_t+q_{s'}+q_{t'}=\omega+\nu-\omega=\nu$, $z_2=q_{s'}+q_{t'}=\nu-\omega$, and $z_3=q_{t'}=-\omega$.}:
\begin{align}
(1)\quad s<t<s'<t'&:\quad (\nu,\nu-\omega,-\omega),
\label{eq:frequency_ordering_one}
\\
(2)\quad s'<t'<s<t&:\quad (-\nu,\omega-\nu,\omega),
\label{eq:frequency_ordering_two}
\\
(3)\quad s<s'<t<t'&:\quad (\nu,0,-\omega),
\label{eq:frequency_ordering_three}
\\
(4)\quad s'<s<t'<t&:\quad (-\nu,0,\omega),
\label{eq:frequency_ordering_four}
\\
(5)\quad s<s'<t'<t&:\quad (\nu,0,\omega),
\label{eq:frequency_ordering_five}
\\
(6)\quad s'<s<t<t'&:\quad (-\nu,0,-\omega).
\label{eq:frequency_ordering_six}
\end{align}
Each ordering $j$ of the time-domain four-endpoint kernel of Eq.~\eqref{eq:four_endpoint_kernel} becomes the Sambe-space resolvent chain\footnote{For ordering~(1), $s<t<s'<t'$, Eq.~\eqref{eq:sector_resolvent_chain} specializes to $\mathbf{A}=\mathbf{A}_{\mathsf{L}}$, $\mathbf{M}_2=\mathbf{D}_{\mathsf{L}}$, $\mathbf{M}_3=\mathbf{A}_{\mathsf{R}}$, and $\mathbf{D}=\mathbf{D}_{\mathsf{R}}$, so that $\mathcal{R}_k^{\mathrm{SH},(1)}(\omega,\nu)=\mathbf{D}_{\mathsf{R}}\mathbb{R}_{\mathsf{R}}(-\omega)\mathbf{A}_{\mathsf{R}}\mathbb{R}_0(\nu-\omega)\mathbf{D}_{\mathsf{L}}\mathbb{R}_{\mathsf{L}}(\nu)\mathbf{A}_{\mathsf{L}}$.}
\begin{align}
\mathcal{R}_k^{\mathrm{SH},(j)}(\omega,\nu)
&=
\mathbf{D}\,
\mathbb{R}_{q_3}(z_3)\,
\mathbf{M}_3\,
\mathbb{R}_{q_2}(z_2)\,
\mathbf{M}_2\,
\mathbb{R}_{q_1}(z_1)\,
\mathbf{A},
\label{eq:sector_resolvent_chain}
\end{align}
where $\mathbf{A}$ and $\mathbf{D}$ are the Sambe-space intersector embeddings realizing the initial activation and final deactivation maps, and $\mathbf{M}_j$ likewise realizes the intermediate endpoint map ($\mathcal{A}$ or $\mathcal{D}$) at $x_{j-1}$, with the sector sequence $(q_1,q_2,q_3)$ and frequencies $(z_1,z_2,z_3)$ fixed by ordering $j$ through Eqs.~\eqref{eq:frequency_ordering_one}--\eqref{eq:frequency_ordering_six}.
The shared-history frequency kernel is the sum of the six orderings,
\begin{align}
\mathcal{R}_{k,mn;l}^{\mathrm{SH}}(\omega,\nu)
&=
\sum_{j=1}^{6}
\left[\mathcal{R}_k^{\mathrm{SH},(j)}(\omega,\nu)\right]_{mn;l}.
\label{eq:R_SH_sum}
\end{align}
Convolving with the bath kernels yields the occupied Sambe components,
\begin{align}
G_{k,mn}^{\gtrless}(\omega)
&=
\sum_{l\in\mathbb{Z}}
\int_{-\varOmega/2}^{\varOmega/2}
\frac{\mathrm{d}\nu}{2\pi}\,
\mathcal{R}_{k,mn;l}^{\mathrm{SH}}(\omega,\nu)
\varSigma_{\mathrm{bath}}^{\gtrless}(\nu+l\varOmega),
\label{eq:keldysh_column_solve}
\end{align}
where the stationary wide-band bath is diagonal in the Sambe injection label,
\begin{align}
\varSigma^{\gtrless}_{ll'}(\nu)
&=
\delta_{ll'}
\varSigma_{\mathrm{bath}}^{\gtrless}(\nu+l\varOmega)\,
\mathbb{I}_{\mathcal{V}_{\mathrm{ph}}},
\label{eq:wbl_sigma_k}
\end{align}
so that only the single sum over $l$ appears in Eq.~\eqref{eq:keldysh_column_solve}.
The observation frequency $\omega$ and the bath injection frequency $\nu$ remain independent variables because a source with finite correlation time redistributes bath-injected weight among Floquet channels.

The prescribed source state admits the Fourier expansion
\begin{align}
\rho_{\mathrm{ph}}^{\mathrm{ps}}(t)
= \sum_m \mathrm{e}^{-\mathrm{i} m\varOmega t} \rho_{\mathrm{ph},m}^{\mathrm{ps}},
\quad
\left[\Lket{\rho_{\mathrm{F}}}\right]_m
\equiv \Lket{\rho_{\mathrm{ph},m}^{\mathrm{ps}}}.
\end{align}
Here, the subscript $\mathrm{F}$ denotes the Floquet--Sambe representation of a time-periodic source object.
For the rotating Bogoliubov-frame Gaussian sources, $\rho_{\mathrm{ph}}^{\mathrm{ps}}(t)=\rho_{\mathrm{ph}}^{\mathrm{ss}}$ is time independent, so only the $m=0$ harmonic is nonzero.

Nonoverlap contributions to $G^{<,>}$ involve the source-only resolvent $\mathbb{R}_0$.
The source-only Floquet generator has a stationary right vector
\begin{align}
\mathbb{K}_0\Lket{\rho_{\mathrm{F}}}
&=
0,
\label{eq:rho_F_null}
\end{align}
with Floquet--Sambe left trace vector
\begin{align}
\left[\Lbra{I_{\mathrm{F}}}\right]_m = \delta_{m0}\Lbra{I_{\mathrm{ph}}},
\quad
\Lbraket{I_{\mathrm{F}}}{\rho_{\mathrm{F}}} = 1.
\label{eq:I_F_normalization}
\end{align}
Every Floquet translation of this stationary mode retained in the Sambe window $m\in[-M,M]$ is also a purely imaginary eigenmode,
\begin{align}
\mathbb{K}_0\Lket{\rho_{\mathrm{F}}^{(m)}}
&=
\mathrm{i} m\varOmega\Lket{\rho_{\mathrm{F}}^{(m)}}.
\end{align}
These modes do not decay, so a naive real-frequency inverse of $-\mathrm{i} zI_0-\mathbb{K}_0$ is singular on the Floquet replica poles.
We therefore isolate them by the projectors
\begin{gather}
P_m = \Lket{\rho_{\mathrm{F}}^{(m)}}\Lbra{I^{(m)}},
\quad
\left[\Lbra{I^{(m)}}\right]_n = \delta_{nm}\Lbra{I_{\mathrm{ph}}},
\label{eq:replica_mode_projector}
\\
P_{\mathrm{rep}} = \sum_{m=-M}^{M}P_m,
\quad
Q_c = I-P_{\mathrm{rep}}.
\label{eq:replica_projector}
\end{gather}
The resulting source-only resolvent is
\begin{align}
\mathbb{R}_0(z)
= \sum_{m=-M}^{M}
\frac{\mathrm{i}}{z+m\varOmega+\mathrm{i}0^+} P_m
+ Q_c \left[
-\mathrm{i} zI_0-\mathbb{K}_0
\right]^{-1} Q_c.
\label{eq:replica_resolvent}
\end{align}
The physical terminal trace remains the $m=0$ bra $\Lbra{I_{\mathrm{F}}}$, even though all replica poles contribute to nonoverlap integrals.
The distributional evaluation of those poles on a discrete bath-frequency grid is given in Appendix~\ref{sec:app_implementation}.

As in the time domain, the retarded and advanced components follow from the equal-time covariance seed.
Fourier transforming Eq.~\eqref{eq:covariance_equation}, the covariance harmonics satisfy
\begin{align}
\sum_l
\left[\mathbb{K}_{\mathsf{LR}}\right]_{ml}
W_{k,l}
&=
-\varGamma\rho_{\mathrm{ph},m}^{\mathrm{ps}}.
\label{eq:covariance_sambe}
\end{align}
The retarded source is the full Toeplitz covariance insertion
\begin{align}
\left[\mathbf{W}_k\right]_{mn}
&=
W_{k,m-n}.
\label{eq:retarded_covariance_source}
\end{align}
It contains all covariance harmonics $W_{k,m}$, is generally not confined to the zero Sambe sector, and is not the bare periodic source state $\rho_{\mathrm{ph}}^{\mathrm{ps}}$.

The Sambe column of the retarded auxiliary~$\mathcal{X}^{\mathrm{R}}$ of Eq.~\eqref{eq:left_branch_propagator} solves
\begin{align}
\sum_l
\left[
-\mathrm{i}(\omega+m\varOmega)\delta_{ml}I_{\mathsf{L}}
-
\mathcal{K}_{\mathsf{L},m-l}
\right]
\Lket{\mathcal{X}_{\mathrm{raw},ln}^{\mathrm{R}}(\omega)}
&=
\Lket{W_{k,m-n}}.
\label{eq:GR_sambe_solve}
\end{align}
The advanced component follows from Hermiticity,
\begin{align}
G_{k,mn}^{\mathrm{A}}(\omega)
&=
\left[
G_{k,nm}^{\mathrm{R}}(\omega)
\right]^*.
\label{eq:advanced_raw_branch}
\end{align}
The raw retarded and advanced columns are denoted by
\begin{gather}
\Lket{\mathcal{X}_{\mathrm{raw}}^{\mathrm{R}}(\omega)}
= \mathbb{R}_{\mathsf{L}}(\omega)\mathbf{W}_k,
\label{eq:raw_column_retarded}
\\
\Lket{\mathcal{X}_{\mathrm{raw}}^{\mathrm{A}}(\omega)}
= \left[ \Lket{\mathcal{X}_{\mathrm{raw}}^{\mathrm{R}}(\omega)} \right]^{\ddagger},
\label{eq:raw_column_advanced}
\end{gather}
where $\ddagger$ denotes the combined Liouville adjoint and Sambe-index exchange required by Eq.~\eqref{eq:RA_hermiticity}.
The physical readouts are
\begin{gather}
G^{\mathrm{R}}_{m0}(k,\omega)
= -\mathrm{i}\Lbraket{I_{\mathrm{ph}}}{\mathcal{X}_{\mathrm{raw},m0}^{\mathrm{R}}(\omega)},
\label{eq:ra_column_readout_R}
\\
G^{\mathrm{A}}_{m0}(k,\omega)
= \left[ G^{\mathrm{R}}_{0m}(k,\omega) \right]^*.
\label{eq:ra_column_readout}
\end{gather}

\paragraph*{Floquet-harmonic components.}
The full period dependence of a component is carried by its Floquet harmonics $G^x_{k,m0}$, related to the Wigner harmonics of Eq.~\eqref{eq:wigner_conv} by Eq.~\eqref{eq:wigner_to_sambe} at $n=0$,
\begin{align}
G^x_{k,m0}(\omega)
&=
G^x_{k,m}\left(\omega+\frac{m}{2}\varOmega\right),
\quad
x\in\{\mathrm{R},\mathrm{A},<,>\}.
\label{eq:m0_wigner_map}
\end{align}
For $G^{\mathrm{R,A}}$, the $m0$ harmonics with $|m|\leq M$ are the corresponding Sambe blocks of Eqs.~\eqref{eq:raw_column_retarded}--\eqref{eq:raw_column_advanced}, read out through Eqs.~\eqref{eq:ra_column_readout_R}--\eqref{eq:ra_column_readout}.
By contrast, the lesser and greater harmonics cannot be read off from a single Sambe block, because the bath convolution in Eq.~\eqref{eq:keldysh_column_solve} couples the observation frequency to the injection frequency and carries the $m=0$ source terminal $\Lbra{I_{\mathrm{F}}}$.
The physical $G^{<,>}_{k,m0}$ is instead assembled by inserting a Sambe translation at the left-leg deactivation boundary of each of the six orderings; the explicit construction is given in Appendix~\ref{sec:app_implementation}.

\subsection{Observables}\label{sec:observables}

The period-averaged spectrum is defined by
\begin{align}
A^{\mathrm{R}}(k,\omega)
&= -\frac{1}{\pi} \im G^{\mathrm{R}}_{00}(k,\omega).
\label{eq:A0_def}
\end{align}
For a branch construction satisfying the Hermiticity and CAR identities stated above, this definition is equivalent to the retarded--advanced difference,
\begin{align}
A^{\mathrm{R}}(k,\omega)
&=
\re
\left[
\frac{\mathrm{i}}{2\pi}
\left(
G^{\mathrm{R}}_{00}(k,\omega)
-
G^{\mathrm{A}}_{00}(k,\omega)
\right)
\right].
\label{eq:A0_RA_equiv}
\end{align}
The period-averaged lesser and greater spectra are defined as
\begin{align}
A^<(k,\omega)
&= \frac{1}{2\pi}\re\left[-\mathrm{i} G^{<}_{00}(k,\omega)\right],
\label{eq:A0_less_def}
\\
A^>(k,\omega)
&= \frac{1}{2\pi} \re\left[+\mathrm{i} G^{>}_{00}(k,\omega)\right].
\label{eq:A0_great_def}
\end{align}
The algebraic component relations imply
\begin{gather}
G^{>}_{00}(k,\omega)-G^{<}_{00}(k,\omega)
=
G^{\mathrm{R}}_{00}(k,\omega)-G^{\mathrm{A}}_{00}(k,\omega),
\label{eq:keldysh_m0_identity}
\\
A^{\mathrm{R}}(k,\omega)
=
A^<(k,\omega)+A^>(k,\omega).
\label{eq:spectral_identity}
\end{gather}
These relations follow analytically from the bath anticommutator and the normalized covariance $W_k$.

The period-averaged occupied momentum distribution is given by
\begin{align}
n_k = \int_{-\infty}^{\infty}\mathrm{d}\omega\,A^<(k,\omega).
\label{eq:nk_def}
\end{align}
The period-averaged density of states (DOS) and occupied spectral weight are defined by
\begin{align}
A^{\mathrm{R}}(\omega) &= \frac{1}{N_k} \sum_{k} A^{\mathrm{R}}(k,\omega),
\\
A^{<}(\omega) &= \frac{1}{N_k} \sum_{k} A^{<}(k,\omega),
\end{align}
respectively.
The momentum-resolved and momentum-integrated effective spectral occupancy ratios are then
\begin{align}
f_{\mathrm{eff}}(k,\omega) = \frac{A^<(k,\omega)}{A^{\mathrm{R}}(k,\omega)},
\quad
f_{\mathrm{eff}}(\omega) = \frac{A^<(\omega)}{A^{\mathrm{R}}(\omega)}.
\label{eq:feff_def}
\end{align}
These ratios give the frequency-resolved fraction of retarded spectral weight that is occupied in the period-averaged nonequilibrium steady state (NESS).

The period-averaged components above are the $m=0$ harmonics $G^{x}_{00}$.
Average-time dependence is carried by the Wigner harmonics of Eq.~\eqref{eq:wigner_conv}.
Because the stored columns $G^{x}_{k,m0}$ are reduced-zone Sambe elements, each harmonic must first be unfolded to its own relative frequency by inverting Eq.~\eqref{eq:m0_wigner_map},
\begin{align}
G^{x}_{k,m}(\omega)
&=
G^{x}_{k,m0}\left(\omega-\frac{m}{2}\varOmega\right),
\label{eq:sambe_to_wigner_unfold}
\end{align}
after which
\begin{align}
G^{x}(k,\omega,t_{\mathrm{p}})
&=
\sum_m \mathrm{e}^{-\mathrm{i} m\varOmega t_{\mathrm{p}}}
G^{x}_{k,m}(\omega),
\label{eq:Gm0_tav_recon}
\end{align}
where $x\in\{\mathrm{R},<\}$ and $\omega$ is the Wigner relative frequency on the extended axis.

A Gaussian probe envelope $\exp[-(t-t_{\mathrm{p}})^2/2\sigma_t^2]$ applied to both time arguments factorizes in Wigner coordinates into $\exp[-(t_{\mathrm{av}}-t_{\mathrm{p}})^2/\sigma_t^2]$ and $\exp[-\tau^2/4\sigma_t^2]$, as in the standard probe-weighted reconstruction of time-resolved photoemission from nonequilibrium Green's functions \cite{Eckstein2008, Freericks2009, Sentef2015, Randi2017}.
The first damps each average-time harmonic by $w_m=\exp[-(m\varOmega\sigma_t/2)^2]$, and the second convolves the relative-frequency spectrum with
\begin{align}
K_{\sigma}(\omega-\nu)
&=
\frac{\sigma_t}{\sqrt{\pi}}
\exp\left[-\sigma_t^2(\omega-\nu)^2\right],
\label{eq:gaussian_probe_kernel}
\end{align}
yielding probe-filtered harmonics
\begin{align}
G^{x}_{k,m,\mathrm{probe}}(\omega)=w_m\int\mathrm{d}\nu\,K_{\sigma}(\omega-\nu)\,G^{x}_{k,m}(\nu).
\end{align}
The corresponding time-resolved spectra are
\begin{gather}
A^{\mathrm{R}}_{\mathrm{probe}}(k,\omega,t_{\mathrm{p}})
= -\frac{1}{\pi} \im \sum_m
\mathrm{e}^{-\mathrm{i} m\varOmega t_{\mathrm{p}}}
G^{\mathrm{R}}_{k,m,\mathrm{probe}}(\omega),
\label{eq:A_probe_R}
\\
A^{<}_{\mathrm{probe}}(k,\omega,t_{\mathrm{p}})
= \frac{1}{2\pi} \re \left[ -\mathrm{i} \sum_m
\mathrm{e}^{-\mathrm{i} m\varOmega t_{\mathrm{p}}}
G^{<}_{k,m,\mathrm{probe}}(\omega)
\right].
\label{eq:A_probe_L}
\end{gather}
Because both factors derive from the same envelope, $\sigma_t$ sets the usual time--frequency trade-off rather than serving as an independently adjustable smoothing parameter \cite{Eckstein2008, Freericks2009, Sentef2015, Randi2017}.
Increasing $\sigma_t$ sharpens the frequency resolution, $K_\sigma\to\delta$, but suppresses the average-time harmonics, $w_m\to\delta_{m0}$, so that the period-averaged spectrum is recovered.
Decreasing $\sigma_t$ retains all harmonics at a frequency resolution of order $1/\sigma_t$.
No single limit returns the unfiltered reconstruction of Eq.~\eqref{eq:Gm0_tav_recon}.
Since $A^{<}_{\mathrm{probe}}$ is a Gaussian-windowed photoemission intensity, it is nonnegative by construction \cite{Freericks2009}.

The momentum-resolved spectra defined above are gauge-fixed quantities in a chosen temporal gauge.
Gauge-invariant photoemission-like observables can be defined by dressing real-space Green's functions with Wilson lines \cite{Freericks2015, Bertoncini1991}, but we do not pursue such constructions here and instead focus on gauge-fixed Floquet spectra that are directly connected to the Peierls-substituted lattice Hamiltonian.
Such a construction dresses the real-space Green's function with a path-ordered phase factor of the vector potential along the separation between the two operators, which reduces to a product of inverse Peierls links for a lattice path in temporal gauge; it becomes essential only when one aims at strictly gauge-invariant momentum-resolved observables.

\subsection{Relation to existing limits and approximations}

The classical coherent Floquet limit should be understood as the joint limit
\begin{align}
A_{\mathrm{cl}}=2\lambda|\alpha_0|=\mathrm{const.},
\quad
\lambda\to0,
\quad
|\alpha_0|\to\infty.
\label{eq:classical_limit}
\end{align}
In this limit, the classical Peierls phase $\lambda_\ell X_{\mathrm{cl}}(t)$ remains finite, whereas the quantum phase fluctuation $\lambda_\ell \delta X_{\mathrm{lab}}(t)$ vanishes.
Here, $\delta X_{\mathrm{lab}}(t)$ is the fluctuation defined in Eq.~\eqref{eq:deltaX_def}.
Equivalently, the link operator becomes a c-number only when $\lambda_\ell^2\langle \delta X_{\mathrm{lab}}^2\rangle_{\mathrm{ss}} \to 0$.
Thus, the classical Floquet limit is not obtained by increasing $|\alpha_0|$ at fixed $\lambda_\ell$; it is the large-mode-volume limit in which $A_{\mathrm{zpf}}$ goes to zero, and hence so does $\lambda_\ell$.
For squeezed backgrounds, the corresponding small parameter is the largest phase variance carried by the link, $\lambda_\ell^2\max_t\langle \delta X_{\mathrm{lab}}(t)^2 \rangle_{\mathrm{ss}}$.
For squeezed vacuum, where the coherent mean vanishes, it is therefore more natural to organize comparisons by the fluctuation scale, for example $\lambda^2\mathrm{e}^{2r}$, rather than by a classical drive amplitude.
Strong squeezing can invalidate a classical-link approximation even when $\lambda_\ell$ itself is small.

\paragraph*{Classical coherent Floquet Green's function in Sambe form.}
In the standard classical Floquet Green's-function formulation \cite{Tsuji2008, Aoki2014}, $D_\ell(t)$ is replaced by a prescribed periodic c-number link $D_{\ell,{\mathrm{cl}}}(t)=\sum_p \mathrm{e}^{-\mathrm{i} p\varOmega t}D_{\ell,p}^{\mathrm{cl}}$.
With the Hamiltonian harmonics $h_{p}$ defined from Eq.~\eqref{eq:hk_peierls_general}, the retarded Floquet Green's function $G^{\mathrm{R}}(\omega)$ is specified in Sambe space through its inverse,
\begin{align}
\left[G^{\mathrm{R},-1}(\omega)\right]_{mn} = (\omega+m\varOmega)\delta_{mn} - h_{m-n} - \varSigma_{\mathrm{bath},mn}^{\mathrm{R}},
\label{eq:classical_sambe_matrix}
\end{align}
with the wide-band bath kernels
\begin{gather}
\varSigma_{\mathrm{bath},mn}^{\mathrm{R}}
= -\frac{\mathrm{i}\varGamma}{2}\delta_{mn},
\label{eq:wideband_retarded_classical}
\\
\varSigma_{\mathrm{bath},mn}^{<}
= \mathrm{i}\varGamma f_{\mathrm{bath}}(\omega+m\varOmega)\,\delta_{mn},
\label{eq:wideband_lesser_classical}
\\
\varSigma_{\mathrm{bath},mn}^{>}
= -\mathrm{i}\varGamma\left[1-f_{\mathrm{bath}}(\omega+m\varOmega)\right]\delta_{mn}.
\label{eq:wideband_greater_classical}
\end{gather}
The advanced component follows from Hermiticity, $G^{\mathrm{A}}(\omega)=[G^{\mathrm{R}}(\omega)]^\dagger$, and the lesser and greater components are obtained by the Keldysh equations
\begin{align}
G^{<}(\omega)
&= G^{\mathrm{R}}(\omega) \varSigma_{\mathrm{bath}}^{<}(\omega) G^{\mathrm{A}}(\omega),
\label{eq:classical_lesser_equation}
\\
G^{>}(\omega)
&= G^{\mathrm{R}}(\omega) \varSigma_{\mathrm{bath}}^{>}(\omega) G^{\mathrm{A}}(\omega).
\label{eq:classical_lg_equation}
\end{align}
These bath-only lesser and greater equations apply to classical coherent Floquet calculations of the type used in dissipative Floquet steady-state studies \cite{Tsuji2009, Aoki2014, Ono2018, Ono2019}, whereas the Born and SCBA comparisons use the conventional contour Dyson equations with the fluctuation-induced self-energy $\varSigma_D$ (Appendix~\ref{sec:app_born_scba}).
Neither construction is the lesser/greater prescription used in the shared-history two-leg formulation of Sec.~\ref{sec:bath_car}, except in the deterministic or source-decoupled limits where the two-leg kernel factorizes.
For a pure classical coherent drive, the c-number link is $D_{\mathrm{cl}}(t)=\exp[\mathrm{i}\lambda X_{\mathrm{cl}}(t)]$, the single-band specialization of $D_{\ell,\mathrm{cl}}(t)$ above.
For squeezed sources, the corresponding one-time mean is instead the full period link $\bar{D}(t)$ of Eq.~\eqref{eq:mean_link_instantaneous}, which retains the Gaussian envelope generated by the source variance; the deterministic Floquet spectrum generated by that mean link is introduced in Sec.~\ref{sec:benchmark} and compared with the shared-history spectra.

\paragraph*{Born and self-consistent Born approximations.}
To connect to diagrammatic approximations, it is useful to decompose the link operator into its mean and fluctuation parts,
\begin{align}
D_\ell(t) = \bar{D}_\ell(t) + \delta D_\ell(t),
\label{eq:D_split}
\end{align}
where $\bar{D}_\ell(t)=\langle D_\ell(t)\rangle_{\mathrm{ss}}$ is the mean part evaluated in the prescribed photon source and $\delta D_\ell(t)=D_\ell(t)-\bar{D}_\ell(t)$ is the fluctuation.
Expanding the electron self-energy in powers of $\delta D_\ell$ leads to a Born approximation controlled by the connected displacement correlator.
A convenient quantity is the connected characteristic function, or equivalently the displacement propagator,
\begin{align}
\mathscr{D}_{\ell\ell'}(z,z')
&= -\mathrm{i} \left[
\left\langle \mathcal{T}_{C} D_\ell(z) D_{\ell'}(z') \right\rangle_{\mathrm{ss}} - \bar{D}_\ell(z) \bar{D}_{\ell'}(z')
\right],
\label{eq:Dscr_def}
\end{align}
which reduces to the standard photon-coordinate Green's function in the weak-field expansion once the directed Peierls vertex factors are included with a consistent sign convention.
This c-number correlator enters only the Born and SCBA comparison schemes; the main shared-history construction retains the source operator sector until the final source trace.
To make the connection explicit, introduce a directed bond index $s=\pm 1$ and a positive light--matter coupling strength $\lambda$ such that
\begin{align}
D_s(z) &\equiv \exp\left[+\mathrm{i} s\lambda X_{\mathrm{lab}}(z)\right],
\quad
D_\ell \equiv D_{s(\ell)},
\label{eq:Ds_def}
\end{align}
where $s(\ell)$ encodes the directed-bond orientation.
With the single-band convention of Eq.~\eqref{eq:hk_1d}, $D_{s=+1}=D$ and $D_{s=-1}=D^\dagger$.
With
\begin{align}
D_X(z,z')
&\equiv
-\mathrm{i}\left\langle \mathcal{T}_{C}\,\delta X_{\mathrm{lab}}(z)\,\delta X_{\mathrm{lab}}(z')\right\rangle_{\mathrm{ss}},
\label{eq:DX_def}
\end{align}
the directed connected displacement propagator
\begin{align}
\mathscr{D}_{ss'}(z,z')
&\equiv
-\mathrm{i}\left\langle \mathcal{T}_{C}\,\delta D_s(z)\,\delta D_{s'}(z')\right\rangle_{\mathrm{ss}}
\label{eq:Dscr_ss_def}
\end{align}
has the weak-field expansion
\begin{gather}
\delta D_s(z)
=
\mathrm{i} s\lambda\,\delta X_{\mathrm{lab}}(z)
+
\mathcal{O}(\lambda^2),
\label{eq:Ds_weakfield}
\\
\mathscr{D}_{ss'}(z,z')
=
-ss'\lambda^2 D_X(z,z')
+\mathcal{O}(\lambda^3).
\label{eq:Dscr_weakfield}
\end{gather}
This expansion takes $\lambda\to0$ at fixed $X_{\mathrm{cl}}$, so that the full Peierls phase is small.
In the joint classical-drive limit of Eq.~\eqref{eq:classical_limit}, where $\lambda X_{\mathrm{cl}}$ remains finite while $\lambda\delta X_{\mathrm{lab}}\to 0$, the corresponding fluctuation expansion instead retains the classical phase,
\begin{align}
\delta D_s(t)
&=
\mathrm{e}^{\mathrm{i} s\lambda X_{\mathrm{cl}}(t)}
\left[\mathrm{i} s\lambda\,\delta X_{\mathrm{lab}}(t)
+\mathcal{O}\left(\lambda^2\,\delta X_{\mathrm{lab}}^2\right)\right].
\label{eq:Dscr_classical_phase_expansion}
\end{align}
Thus, the Peierls-displacement formulation connects to the usual coordinate-propagator self-energy only after including the directed vertex factors and the above sign convention.
The small parameter of the Born expansion is not only $\lambda$, but the phase variance $\lambda^2\langle \delta X_{\mathrm{lab}}^2 \rangle_{\mathrm{ss}}$, which can be enhanced by squeezing.
Keeping $\mathscr{D}$ as an externally prescribed input and constructing $\varSigma_D$ from it yields a Born self-energy.
Schematically, iterating the resulting Dyson equation self-consistently gives the SCBA,
\begin{align}
G^{-1} = G_{\mathrm{ML}}^{-1} - \varSigma_D[G,\mathscr{D}],
\label{eq:dyson_schematic}
\end{align}
where $G_{\mathrm{ML}}$ is the deterministic mean-link Floquet Green's function generated by replacing $D_\ell(t)$ with $\bar{D}_\ell(t)$ and retaining the wide-band bath.
These Born-type constructions are perturbative comparison schemes for the regime of small $\lambda^2\langle\delta X_{\mathrm{lab}}^2\rangle_{\mathrm{ss}}$.
The component Keldysh--Dyson equations solved in the numerical comparisons, the iteration controls, and the closed-form construction of $\mathscr{D}_{ss'}$, including how the photon damping rate $\kappa$ enters the two-time field covariance through the quantum regression theorem, are given in Appendix~\ref{sec:app_born_scba}.

The diagrammatic structure of Eq.~\eqref{eq:dyson_schematic} is formally analogous to Migdal--Eliashberg theory \cite{Migdal1958, Eliashberg1960}, with $\delta D_\ell$ playing the role of the boson field, $\mathscr{D}_{\ell\ell'}$ the role of the phonon propagator, and the vertex corrections dropped.
Unlike conventional Migdal--Eliashberg theory, however, no adiabatic frequency ratio protects this truncation in the regime of Sec.~\ref{sec:benchmark}, where $\varOmega=t_{\mathrm{h}}$; Eq.~\eqref{eq:dyson_schematic} is a weak-coupling truncation controlled by $\lambda^2\langle\delta X_{\mathrm{lab}}^2\rangle_{\mathrm{ss}}\ll1$, and the omitted vertex corrections are not expected to be parametrically small.

\subsection{Numerical method}\label{sec:numerical_notes}

We summarize the numerical realization of the four-sector and six-endpoint equations, Eqs.~\eqref{eq:sector_generator_zero}--\eqref{eq:branch_generator_overlap} and \eqref{eq:four_endpoint_kernel}--\eqref{eq:six_ordering_decomposition}.
The source Hilbert space is truncated in the Bogoliubov--Fock basis to
\begin{align}
P_{N_b} = \sum_{n=0}^{N_b-1}|n_b\rangle\langle n_b|,
\end{align}
where $|n_b\rangle$ is the number state of the Bogoliubov mode $b$ with $b^\dagger b|n_b\rangle=n|n_b\rangle$, and $N_b$ is the dimension of the truncated photon space.
The finite representation of the canonical displacement is the projected displacement operator
\begin{align}
\mathcal{D}_{b,N_b}(\beta)
= P_{N_b} \exp\left( \beta b^\dagger-\beta^*b \right) P_{N_b}.
\label{eq:link_matrix_exp}
\end{align}
The corresponding truncated Peierls link is
\begin{align}
D_{\ell,N_b}(t)
=
\exp\bigl[\mathrm{i}\lambda_\ell X_{\mathrm{cl}}(t)\bigr]
\mathcal{D}_{b,N_b}\bigl(\beta_\ell^{(b)}(t)\bigr),
\label{eq:truncated_peierls_link}
\end{align}
with $\beta_\ell^{(b)}(t)$ from Eq.~\eqref{eq:bogoliubov_displacement_amplitude}.
At a finite cutoff, $D_{\ell,N_b}(t)$ is not exactly unitary in general.
The Bloch Hamiltonian is nevertheless Hermitian because the directed link and its explicit adjoint are assembled as the pair appearing in Eq.~\eqref{eq:hk_1d}.

On a time grid $t_j=jT/N_t$, the projected quantum displacement of Eq.~\eqref{eq:link_matrix_exp} is multiplied by the scalar phase $\exp[\mathrm{i}\lambda_\ell X_{\mathrm{cl}}(t_j)]$ from Eq.~\eqref{eq:D_factorization_bogoliubov}.
The link harmonics are obtained by
\begin{align}
D_{\ell,p}
&=
\frac{1}{N_t}
\sum_{j=0}^{N_t-1}
\mathrm{e}^{+\mathrm{i} p\varOmega t_j}
D_\ell(t_j),
\label{eq:D_harmonics_nt}
\end{align}
and satisfy $(D^\dagger)_p=D_{-p}^\dagger$.
For a Sambe-space cutoff $m\in[-M,M]$, the difference $m-n$ spans $[-2M,2M]$.
The Peierls-generator harmonic cutoff is therefore kept distinct from the Sambe-space cutoff and is set to $M_D=2M$ so that all couplings internal to the retained Sambe window are present.
The source cutoff $N_b$, Sambe cutoff $M$, time grid $N_t$, bath-frequency grid, and memory-time grid are refined independently.

The link harmonics determine the sector generators $\mathcal{K}_q$ of Eqs.~\eqref{eq:sector_generator_zero}--\eqref{eq:branch_generator_overlap} and their Floquet operators $\mathbb{K}_q$ of Eq.~\eqref{eq:branch_sambe_matrix}.
The periodic source vector $\Lket{\rho_{\mathrm{F}}}$ is obtained from the null space of $\mathbb{K}_0$, Eq.~\eqref{eq:rho_F_null}, with the trace normalization of Eq.~\eqref{eq:I_F_normalization} imposed explicitly.
The covariance harmonics $W_{k,m}$ are then found from Eq.~\eqref{eq:covariance_sambe}.
The covariance solution is checked against its defining equation in Eq.~\eqref{eq:covariance_equation} and the normalization condition in Eq.~\eqref{eq:covariance_normalization}.
The retarded component is evaluated independently by solving Eq.~\eqref{eq:GR_sambe_solve} with the Toeplitz covariance source.
The advanced component is obtained from Eq.~\eqref{eq:advanced_raw_branch}, with the explicitly solved adjoint right sector providing an independent verification.

The lesser and greater responses are assembled from factorized one-leg and two-leg building blocks on a common bath-frequency grid; the explicit frequency-domain chains for the six orderings, the left-leg Sambe shifts producing the harmonics $G^{<,>}_{k,m0}$, and the CAR-based correction of the contact quadrature error that defines the reported lesser component are collected in Appendix~\ref{sec:app_implementation}.
The adopted bath-frequency, memory, and time grids and their refinement checks are documented in Appendix~\ref{sec:app_cutoff}.

The lesser and greater components are also constructed independently from Eq.~\eqref{eq:keldysh_column_solve} as a numerical cross-check of the six-ordering and bath-channel implementation; representative residuals of that independent assembly, evaluated prior to CAR reconstruction of the reported components, and a deterministic classical Floquet calculation are reported in Appendix~\ref{sec:app_cutoff}.
The greater component is then obtained from the CAR identity as
\begin{align}
G^> = G^< + G^{\mathrm{R}} - G^{\mathrm{A}},
\label{eq:greater_from_car}
\end{align}
cf.\ Eq.~\eqref{eq:keldysh_m0_identity}, and $G^{\mathrm{K}}=G^>+G^<$ is then derived.

The sector linear solves use the resolvents of Eq.~\eqref{eq:branch_resolvents}.
Dense and sparse direct solves provide reference results for small matrix dimensions, while the large-scale solver applies the Floquet operators without constructing their full dense matrices and uses preconditioned Krylov solves.
For the reported shared-history spectra, each branch and shifted Krylov solve is retained only if the unpreconditioned relative residual is below $10^{-7}$, using the generalized minimal residual (GMRES) method with a harmonic-diagonal preconditioner.

\section{Numerical results}\label{sec:benchmark}

\subsection{Model and numerical parameters}
The model used in this section is the one-dimensional single-band nearest-neighbor model in Eq.~\eqref{eq:hk_1d} with a single-mode Gaussian source specified by $(\varOmega,\alpha_0,r,\phi_0,\kappa)$ and the positive light--matter coupling strength $\lambda$ in the convention of Eq.~\eqref{eq:hk_1d}.
The fermionic bath is specified by $(\varGamma,\mu_{\mathrm{bath}},\beta_{\mathrm{bath}})$.
The bath-only lesser and greater equations are used for the classical Floquet calculations and for the deterministic mean-link Floquet spectra $A_{\mathrm{ML}}^{\mathrm{R}}$ generated by $D_{\mathrm{ML}}(t)\equiv\bar{D}(t)$, whereas the Born and SCBA comparisons include the corresponding fluctuation self-energies.
The mean-link Floquet solve uses the same Sambe and Peierls harmonic cutoffs $(M,M_D)$ and $(k,\omega)$ grids as the shared-history spectra, but requires no photon Fock cutoff.
We evaluate $G^{\mathrm{R}}_{m0}(\omega)$ for $|m|\leq4$ together with the period-averaged components $G^{\mathrm{R},<,>}_{00}(\omega)$ and the corresponding spectra $A^{\mathrm{R},<,>}$ and $f_{\mathrm{eff}}$.
The shared-history calculations in this section use the parameters and cutoffs collected in Table~\ref{tab:benchmark_params}; source-specific parameters such as $\lambda$, $\kappa$, and the Gaussian-state parameters $(\alpha_0,r,\phi_0)$ are given in the corresponding figure captions.

\begin{table}[t]
\centering
\small
\caption{Parameters and numerical settings for the shared-history results.
Convergence with respect to $N_b$, $M$, $M_D$, and the bath-frequency and memory grids is documented in Appendix~\ref{sec:app_cutoff}.}
\label{tab:benchmark_params}
\begin{tabular*}{\columnwidth}{@{\extracolsep{\fill}}p{0.43\columnwidth}p{0.48\columnwidth}@{}}
\hline\hline
Quantity & Value \\
\hline
Nearest-neighbor hopping $t_{\mathrm{h}}$ & $1$ (units of energy) \\
Carrier frequency $\varOmega$ & $1$ \\
Chemical potential $\mu_{\mathrm{bath}}$ & $0$ \\
Bath linewidth $\varGamma$ & $0.2$ \\
Inverse temperature $\beta_{\mathrm{bath}}$ & $10$ \\
Photon Fock cutoff $N_b$ & $12$ \\
Sambe cutoff $M$ & $10$ \\
Peierls harmonic cutoff $M_D$ & $20$ ($=2M$) \\
Time-grid points $N_t$ & $64$ \\
Momentum grid & $k \in [-\pi, \pi)$, $N_k = 256$ \\
Frequency range & $\omega \in [-8,8]$, $N_\omega = 1601$ \\
Bath-frequency grid & $\nu\in[-25,25]$, $\Delta\nu=0.0125$, $N_\nu=4001$ \\
Memory quadrature & $\Delta\tau=0.00625$, $\epsilon_{\mathrm{mem}}=10^{-10}$ \\
Krylov residual & relative residual $<10^{-7}$ \\
\hline\hline
\end{tabular*}
\end{table}

To aid the interpretation of the momentum-resolved spectra, we also introduce the instantaneous and period-averaged mean-link dispersions obtained by replacing the Peierls operators by their source expectation values.
For the single-band model, we define
\begin{align}
\bar{D}(t)
= \Tr_{\mathrm{ph}}\left[\rho_{\mathrm{ph}}^{\mathrm{ss}}D(t)\right],
\label{eq:mean_link_band_reference}
\end{align}
with $\bar{D}^*(t)=\Tr_{\mathrm{ph}}[\rho_{\mathrm{ph}}^{\mathrm{ss}}D^\dagger(t)]$ following from Hermiticity of $\rho_{\mathrm{ph}}^{\mathrm{ss}}$.
These dispersions are not quasienergy bands of the full shared-history construction; they only indicate where the band would lie if the operator-valued link were replaced by its one-time mean value.
For the Gaussian sources considered here, $\rho_{\mathrm{ph}}^{\mathrm{ss}}=|0_b\rangle\langle0_b|$ in the Bogoliubov frame, so the factorization in Eq.~\eqref{eq:D_factorization_bogoliubov} together with the vacuum expectation value of a Bogoliubov displacement operator gives the instantaneous mean link
\begin{align}
\bar{D}(t)
=
\mathrm{e}^{\mathrm{i}\lambda X_{\mathrm{cl}}(t)}\,\mathrm{e}^{-\lambda^2|f(t)|^2/2},
\label{eq:mean_link_instantaneous}
\end{align}
with $f(t)$ from Eq.~\eqref{eq:f_def}.
The corresponding instantaneous mean-link dispersion and static period-averaged dispersion are
\begin{gather}
\bar{\varepsilon}_k(t)
=
-t_{\mathrm{h}}\left(\mathrm{e}^{-\mathrm{i} k}\bar{D}(t)+\mathrm{e}^{+\mathrm{i} k}\bar{D}^*(t)\right),
\label{eq:mean_link_band_instantaneous}
\\
\bar{\varepsilon}_k
= \frac{1}{T} \int_0^T\mathrm{d} t\, \bar{\varepsilon}_k(t)
= -t_{\mathrm{h}}\left(\mathrm{e}^{-\mathrm{i} k}\bar{D}_0+\mathrm{e}^{+\mathrm{i} k}\bar{D}_0^*\right),
\label{eq:mean_link_band_dispersion}
\end{gather}
respectively, where $\bar{D}_0 = T^{-1} \int_0^T\mathrm{d} t\,\bar{D}(t)$.
The static dispersion $\bar{\varepsilon}_k$ serves as the reference curve in the spectral figures and the instantaneous dispersion $\bar{\varepsilon}_k(t)$ as the probe-time reference in Fig.~\ref{fig:time_resolved}.
For squeezed vacuum ($X_{\mathrm{cl}}=0$), Eq.~\eqref{eq:mean_link_instantaneous} is real and positive, so Eq.~\eqref{eq:mean_link_band_instantaneous} reduces to $\bar{\varepsilon}_k(t)=-2t_{\mathrm{h}}\bar{D}(t)\cos k$.
For a general squeezed coherent source with real $\alpha_0$, the period average of Eq.~\eqref{eq:mean_link_instantaneous} follows in closed form.
The two exponential factors are expanded with the Jacobi--Anger identity $\mathrm{e}^{\mathrm{i} z\cos\theta}=\sum_{n}\mathrm{i}^nJ_n(z)\mathrm{e}^{\mathrm{i} n\theta}$ and the generating function $\mathrm{e}^{z\cos\theta}=\sum_{n}I_n(z)\mathrm{e}^{\mathrm{i} n\theta}$ of the ordinary and modified Bessel functions of the first kind.
The period average keeps only the harmonic-matching terms; with the classical Peierls amplitude $A_{\mathrm{cl}}=2\lambda|\alpha_0|$, it reads
\begin{align}
\bar{D}_0
&=\mathrm{e}^{-\lambda^2\cosh(2r)/2}
\biggl[
J_0(A_{\mathrm{cl}})I_0(\zeta)
\notag \\
&\quad
+2\sum_{m=1}^{\infty}(-1)^m J_{2m}(A_{\mathrm{cl}})\,I_m(\zeta)\,\cos(m\phi_0)
\biggr],
\label{eq:mean_link_squeezed_coherent}
\end{align}
where $J_n$ and $I_m$ are the ordinary and modified Bessel functions of the first kind and $\zeta \equiv \lambda^2 \sinh(2r)/2$.
This expression is manifestly real, so $\bar{D}_0^*=\bar{D}_0$, and it reduces to the two limits used elsewhere in this work.
For a pure coherent source ($r=0$, so $\zeta=0$ and only the $m=0$ term survives), Eq.~\eqref{eq:mean_link_squeezed_coherent} gives
\begin{align}
\bar{D}_0=\mathrm{e}^{-\lambda^2/2}J_0(A_{\mathrm{cl}}).
\label{eq:mean_link_coherent}
\end{align}
The period-averaged mean link therefore already contains the familiar dynamical-localization renormalization of the bandwidth by $J_0(A_{\mathrm{cl}})$ \cite{Dunlap1986}, up to a residual quantum suppression factor $\mathrm{e}^{-\lambda^2/2}$ that approaches unity in the classical limit $\lambda\to0$ and $|\alpha_0|\to\infty$ at fixed $A_{\mathrm{cl}}$.
For squeezed vacuum ($\alpha_0=0$, so $A_{\mathrm{cl}}=0$ and $J_{2m}(0)=0$ for $m\neq0$), only the $m=0$ term survives instead, giving
\begin{align}
\bar{D}_0
= \mathrm{e}^{-\lambda^2\cosh(2r)/2}\, I_0(\zeta).
\label{eq:mean_link_squeezed}
\end{align}
This factor is real and strictly positive for all $\lambda$ and $r$, and therefore exhibits no sign-changing band collapse analogous to the zeros of $J_0(A_{\mathrm{cl}})$.
For a general squeezed coherent source, Eq.~\eqref{eq:mean_link_squeezed_coherent} shows that the squeezing phase $\phi_0$ enters $\bar{D}_0$ only through the $\cos(m\phi_0)$ terms weighted by $I_m(\zeta)/I_0(\zeta)$, which is small unless $\zeta$ is large; we use this general result in Sec.~\ref{sec:squeezed_coherent} when comparing squeezed coherent sources with different squeezing phases.

The periodic one-time link expectation value also defines a deterministic Floquet problem through
\begin{align}
D_{\mathrm{ML}}(t) \equiv \bar{D}(t)
= \langle D(t)\rangle_{\mathrm{ss}},
\label{eq:mean_link_drive}
\end{align}
with Fourier harmonics
\begin{align}
\bar{D}_n
=
\frac{1}{T}\int_0^T\mathrm{d} t\, \bar{D}(t)\,\mathrm{e}^{+\mathrm{i} n\varOmega t}.
\label{eq:Dbar_floquet_harmonics}
\end{align}
Replacing the operator-valued Peierls link by $D_{\mathrm{ML}}(t)$ and solving the resulting deterministic Floquet--Keldysh problem with the same wide-band bath as in Eqs.~\eqref{eq:classical_sambe_matrix}--\eqref{eq:classical_lg_equation} yields the mean-link Floquet spectrum $A_{\mathrm{ML}}^{\mathrm{R}}(k,\omega)$.
This construction retains all Fourier harmonics of $\bar{D}(t)$, while the source Hilbert space and its multitime correlations are absent.
It is distinct from the classical coherent Floquet drive $D_{\mathrm{cl}}(t)=\exp[\mathrm{i}\lambda X_{\mathrm{cl}}(t)]$, which omits the Gaussian envelope in Eq.~\eqref{eq:mean_link_instantaneous}, and from the shared-history spectrum $A_{\mathrm{SH}}^{\mathrm{R}}$ obtained from the operator-valued link.

The results are organized from the classical coherent limit to quantum-source effects.
We first test the classical coherent limit and finite-coupling coherent corrections, then examine squeezed-vacuum spectra, occupations, and effective distributions.
We next present harmonic-resolved spectra, before turning to squeezed coherent sources.

\subsection{Coherent sources}

\begin{figure*}[t]\centering
\includegraphics[scale=1]{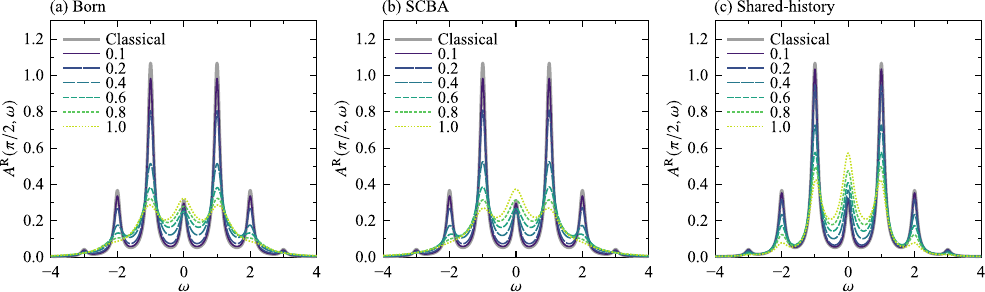}
\caption{Period-averaged spectral function, $A^{\mathrm{R}}(k,\omega)$, at $k = \pi/2$ for coherent Gaussian sources with fixed classical Peierls amplitude $A_{\mathrm{cl}}=1.1$ and decreasing quantum coupling $\lambda$, compared across three descriptions: (a) Born, (b) SCBA, and (c) the shared-history construction.
The gray reference in each panel shows the classical Floquet result with the same $A_{\mathrm{cl}}$ and wide-band bath.
The source damping is $\kappa = 1$.
}
\label{fig:coherent_lambda}
\end{figure*}

\begin{figure*}[t]\centering
\includegraphics[scale=1]{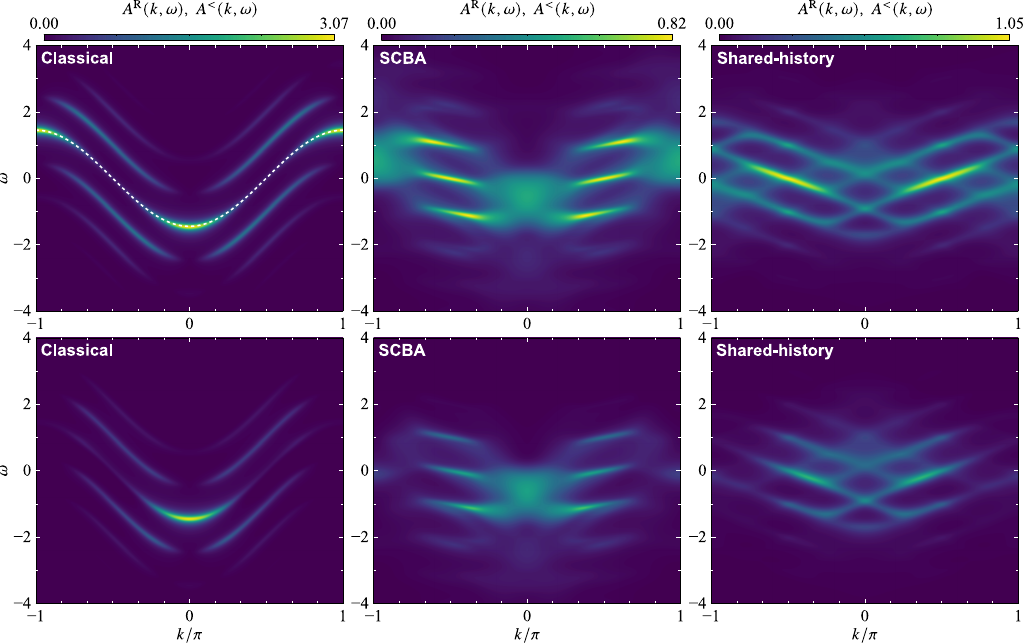}
\caption{Period-averaged retarded and occupied spectral weights for a finite-$\lambda$ coherent source with $A_{\mathrm{cl}} = 1.1$ and $\lambda=1$, computed with (left) classical Floquet theory, (middle) SCBA, and (right) the shared-history construction.
The upper panels show $A^{\mathrm{R}}(k,\omega)$, and the lower panels show $A^{<}(k,\omega)$.
In each column, the color scale is shared between the two panels and its maximum is set to the maximum of $A^{\mathrm{R}}(k,\omega)$ in that column.
The dotted lines indicate the corresponding static dispersion reference $\bar{\varepsilon}_k$.
The source damping is $\kappa = 0.1$.
}
\label{fig:coherent_lam1}
\end{figure*}

Before discussing genuinely nonclassical sources, we first verify the classical coherent limit.
For a coherent Gaussian source, the laboratory coordinate separates into a classical component and fluctuations, and the classical Peierls amplitude is $A_{\mathrm{cl}}=2\lambda|\alpha_0|$.
The classical Floquet limit is reached by taking $\lambda\to0$ and $|\alpha_0|\to\infty$ at fixed $A_{\mathrm{cl}}$, so that the operator-valued fluctuation $\lambda\delta X_{\mathrm{lab}}$ becomes negligible.
Figure~\ref{fig:coherent_lambda} shows the period-averaged retarded spectrum at $k=\pi/2$ for fixed $A_{\mathrm{cl}}=1.1$ and $\lambda = 0.1$--$1$, comparing (a) Born, (b) SCBA, and (c) the shared-history construction against the same classical Floquet reference.
All three descriptions approach the classical result with the same wide-band bath in this joint limit, providing a direct coherent-limit check of each approximation.
At finite $\lambda$, the deviation from the gray classical reference is a finite-$\lambda$ correction within each scheme: the shared-history link retains the full displacement operator generated by the fluctuation phase $\lambda\delta X_{\mathrm{lab}}(t)$, whereas Born and SCBA retain this fluctuation only through the two-point displacement correlator entering $\varSigma_D$, so the line shape of the deviation at $\lambda\gtrsim0.4$ depends on the approximation.
For $\lambda\gtrsim0.2$ at fixed $A_{\mathrm{cl}}$, the departure from the classical Floquet reference grows: spectral weight is redistributed among the Floquet sidebands, and the overall multi-peak profile becomes smoother, an effect that is more pronounced in Born and SCBA than in the shared-history construction.

\begin{figure*}[t]\centering
\includegraphics[scale=1]{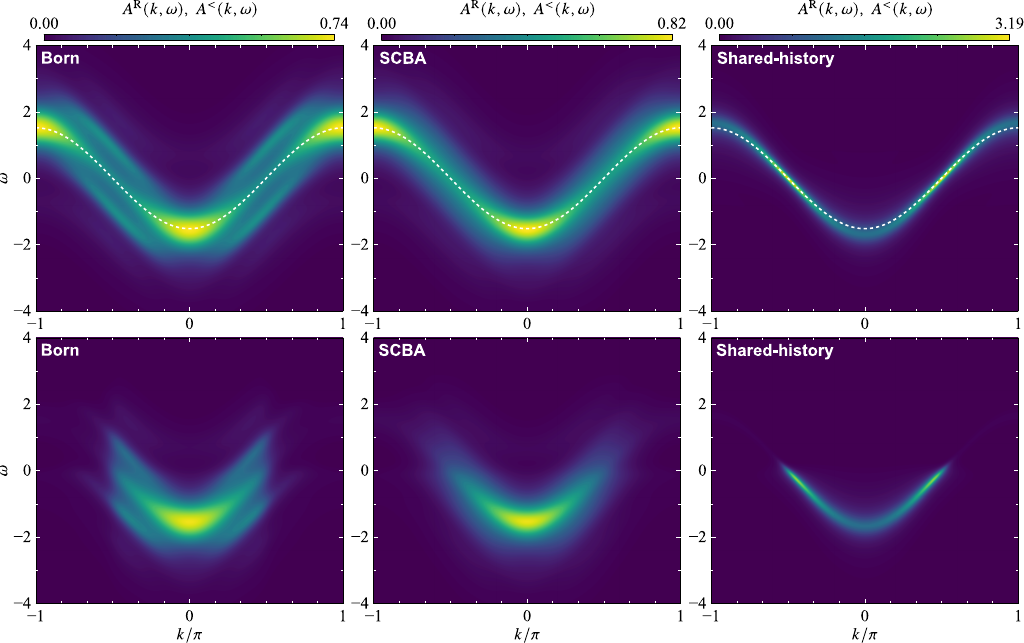}
\caption{Period-averaged retarded and occupied spectral weights for squeezed vacuum with $r=4$ and $\kappa=1$, computed using (left) the Born approximation, (middle) SCBA, and (right) the shared-history construction.
The upper panels show $A^{\mathrm{R}}(k,\omega)$, and the lower panels show $A^{<}(k,\omega)$.
In each column, the color scale is shared between the two panels and its maximum is set to the maximum of $A^{\mathrm{R}}(k,\omega)$ in that column.
The dotted lines indicate the corresponding static dispersion reference $\bar{\varepsilon}_k$.
The coupling is $\lambda = 0.02$.
}
\label{fig:squeezed_r4}
\end{figure*}

Having established the classical limit, we next examine a finite-$\lambda$ coherent source.
Figure~\ref{fig:coherent_lam1} compares the classical Floquet, SCBA, and shared-history calculations at $A_{\mathrm{cl}}=1.1$ and $\lambda=1$.
The classical calculation produces sharp Floquet replicas following the static dispersion reference.
In the SCBA panel, strong broadening and incoherent weight accumulate around $k=0$ and $k=\pi$, whereas near $k=\pm\pi/2$ the main peak and Floquet sidebands remain visible, with a dispersion slope smaller than that of the classical Floquet replicas.
In the shared-history spectrum, the original cosine-like band is no longer apparent: as in the SCBA panel, a main peak and sideband structure persist near $|k|\sim\pi/2$, while comparatively sharp peaks also remain around $k=0$ and $k=\pi$, so that the overall weight appears arranged in a more linear, diamond-like dispersion that does not track the static dispersion reference.
That diamond-like dispersion is likewise flatter than the classical Floquet dispersion, but steeper than the corresponding SCBA features.
At these parameters, the Born and SCBA control parameter $\lambda^2\langle\delta X_{\mathrm{lab}}^2\rangle_{\mathrm{ss}}$ is not small, so SCBA should be read as a weak-coupling comparison rather than a controlled expansion.
Within that comparison, SCBA already retains the Gaussian mean link, the closed connected Peierls correlator $\mathscr{D}$, and the self-consistent rainbow resummation of $\varSigma_D$, but it still truncates the electronic self-energy at the two-point fluctuation skeleton and omits vertex corrections.
The shared-history construction instead keeps the full operator-valued Peierls link in the enlarged-space resolvent; within the prescribed-source approximation, its overall spectral structure is therefore more reliable than that of SCBA.

\subsection{Squeezed vacuum sources}

\begin{figure*}[t]\centering
\includegraphics[scale=1]{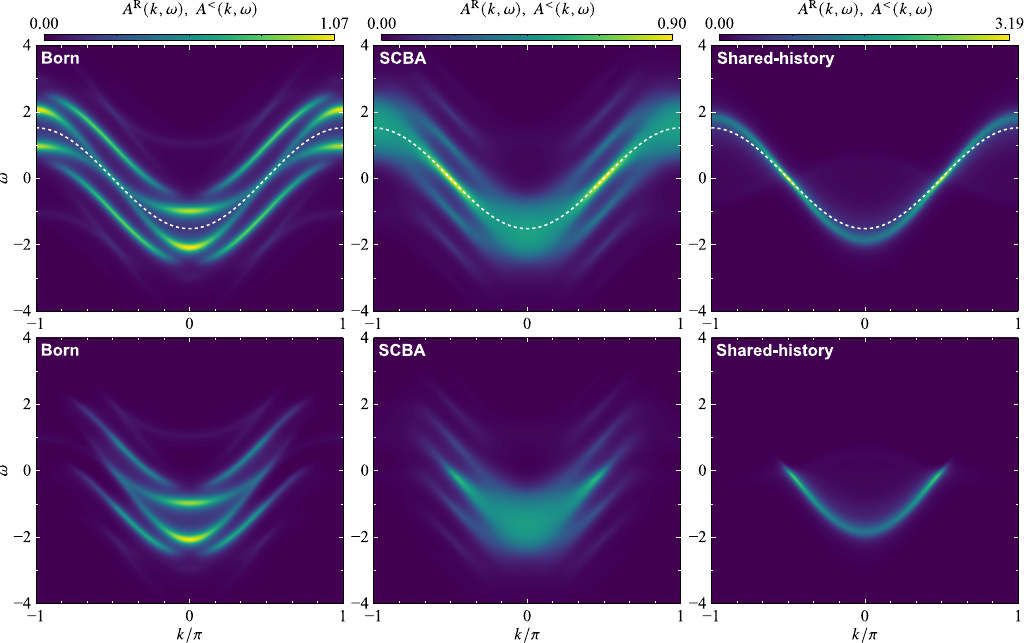}
\caption{Same squeezed-vacuum comparison as Fig.~\ref{fig:squeezed_r4}, but with the smaller source damping rate $\kappa=0.1$.
}
\label{fig:squeezed_r4_kappa01}
\end{figure*}

\begin{figure}[t]\centering
\includegraphics[scale=1]{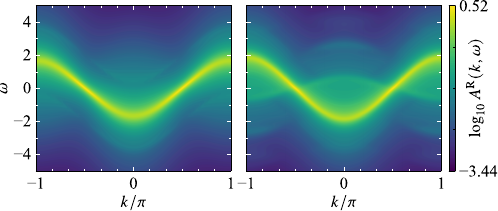}
\caption{Shared-history period-averaged retarded spectral weight for squeezed vacuum with $r=4$, displayed as $\log_{10}A^{\mathrm{R}}(k,\omega)$ for (left) $\kappa=1$ and (right) $\kappa=0.1$.
Both panels use the same logarithmic color scale.
The remaining parameter is $\lambda=0.02$.
}
\label{fig:squeezed_r4_log}
\end{figure}

\begin{figure*}[t]\centering
\includegraphics[scale=1]{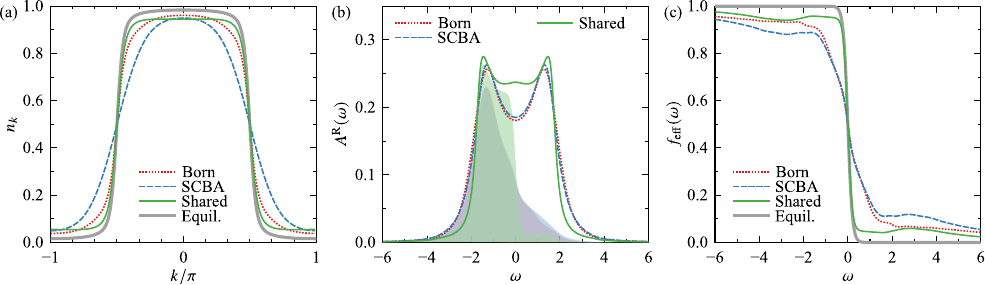}
\caption{Period-averaged occupation and distribution quantities for squeezed vacuum with $r=4$.
(a) Occupied spectral weight integrated over frequency, $n_k$.
(b) Density of states $A^{\mathrm{R}}(\omega)$, with shading indicating the occupied spectral weight $A^{<}(\omega)$.
(c) Effective spectral occupancy ratio $f_{\mathrm{eff}}(\omega)=A^</A^{\mathrm{R}}$.
In every panel, the Born (red dotted), SCBA (blue dashed), and shared-history (green solid) results are overlaid.
In panel (a), the gray curve denotes the undriven equilibrium occupation obtained from Eq.~\eqref{eq:nk_def} with $A^{<}$ evaluated at $\alpha_0=0$ and $r=0$; in panel (c), the gray curve is the bath Fermi--Dirac function $f_{\mathrm{bath}}(\omega)$.
The parameters are $\lambda = 0.02$ and $\kappa = 1$.
}
\label{fig:squeezed_r4_nk_dos_feff}
\end{figure*}

We now turn to squeezed vacuum, for which $\alpha_0=0$ and $\phi_0=0$, and hence $X_{\mathrm{cl}}(t)=0$.
The source nevertheless has large Gaussian fluctuations and an anomalous covariance, so the Peierls link remains a genuinely operator-valued object.
Although the bare coupling used below is small, $\lambda=0.02$, the relevant squeezed fluctuation scale is not $\lambda$ alone.
In the Bogoliubov representation, the displacement part of the link is controlled by $|\lambda f(t)|$, whose maximum is $\lambda\mathrm{e}^r$.
For $r=4$, this gives $\lambda\mathrm{e}^r\approx1.1$.
Thus, the Peierls exponential samples a nonperturbative fluctuation range even though $\lambda\ll1$.
The squeezed-vacuum spectra in what follows should therefore be interpreted as small bare-coupling but large squeezed-phase-fluctuation results.

Figure~\ref{fig:squeezed_r4} compares the Born, SCBA, and shared-history calculations for $r=4$ at $\kappa=1$.
The Born and SCBA retarded spectra retain a broad cosine-like band close to the static dispersion reference, whereas the shared-history $A^{\mathrm{R}}$ remains substantially sharper across the Brillouin zone.
Near $|k|\sim\pi/2$, the shared-history weight tracks the static dispersion reference closely and shows no resolved side peaks in the period-averaged spectrum; around $k=0$ and $k=\pi$, it exhibits comparatively weak additional broadening and a slight departure from that reference.
For the shared-history spectra, this $k$-dependent broadening need not be attributed entirely to dynamics beyond the one-time mean link.
For the squeezed vacuum considered here, $\bar{D}(t)$ is real, so Eq.~\eqref{eq:mean_link_band_instantaneous} reduces to $\bar{\varepsilon}_k(t)=-2t_{\mathrm{h}}\bar{D}(t)\cos k$.
Consequently, the periodic mean-link motion vanishes at $|k|=\pi/2$ and is largest near $k=0$ and $k=\pi$.
Even at the mean-link level, this breathing can therefore contribute to the broadening of the period-averaged spectrum away from $|k|=\pi/2$, in addition to shared-history dynamics beyond the one-time mean link; this is a kinematic reading of that weight, not an additive decomposition of the $m=0$ spectrum.
This contrast between $|k|\sim\pi/2$ and the zone edges ($k=0$ and $k=\pi$) in the shared-history spectra is opposite to the SCBA pattern at the same $\kappa$: the SCBA weight is most strongly broadened near $|k|\sim\pi/2$.
That SCBA momentum pattern is consistent with a $\kappa$-sensitive fluctuation contribution controlled by the Peierls vertex: in the weak-field limit of Appendix~\ref{sec:app_born_scba}, the two vertex factors contribute a prefactor $4t_{\mathrm{h}}^2\lambda^2\sin^2 k$, which is maximal near $|k|=\pi/2$ and vanishes at $k=0$ and $k=\pi$.
Broadening near $k=0$ and $k=\pi$ can instead be read as one-dimensional Van Hove kinematics associated with the band edges.
The instantaneous mean-link displacement is proportional to $\cos k$, so that its frequency excursion scales as $|\cos k|$ and vanishes at $|k|=\pi/2$, providing a kinematic contribution to the shared-history pattern.
Because the present squeezed fluctuation scale $\lambda\mathrm{e}^{r}\approx1.1$ is not perturbatively small, the weak-field vertex argument is used only as a qualitative interpretation of the SCBA momentum pattern.

The lesser spectrum likewise depends on the approximation: the shared-history $A^{<}(k,\omega)$ remains concentrated on a sharp occupied branch at $\omega \lesssim \mu_{\mathrm{bath}} = 0$, whereas the Born and SCBA weights, following their broadened retarded dispersions, also spread to some extent into the $\omega \gtrsim \mu_{\mathrm{bath}}$ region.
This contrast parallels that already seen in $A^{\mathrm{R}}$ and follows from the same difference between the approximations: Born and SCBA redistribute occupied weight through the two-point fluctuation self-energy $\varSigma_D$, whereas the shared-history $A^{<}$ is assembled from the two-leg bath convolution with the full operator-valued Peierls link.
Taken together, the squeezed source affects both the breadth of $A^{\mathrm{R}}$ and the redistribution of occupied spectral weight, with the quantitative pattern depending on the approximation.

To examine the dependence on the source correlation time, Fig.~\ref{fig:squeezed_r4_kappa01} repeats the squeezed-vacuum comparison for a smaller damping rate, $\kappa=0.1$.
Reducing $\kappa$ increases the photon correlation time, and the spectral structures associated with the squeezed source become more pronounced.
The Born result shows the clearest multi-branch sideband structure.
The SCBA smears this structure through self-consistency in a momentum-dependent way: near $|k|\sim\pi/2$ the Born sidebands remain visible though broadened, whereas around $k=0$ and $k=\pi$ the multi-branch features are washed out more strongly.
Relative to Fig.~\ref{fig:squeezed_r4}, reducing $\kappa$ therefore restores sideband structure near $|k|\sim\pi/2$ where the $\kappa$-sensitive broadening was strongest, while the residual washout near $k=0$ and $k=\pi$ remains consistent with one-dimensional Van Hove kinematics.

These Born and SCBA trends do not carry over to the shared-history construction.
Even at $\kappa=0.1$, the shared-history $A^{\mathrm{R}}$ remains essentially unbroadened near $|k|=\pi/2$ and tracks the static dispersion reference without resolved side peaks on the linear color scale of Fig.~\ref{fig:squeezed_r4_kappa01}, while around $k=0$ and $k=\pi$ it shows modest broadening and a visible departure from that reference, again consistent with a mean-link breathing contribution that is absent at $|k|=\pi/2$.
Relative to Fig.~\ref{fig:squeezed_r4}, the departure from the static dispersion reference is larger at $\kappa=0.1$.
Because the prescribed squeezed-vacuum steady state, and hence the complete periodic mean link $\bar{D}(t)$, is unchanged when only $\kappa$ is varied, the deterministic mean-link Floquet spectra are identical, i.e.,
\begin{align}
A_{\mathrm{ML}}^{\mathrm{R}}(k,\omega) \big\vert_{\kappa=0.1}
= A_{\mathrm{ML}}^{\mathrm{R}}(k,\omega) \big\vert_{\kappa=1}.
\label{eq:ml_kappa_identity}
\end{align}
The shared-history spectra nevertheless differ, as seen by comparing Figs.~\ref{fig:squeezed_r4} and \ref{fig:squeezed_r4_kappa01}.
The $\kappa$ dependence is therefore absent from the one-time mean-link dynamics and originates from the $\kappa$-dependent multitime source dynamics retained in the shared-history propagation.
This comparison distinguishes the squeezed fluctuation amplitude, which enters $\bar{D}(t)$, from the source correlation time, which does not.

Figure~\ref{fig:squeezed_r4_log} replots the shared-history retarded spectra on a logarithmic color scale to expose spectral weight hidden by the linear scales of Figs.~\ref{fig:squeezed_r4} and \ref{fig:squeezed_r4_kappa01}.
At fixed momentum, the weak structures form nonmonotonic satellite ridges away from the dominant spectral peak and therefore cannot be accounted for by the monotonically decaying Lorentzian tail of that peak.
These features are far more clearly resolved at $\kappa=0.1$ than at $\kappa=1$.
Their approximate $2\varOmega$ spacing is consistent with the even-harmonic selection rule for squeezed vacuum discussed in Sec.~\ref{sec:sambe_harmonics}, but that spacing alone does not distinguish multitime source dynamics from deterministic mean-link Floquet replicas, because the complete mean-link drive $D_{\mathrm{ML}}(t)$ also contains even harmonics.
Separately, at $\kappa=0.1$, the same figure shows a comparatively strong ridge near $\omega\approx +0.75\cos k$, with a momentum dispersion opposite to that of the dominant $-\cos k$ branch.
Neither the instantaneous nor the period-averaged mean-link dispersion can produce such an inverted ridge: $\bar{D}(t)$ remains real and positive, so $\bar{\varepsilon}_k(t)=-2t_{\mathrm{h}}\bar{D}(t)\cos k$ only breathes a dispersion of the form $-\cos k$, and Floquet replicas of that dispersion retain the same momentum shape.
Because the mean-link Floquet spectrum is identical for the two damping rates by Eq.~\eqref{eq:ml_kappa_identity}, the $\kappa$-dependent visibility of both the satellite weight and the inverted ridge originates from multitime source dynamics beyond the complete deterministic mean-link Floquet calculations; the microscopic mechanism of the inverted ridge is not identified here.

The spectral changes induced by squeezed vacuum are also visible in integrated observables.
Figure~\ref{fig:squeezed_r4_nk_dos_feff} shows the momentum distribution $n_k$, the DOS $A^{\mathrm{R}}(\omega)$ with the occupied weight $A^<(\omega)$, and the effective spectral occupancy ratio $f_{\mathrm{eff}}(\omega)$.
In panel (a), the shared-history $n_k$ remains comparatively close to the undriven equilibrium step, retaining a relatively sharp Fermi-level jump, whereas Born and SCBA yield smoother occupations over a wider portion of the Brillouin zone.
In panel (b), the Born and SCBA densities of states retain a one-dimensional cosine-band profile with Van Hove peaks near the band edges, whereas the shared-history DOS is relatively enhanced over $|\omega|\lesssim1.5$.
That enhancement is consistent with the shared-history spectra remaining essentially unbroadened near $|k|\sim\pi/2$.
The shaded $A^{<}(\omega)$ in panel (b) and $f_{\mathrm{eff}}(\omega)$ in panel (c) together show a transfer of occupied weight into $\omega>\mu_{\mathrm{bath}}$.
This transfer is weaker for the shared-history construction than for Born and SCBA.
Relative to the gray bath reference, Born and SCBA yield systematically gentler slopes of $f_{\mathrm{eff}}$ over a broad frequency range, while the shared-history $f_{\mathrm{eff}}$ stays close to $f_{\mathrm{bath}}$ for $|\omega|\lesssim0.1$ and shows a milder redistribution at larger $|\omega|$.

\subsection{Floquet harmonic structure and time-resolved spectra}\label{sec:sambe_harmonics}

\begin{figure*}[t]\centering
\includegraphics[scale=1]{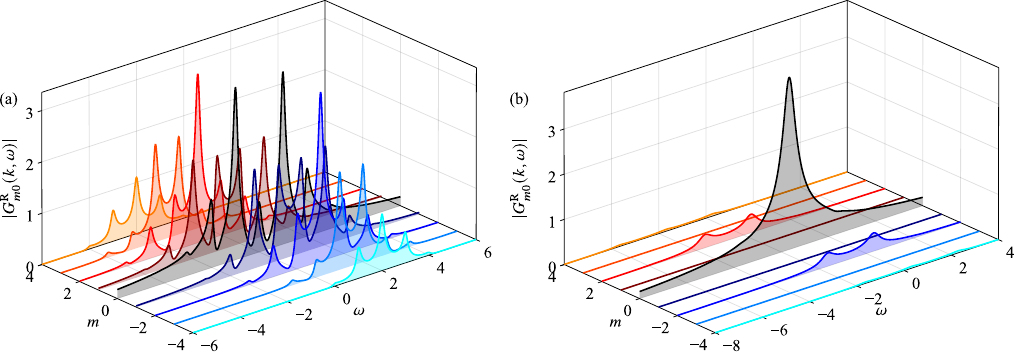}
\caption{Absolute value of the harmonic-resolved retarded Green's function, $|G_{m0}^{\mathrm{R}}(k, \omega)|$, for $m = 0, \pm 1, \dots, \pm 4$: (a) a coherent source with $A_{\mathrm{cl}} = 1.1$ at $k = \pi/2$ and (b) a squeezed vacuum source with $r = 4$ at $k = 0$.
The parameters are $\lambda = 0.02$ and $\kappa = 1$.
}
\label{fig:gr_m0}
\end{figure*}

\begin{figure}[t]\centering
\includegraphics[scale=1]{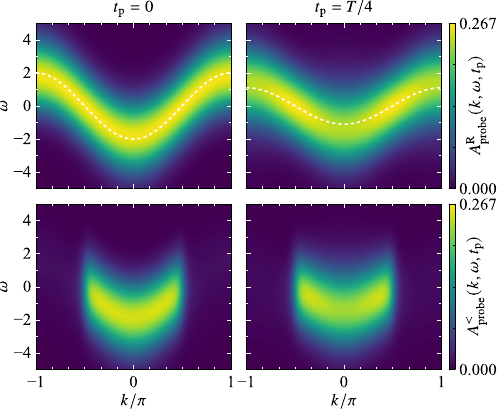}
\caption{Time-resolved Gaussian-probe spectra $A^{\mathrm{R}}_{\mathrm{probe}}(k,\omega,t_{\mathrm{p}})$ (top) and $A^{<}_{\mathrm{probe}}(k,\omega,t_{\mathrm{p}})$ (bottom) for squeezed vacuum with $r=4$ and $\kappa=1$ at probe times $t_{\mathrm{p}}=0$ (left) and $t_{\mathrm{p}}=T/4$ (right), reconstructed from $G^{\mathrm{R},<}_{m0}$ with probe width $\sigma_t=0.5$.
All four panels share one color scale whose maximum is the maximum of $A^{\mathrm{R}}_{\mathrm{probe}}$.
The dotted lines indicate the instantaneous mean-link dispersion reference $\bar{\varepsilon}_k(t_{\mathrm{p}})$ of Eq.~\eqref{eq:mean_link_band_instantaneous}.
The coupling is $\lambda=0.02$.
}
\label{fig:time_resolved}
\end{figure}

The preceding figures focus mainly on the period-averaged $m=0$ sector and on integrated quantities.
To clarify the Floquet-harmonic structure behind those spectra, Fig.~\ref{fig:gr_m0} displays $|G^{\mathrm{R}}_{m0}(k,\omega)|$ for a coherent source at $k=\pi/2$ and a squeezed vacuum source at $k=0$.
The sharp comb in the coherent case reflects coherent periodic phase modulation.
In the squeezed-vacuum panel, without a classical coherent field, the $m=0$ component is dominant and only the even-$m$ components are appreciable, consistent with the selection rule below.
This even--odd selection follows from a combined half-period translation and photon-parity symmetry.
For squeezed vacuum, the coefficient $f(t)$ of Eq.~\eqref{eq:f_def} satisfies $f(t+T/2)=-f(t)$, so that $D(t+T/2)=\varPi_b D(t)\varPi_b^\dagger$, where $\varPi_b=\exp(\mathrm{i}\pi b^\dagger b)$.
Because the source Liouvillian, its vacuum steady state, and the source trace are invariant under this parity transformation, the source-traced response is $T/2$-periodic in the average time and therefore contains only even Floquet harmonics.
The comparison demonstrates that the distinction between coherent and squeezed-vacuum Gaussian sources is not only a change in the $m=0$ spectrum, but also a qualitative change in the harmonic mixing encoded in the off-diagonal amplitudes.

\begin{figure*}[t]\centering
\includegraphics[scale=1]{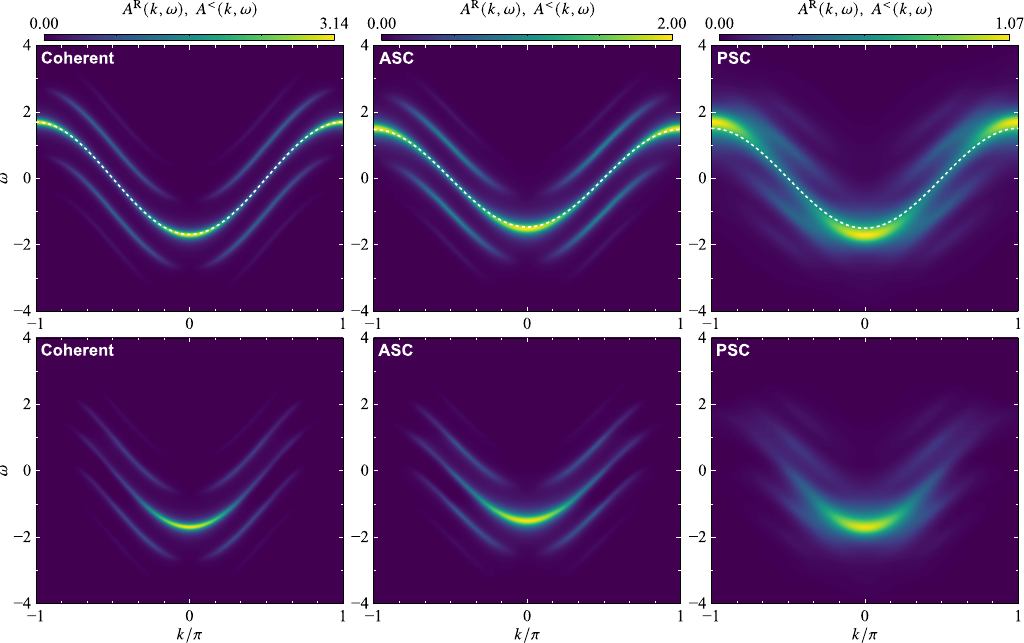}
\caption{Period-averaged retarded and occupied spectral weights for (left) a coherent source with $\alpha_0 = 20$, (middle) an ASC source with $\alpha_0 = 20$, $r = 3.6$, and $\phi_0 = 0$, and (right) a PSC source with $\alpha_0 = 20$, $r = 3.6$, and $\phi_0 = \pi$, computed with the shared-history construction.
The upper panels show $A^{\mathrm{R}}(k,\omega)$, and the lower panels show $A^{<}(k,\omega)$.
In each column, the color scale is shared between the two panels and its maximum is set to the maximum of $A^{\mathrm{R}}(k,\omega)$ in that column.
The dotted lines indicate the corresponding static dispersion reference $\bar{\varepsilon}_k$, evaluated separately for each source from Eq.~\eqref{eq:mean_link_squeezed_coherent}.
The parameters are $\lambda = 0.02$ and $\kappa = 1$.
}
\label{fig:squeezed_coherent}
\end{figure*}

The same harmonics determine the average-time dependence through Eqs.~\eqref{eq:sambe_to_wigner_unfold}--\eqref{eq:A_probe_L}.
Figure~\ref{fig:time_resolved} shows the time-resolved Gaussian-probe spectra for the squeezed-vacuum shared-history calculation of Fig.~\ref{fig:squeezed_r4} at $\sigma_t=0.5$, reconstructed from $|m|\leq6$; this width retains the harmonic weights $w_1=0.94$ and $w_4=0.37$.
Because only even $m$ contribute, the probe spectra are $T/2$-periodic in $t_{\mathrm{p}}$: the probe times $t_{\mathrm{p}}=0$, $T/2$, and $T$ give the same spectrum, as do $t_{\mathrm{p}}=T/4$ and $3T/4$.
The two phases displayed in Fig.~\ref{fig:time_resolved} are therefore separated by half of the reduced period.
Between them, the band breathes about $\omega=0$ rather than shifting rigidly: the retarded peak at $k=0$ moves from $\omega=-1.81$ to $-1.34$ and the peak at $k=\pi$ moves symmetrically from $+1.81$ to $+1.34$, while the peak at $k=\pi/2$ remains at $\omega=0$, so the modulation changes the probe-resolved bandwidth.
Relative to the instantaneous mean-link reference (dotted lines), the retarded weight at $t_{\mathrm{p}}=0$ tracks that cosine band closely across the zone, whereas at $t_{\mathrm{p}}=T/4$ a moderate departure appears near $k=0$ and $k=\pi$.
The probe-resolved profile around $k=0$ and $k=\pi$ is modestly broader at $t_{\mathrm{p}}=T/4$ than at $t_{\mathrm{p}}=0$.
Because these are finite-window Gaussian-probe spectra, that width includes the probe-frequency convolution and the mean-link motion within the probe window, in addition to dynamics beyond the one-time mean link, and should not be identified with an intrinsic instantaneous linewidth.
Taken together, the time-resolved spectra are consistent with the interpretation of the period-averaged shared-history weights in Figs.~\ref{fig:squeezed_r4} and \ref{fig:squeezed_r4_kappa01}: one-time mean-link breathing contributes only away from $|k|=\pi/2$, while multitime shared-history correlations can produce additional phase-dependent spectral reconstruction.
The occupied spectral weight remains concentrated near the zone center and changes only weakly between the two probe phases.
Restricting the reconstruction from $|m|\leq6$ to $|m|\leq4$ changes $A^{\mathrm{R}}_{\mathrm{probe}}$ and $A^{<}_{\mathrm{probe}}$ by at most $1.1\times10^{-4}$ and $9.7\times10^{-5}$, respectively, compared with peak values of order $0.27$, so the retained harmonic window is not the limiting error, and both spectra remain nonnegative over the displayed range.
Movies of the full period for $A^{\mathrm{R}}_{\mathrm{probe}}$ and $A^{<}_{\mathrm{probe}}$ at the same $\sigma_t$ are provided in the Supplemental Material~\cite{SupplementalMaterial}.

\subsection{Squeezed coherent sources}\label{sec:squeezed_coherent}

\begin{figure*}[t]\centering
\includegraphics[scale=1]{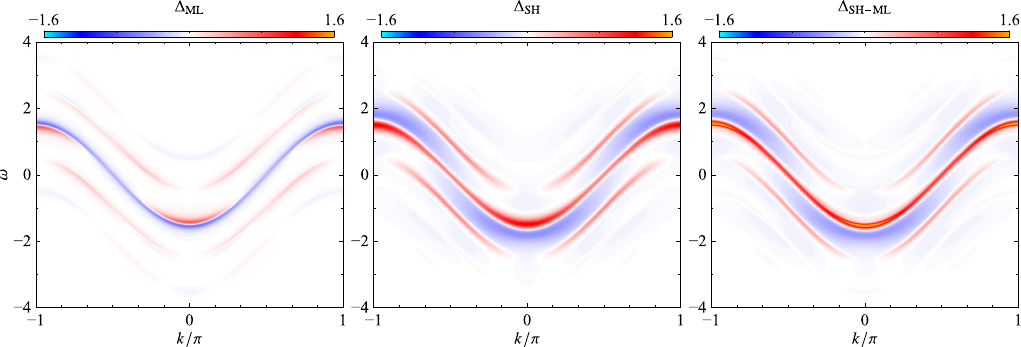}
\caption{Momentum- and frequency-resolved ASC--PSC spectral contrasts on a common diverging color scale.
(a) Deterministic mean-link Floquet difference $\Delta_{\mathrm{ML}}$.
(b) Shared-history difference $\Delta_{\mathrm{SH}}$.
(c) Difference of the two contrasts, $\Delta_{\mathrm{SH-ML}}=\Delta_{\mathrm{SH}}-\Delta_{\mathrm{ML}}$.
The parameters match Fig.~\ref{fig:squeezed_coherent}.
}
\label{fig:mean_link_contrasts}
\end{figure*}

\begin{figure*}[t]\centering
\includegraphics[scale=1]{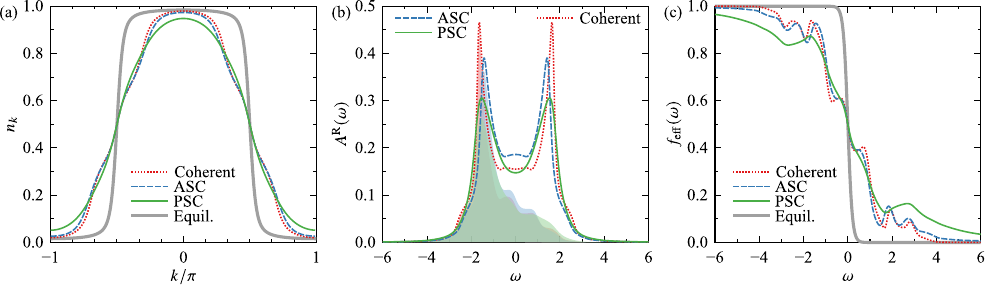}
\caption{Period-averaged occupation and distribution quantities for a coherent source with $\alpha_0 = 20$ (red dotted), an ASC source with $\alpha_0 = 20$, $r = 3.6$, and $\phi_0 = 0$ (blue dashed), and a PSC source with $\alpha_0 = 20$, $r = 3.6$, and $\phi_0 = \pi$ (green solid), obtained from the shared-history construction.
(a) Occupied spectral weight integrated over frequency, $n_k$.
(b) Density of states $A^{\mathrm{R}}(\omega)$, with shading indicating the occupied spectral weight $A^{<}(\omega)$.
(c) Effective spectral occupancy ratio $f_{\mathrm{eff}}(\omega)=A^</A^{\mathrm{R}}$.
In panel (a), the gray curve denotes the undriven equilibrium occupation obtained from Eq.~\eqref{eq:nk_def} with $A^{<}$ evaluated at $\alpha_0=0$ and $r=0$; in panel (c), the gray curve is the bath Fermi--Dirac function $f_{\mathrm{bath}}(\omega)$.
The parameters are $\lambda = 0.02$ and $\kappa = 1$.
}
\label{fig:squeezed_coherent_nk_dos_feff}
\end{figure*}

We finally consider squeezed coherent sources, where a finite coherent amplitude coexists with squeezed fluctuations.
For real $\alpha_0$, the coherent coordinate is
\begin{gather}
X_{\mathrm{cl}}(t)=2\alpha_0\cos(\varOmega t).
\end{gather}
The fluctuation entering the Peierls link is controlled by
\begin{align}
&\left\langle \delta X_{\mathrm{lab}}(t)^2\right\rangle_{\mathrm{ss}} \notag \\
&= |f(t)|^2
= \cosh(2r) - \sinh(2r) \cos(\phi_0-2\varOmega t).
\label{eq:squeezed_coherent_variance}
\end{align}
Thus, $\phi_0=0$ squeezes the quadrature at the maxima of the coherent coordinate, while $\phi_0=\pi$ anti-squeezes it at those times and squeezes the orthogonal phase-shifted quadrature.
We refer to these choices as amplitude-squeezed coherent (ASC) and phase-squeezed coherent (PSC) sources, respectively.

Figure~\ref{fig:squeezed_coherent} shows the period-averaged retarded and occupied spectral weights $A^{\mathrm{R}}(k,\omega)$ and $A^{<}(k,\omega)$ for a coherent source, an ASC source, and a PSC source with the same $\alpha_0=20$ and, for the squeezed sources, $r=3.6$.
The coherent source retains sharp Floquet-like replicas and a well-defined occupied sideband structure.
The ASC spectrum remains close to the coherent result, with a modest overall broadening of the peaks, and still tracks the static dispersion reference $\bar{\varepsilon}_k$ closely while retaining visible sidebands.
By contrast, the PSC spectrum changes more strongly: the weight is broadly redistributed, the sideband structure is largely smeared out, and the dispersion departs measurably from that static reference.
In particular, the occupied weight $A^{<}$ of the PSC source is visibly stronger on the positive-energy ($\omega > \mu_{\mathrm{bath}}$) side of the renormalized band than that of the ASC source.
This difference reflects the relative timing between the coherent Peierls modulation and the squeezed quadrature noise.

At the one-time mean-link level, the ASC--PSC distinction is carried primarily by the nonzero Floquet harmonics of $\bar{D}(t)$, not by the period average $\bar{D}_0$.
Evaluating Eq.~\eqref{eq:mean_link_squeezed_coherent} at these parameters ($\alpha_0=20$, $\lambda=0.02$, $r=3.6$, so $A_{\mathrm{cl}}=0.8$ and $\zeta\approx0.13$) gives $\bar{D}_{0,\mathrm{ASC}}=0.734622$ and $\bar{D}_{0,\mathrm{PSC}}=0.752426$, which differ by only approximately $2.4\%$ relative to their mean value.
Both differ from the pure-coherent value by roughly $10$--$15\%$ because of the overall $\mathrm{e}^{-\lambda^2\cosh(2r)/2}I_0(\zeta)$ suppression relative to $\mathrm{e}^{-\lambda^2/2}$.
The dotted lines in Fig.~\ref{fig:squeezed_coherent} use these exact, source-specific static dispersion references $\bar{\varepsilon}_k$ rather than the pure-coherent reference.
Among the retained harmonics $|n|\leq6$, the largest ASC--PSC difference occurs at $n=2$,
\begin{align}
|\bar{D}_{2,\mathrm{ASC}}-\bar{D}_{2,\mathrm{PSC}}|
\approx 0.0995,
\label{eq:Dbar2_asc_psc}
\end{align}
whereas the $n=0$ difference is approximately $0.018$.
In magnitude,
\begin{align}
|\bar{D}_{2,\mathrm{ASC}}| \approx 0.017,
\quad
|\bar{D}_{2,\mathrm{PSC}}| \approx 0.117.
\label{eq:Dbar2_magnitudes}
\end{align}
The dominant $n=2$ contrast reflects the $2\varOmega$ modulation of the squeezed quadrature variance in Eq.~\eqref{eq:squeezed_coherent_variance}.

For comparison, we define the shared-history (SH) versus mean-link (ML) spectral residuals
\begin{align}
R_{\mathrm{ASC}}(k,\omega)
&=
A_{\mathrm{SH},\mathrm{ASC}}^{\mathrm{R}}(k,\omega)
-
A_{\mathrm{ML},\mathrm{ASC}}^{\mathrm{R}}(k,\omega),
\label{eq:R_ASC_def}
\\
R_{\mathrm{PSC}}(k,\omega)
&=
A_{\mathrm{SH},\mathrm{PSC}}^{\mathrm{R}}(k,\omega)
-
A_{\mathrm{ML},\mathrm{PSC}}^{\mathrm{R}}(k,\omega),
\label{eq:R_PSC_def}
\end{align}
and likewise $R_{\mathrm{coh}}$ for the coherent twin at the same $(\lambda,A_{\mathrm{cl}},\kappa)$.
Each $R_s$ is the pointwise difference between the full shared-history spectrum and the deterministic mean-link spectrum for source $s$.
$R_s$ quantifies the net spectral difference and is not resolved into connected-correlation or cumulant orders.
For the ASC--PSC contrast,
\begin{align}
\Delta_{\mathrm{ML}}(k,\omega)
&=
A_{\mathrm{ML},\mathrm{ASC}}^{\mathrm{R}}(k,\omega)
-
A_{\mathrm{ML},\mathrm{PSC}}^{\mathrm{R}}(k,\omega),
\label{eq:Delta_ML}
\\
\Delta_{\mathrm{SH}}(k,\omega)
&=
A_{\mathrm{SH},\mathrm{ASC}}^{\mathrm{R}}(k,\omega)
-
A_{\mathrm{SH},\mathrm{PSC}}^{\mathrm{R}}(k,\omega),
\label{eq:Delta_SH}
\\
\Delta_{\mathrm{SH-ML}}(k,\omega)
&=
\Delta_{\mathrm{SH}}(k,\omega)
-
\Delta_{\mathrm{ML}}(k,\omega)
\notag\\
&=
R_{\mathrm{ASC}}(k,\omega)
-
R_{\mathrm{PSC}}(k,\omega).
\label{eq:Delta_SH_ML}
\end{align}

Figure~\ref{fig:mean_link_contrasts} displays the ASC--PSC contrasts $\Delta_{\mathrm{ML}}$, $\Delta_{\mathrm{SH}}$, and their difference $\Delta_{\mathrm{SH-ML}}$.
Panel (a) is nonzero wherever the deterministic mean-link Floquet spectra of ASC and PSC differ at a given $(k,\omega)$, even though their period averages $\bar{D}_0$ are nearly equal; that structure may reflect weight redistribution or relative shifts of Floquet peaks between the two sources.
The structured nonzero pattern in panel (a) therefore shows that the ASC--PSC spectral contrast is already present in the nonzero Floquet harmonics of the one-time mean link, so the dotted static dispersion references in Fig.~\ref{fig:squeezed_coherent} understate the mean-link content of the drive.
Panel (b) shows the corresponding shared-history contrast, which is not a uniform rescaling of panel (a).
Panel (c) isolates where that contrast itself changes between the shared-history and mean-link constructions, equivalently where the residuals $R_{\mathrm{ASC}}$ and $R_{\mathrm{PSC}}$ differ; a sign change relative to panel (a) marks a local reversal of the ASC--PSC ordering.
The reconstruction is also momentum dependent: at $|k|=\pi/2$, $\max_\omega |R_{\mathrm{ASC}}(\pi/2,\omega)|=0.047$ and $\max_\omega |R_{\mathrm{PSC}}(\pi/2,\omega)|=0.989$, so the ASC mean-link and shared-history spectra remain close at $|k|=\pi/2$, whereas the PSC spectra exhibit a pronounced reconstruction at the same momentum.
At the weak-coupling coherent control parameters $\lambda=0.02$, $A_{\mathrm{cl}}=0.8$, chosen to match the coherent amplitude of the ASC and PSC sources, and $\kappa=1$, the shared-history and mean-link spectra are nearly identical, with $\max_{k,\omega}|R_{\mathrm{coh}}(k,\omega)|=0.0118$.
A natural interpretation of the sign of the ASC--PSC effect is suggested by the timing structure in Eq.~\eqref{eq:squeezed_coherent_variance}: a sinusoidal drive spends more time near its turning points $X_{\mathrm{cl}}=\pm2\alpha_0$, where $\dot X_{\mathrm{cl}}=0$, than near its zero crossings.
For ASC, the quadrature noise is squeezed at those turning points, which is consistent with the spectrum staying comparatively close to the mean-link Floquet result near $|k|=\pi/2$; for PSC, the same turning points are anti-squeezed instead, consistent with the stronger reconstruction relative to both the static dispersion reference $\bar{\varepsilon}_k$ and $A_{\mathrm{ML}}^{\mathrm{R}}$.

Figure~\ref{fig:squeezed_coherent_nk_dos_feff} shows the corresponding occupation and distribution quantities for the same three sources.
In panel (a), the coherent source retains sideband-induced shoulders in $n_k$, while ASC remains comparatively close to that profile and PSC redistributes $n_k$ more strongly toward larger momenta near $k\sim\pi$.
Between the two squeezed sources, $n_k$ is larger for ASC than for PSC near $k=0$, while the ordering reverses near $k=\pi$, consistent with the stronger positive-energy occupied weight of PSC in Fig.~\ref{fig:squeezed_coherent}.
In panel (b), the coherent DOS retains sharp Van Hove peaks near the band edges, while ASC and PSC suppress and broaden these features and redistribute the occupied weight.
The band-edge DOS is lower for PSC than for ASC, consistent with the broader redistribution of spectral weight for PSC in Fig.~\ref{fig:squeezed_coherent}.
In panel (c), the coherent and ASC $f_{\mathrm{eff}}$ retain multi-step features associated with Floquet sidebands, whereas for PSC those steps are smeared into a smoother effective spectral occupancy.
Near the band edges, the same energy-asymmetric ASC--PSC ordering appears in $f_{\mathrm{eff}}$: around $\omega\approx-1.8$, $f_{\mathrm{eff}}$ is larger for ASC than for PSC, while around $\omega\approx+1.8$ the ordering is reversed.
Together with Fig.~\ref{fig:squeezed_coherent}, this establishes that squeezing-angle control affects not only the retarded spectral function but also the occupied spectral weight and the effective spectral occupancy.

\section{Discussion}\label{sec:discussion}
The numerical results support three physical conclusions.
First, the shared-history construction has the correct classical coherent limit of Eq.~\eqref{eq:classical_limit} (Fig.~\ref{fig:coherent_lambda}), but finite-$\lambda$ coherent sources retain quantum-source corrections beyond the purely classical coherent Floquet drive $D_{\mathrm{cl}}(t)=\exp[\mathrm{i}\lambda X_{\mathrm{cl}}(t)]$, including a rearrangement of the retarded weight away from the cosine-like static dispersion reference $\bar{\varepsilon}_k$ (Fig.~\ref{fig:coherent_lam1}).
Second, squeezed vacuum reconstructs the retarded spectrum even without a coherent mean field (Fig.~\ref{fig:squeezed_r4}); the occupancy redistribution is modest, with $f_{\mathrm{eff}}$ staying comparatively close to $f_{\mathrm{bath}}$, and with a milder transfer of occupied weight into $\omega>\mu_{\mathrm{bath}}$ than in Born and SCBA (Fig.~\ref{fig:squeezed_r4_nk_dos_feff}).
The relevant fluctuation scale is the squeezed phase fluctuation $\lambda\mathrm{e}^r$, not the bare coupling $\lambda$ alone.
Here, $f_{\mathrm{eff}}$ of Eq.~\eqref{eq:feff_def} measures the occupied weight of the period-averaged NESS after convolution of the bath kernels with the shared-history two-leg response, so the photon source redistributes bath-injected spectral weight even though the fermionic bath remains the only source of electronic occupation.
Third, a deterministic Floquet calculation generated by the complete periodic mean link reveals two distinct levels of squeezed-state control.
For squeezed coherent sources, nonzero harmonics of the one-time mean link already produce the ASC--PSC spectral contrast $\Delta_{\mathrm{ML}}$ of Eq.~\eqref{eq:Delta_ML} [Fig.~\ref{fig:mean_link_contrasts}(a)], while the shared-history dynamics further reshape that contrast into $\Delta_{\mathrm{SH}}$ of Eq.~\eqref{eq:Delta_SH} [Fig.~\ref{fig:mean_link_contrasts}(b)], including local sign reversals of the ASC--PSC ordering visible in Fig.~\ref{fig:mean_link_contrasts}(c).
For squeezed vacuum, varying the source damping at fixed mean link leaves the deterministic spectrum unchanged, as in Eq.~\eqref{eq:ml_kappa_identity}, but modifies the shared-history spectrum through the multitime source dynamics (Figs.~\ref{fig:squeezed_r4} and \ref{fig:squeezed_r4_kappa01}).
Together these results offer a route to quantum-state control of Floquet spectra beyond control by the coherent displacement amplitude alone.

The conclusions above are scoped to the prescribed-source shared-history framework: the reported $A^{\mathrm{R}}$, $A^{<}$, $n_k$, and $f_{\mathrm{eff}}$ are two-point spectral readouts of that process and are not identified with Heisenberg correlators of an independently specified joint electron--photon Lindblad dynamics (Sec.~\ref{sec:operator_reconstruction}).
The numerical source model uses the Markovian damping generator in Eq.~\eqref{eq:lindblad_bogoliubov_main} and holds the reservoir-stabilized photon state fixed against many-electron polarization, while the shared-history construction specifies how the continuous bath kernels enter the occupied electronic sector; Born and SCBA provide weak-coupling comparisons within the same setting, rather than an exact closed-cavity solution.
Relative to phase-correlator transport and diagrammatic electron--photon approaches \cite{Souquet2014, Gao2016, Agarwalla2016, Bi2019}, the present framework is not uniformly more complete: it presently omits electron-induced photon polarization, but retains the dissipative source and the full operator-valued Peierls link through the enlarged-space construction of the electronic Green's functions.
A natural extension is to restore that source backaction in stages, for example by updating a classical mean field or mean link from electronic expectation values, and next by dressing the photon propagator with a two-point electronic polarization, along the lines of those diagrammatic treatments \cite{Gao2016, Agarwalla2016}.

The numerical realizations and analytical mean-link formulas of this work specialize to reservoir-stabilized Gaussian photon sources, but the branch and Sambe construction itself only requires that $\mathcal{L}_{\mathrm{ph}}$ be Markovian and that $\rho_{\mathrm{ph}}^{\mathrm{ss}}$ be representable in a truncated photon Fock space.
Gaussianity enters the present calculations mainly through closed one-time Peierls averages obtained from Gaussian characteristic functions, as in the mean-link formulas of Sec.~\ref{sec:benchmark}, and through the Bogoliubov-frame damping generator of Eq.~\eqref{eq:lindblad_bogoliubov_main}, $\mathcal{L}_{\mathrm{ph}}=\kappa\mathcal{L}_{\mathrm{diss}}[b]$, which uniquely stabilizes the pure Gaussian steady states used numerically; the shared-history propagation already treats the link as an operator in Fock space and does not rely on a Wick closure of photon correlators.
Extending the study to non-Gaussian light therefore requires (i) a Markovian source channel whose unique steady state is the target non-Gaussian state, itself a nontrivial reservoir-engineering step beyond replacing $\mathcal{L}_{\mathrm{diss}}[b]$, and (ii) recomputation of one-time references without Gaussian characteristic-function identities; the practical bottlenecks are then the construction of that dissipator, the photon Fock truncation for highly nonclassical states, and the loss of closed mean-link formulas, rather than a reformulation of the shared-history kernels.

A related open direction is the multiorbital extension.
The gauge-covariant Peierls structure of Eq.~\eqref{eq:hk_peierls_general} already admits matrix-valued hopping channels $\bm{\gamma}_\ell(k)$, and the four-sector active-leg topology together with the six chronological orderings remains available because it tracks left/right activation rather than orbital labels, although the internal sector spaces and endpoint maps must in general be enlarged to retain orbital indices.
Independent bands reduce to decoupled copies of the present single-orbital theory when a fixed basis simultaneously diagonalizes the electronic Hamiltonian, the wide-band bath self-energy, and the Peierls channels, so that neither the bath nor the light--matter coupling reintroduces orbital mixing.
Genuine interorbital mixing, however, replaces the scalar shared-history kernel by an orbital-superoperator-valued kernel and requires a matrix equal-time covariance whose photon trace equals the identity on the orbital space.
A naive promotion of $h_k$ and $G_k$ to matrices is not generically guaranteed to preserve this normalization: trace preservation of the source evolution alone does not enforce the required identity normalization after the photon partial trace in orbital space.
The positivity, Pauli bound, and CAR identities of Sec.~\ref{sec:operator_reconstruction} must therefore be re-established before the present Sambe-space construction can be reused with the same guarantees.

It is also useful to contrast the present formulation with phase-space and external-field approaches based on decompositions over c-number field histories \cite{Gorlach2023, EvenTzur2023, Gothelf2025, Li2026, Imai2026a, Imai2026, Imai2026b}.
Both strategies prescribe the light source externally and neglect self-consistent backaction, but they retain different parts of the photon-sector physics.
The cited approaches range from exact or generalized quasiprobability decompositions to approximate incoherent-mixture prescriptions; both classes are efficient for bright fields with macroscopic photon numbers, whereas the present framework keeps the photon field operator-valued, so that noncommuting link correlators, the vacuum phase variance, and the reservoir linewidth $\kappa$ enter the electron Green's functions directly.
This is advantageous for squeezed and few-photon sources, and for two-time steady-state spectra with all Keldysh components obtained within one shared-history construction; conversely, phase-space methods scale more naturally to multimode pulsed fields and bright non-Gaussian states.
The two strategies are complementary, and the coherent-state-ensemble limit could serve as an additional bright-field benchmark for the present framework.

\section{Summary}\label{sec:summary}
We have developed a Floquet Green's-function framework for lattice electrons driven by a reservoir-stabilized Gaussian quantum light source, retaining the full Peierls coupling nonperturbatively within a prescribed-source approximation.
Shared-history convolutions with continuous wide-band bath kernels define the lesser and greater components, and the bath CAR identity closes the retarded and advanced components through the equal-time covariance $W_k$.
On a minimal one-dimensional single-band model, finite-coupling coherent sources deviate from their purely classical coherent Floquet limit, squeezed vacuum reconstructs the retarded spectrum with a modest occupancy redistribution, and the squeezing parameter and phase provide additional control over sideband structure and occupied weight.
Comparing shared-history spectra with the deterministic Floquet problem generated by the complete periodic mean link $\bar{D}(t)$ shows that the ASC--PSC contrast $\Delta_{\mathrm{ML}}$ is already present in the nonzero harmonics of $\bar{D}(t)$, that shared-history dynamics further reshape that contrast into the shared-history contrast $\Delta_{\mathrm{SH}}$ with local sign reversals, and that $\kappa$-dependent changes in squeezed-vacuum spectra at fixed mean link arise from multitime source dynamics.
The numerical results satisfy spectral positivity, fermionic occupation bounds, and the retarded sum rule within the reported finite-window numerical accuracy.
Natural next steps include multiorbital extensions that re-establish the matrix CAR and Pauli bounds under interorbital mixing, staged restoration of electron-induced photon backaction from mean-field updates to two-point polarization feedback, and non-Gaussian sources with engineered Markovian dissipators beyond the Bogoliubov damping dissipator $\mathcal{L}_{\mathrm{diss}}[b]$.

\begin{acknowledgments}
The author is grateful to Shohei Imai for valuable discussions.
This work was supported by the Japan Society for the Promotion of Science (JSPS) KAKENHI Grants No.\ JP23K13052, No.\ JP24K00563, No.\ JP26K06993, and No.\ JP26K00646.
The numerical calculations were performed using the facilities of the Supercomputer Center, the Institute for Solid State Physics, the University of Tokyo.
\end{acknowledgments}

\appendix
\section{Born and SCBA self-energies}\label{sec:app_born_scba}

This appendix gives the explicit closed-form construction of the displacement propagator $\mathscr{D}_{ss'}$ for the Born approximation and SCBA, and summarizes how these self-energies are assembled and, for SCBA, iterated numerically.

\paragraph*{Closed-form connected correlator.}
Introduce the real-time (non-contour-ordered) field covariance
\begin{align}
\mathcal{W}(t,t') \equiv \left\langle \delta X_{\mathrm{lab}}(t)\,\delta X_{\mathrm{lab}}(t')\right\rangle_{\mathrm{ss}},
\label{eq:app_Wcal_def}
\end{align}
where $\delta X_{\mathrm{lab}}$ is defined in Eq.~\eqref{eq:deltaX_def}.
Its greater and lesser orderings are $\mathcal{W}^{>}(t,t')=\mathcal{W}(t,t')$ and $\mathcal{W}^{<}(t,t')=\mathcal{W}(t',t)$.
The contour-ordered correlator of Eq.~\eqref{eq:DX_def} is $D_X(z,z')=-\mathrm{i}\left\langle\mathcal{T}_{C}\,\delta X_{\mathrm{lab}}(z)\delta X_{\mathrm{lab}}(z')\right\rangle_{\mathrm{ss}}$, with Langreth components $D_X^{\gtrless}(t,t')=-\mathrm{i}\,\mathcal{W}^{\gtrless}(t,t')$.
Because both the steady state in Eq.~\eqref{eq:gaussian_state_def} and the linear source dynamics in Eq.~\eqref{eq:lindblad_bogoliubov_main} are Gaussian, all cumulants of $\delta X_{\mathrm{lab}}$ above second order vanish, and expectation values of exponentials linear in $\delta X_{\mathrm{lab}}$ close at second order.
With the decomposition $X_{\mathrm{lab}}=X_{\mathrm{cl}}+\delta X_{\mathrm{lab}}$ and the directed Peierls factors of Eq.~\eqref{eq:Ds_def},
\begin{align}
D_s(t)
= \exp\bigl[+\mathrm{i} s\lambda X_{\mathrm{cl}}(t)\bigr] \exp\bigl[+\mathrm{i} s\lambda\,\delta X_{\mathrm{lab}}(t)\bigr],
\label{eq:app_Ds_split}
\end{align}
where the classical mean factors out because it is a c-number and therefore commutes with $\delta X_{\mathrm{lab}}$.
The one-point Gaussian identity
\begin{align}
\langle \exp[+\mathrm{i}\xi\,\delta X_{\mathrm{lab}}(t)] \rangle_{\mathrm{ss}}
= \exp\left[-\frac{1}{2}\xi^2\mathcal{W}(t,t)\right]
\label{eq:app_gauss_onept}
\end{align}
with $\xi=s\lambda$ (hence $\xi^2=\lambda^2$) then gives the mean link
\begin{align}
\bar{D}_s(t)
= \exp\left[ \mathrm{i} s\lambda X_{\mathrm{cl}}(t)
- \frac{1}{2}\lambda^2\mathcal{W}(t,t) \right].
\label{eq:app_Dbar_closed}
\end{align}
For the displayed operator ordering, the Baker--Campbell--Hausdorff commutator combines with the symmetrized Gaussian covariance to give the ordered covariance $\mathcal{W}^{>}(t,t')$.
The corresponding two-point identity is
\begin{align}
&\langle
\exp[+\mathrm{i}\xi\,\delta X_{\mathrm{lab}}(t)]
\exp[+\mathrm{i}\xi'\,\delta X_{\mathrm{lab}}(t')]
\rangle_{\mathrm{ss}}
\notag\\
&= \exp\left[
-\frac{1}{2}\xi^2\mathcal{W}(t,t)
-\frac{1}{2}\xi'^2\mathcal{W}(t',t')
-\xi\xi'\,\mathcal{W}^{>}(t,t')
\right],
\label{eq:app_gauss_twopt}
\end{align}
with $(\xi,\xi')=(s\lambda,s'\lambda)$.
This yields the two-point characteristic function
\begin{align}
C_{ss'}^{>}(t,t')
&\equiv
\langle D_s(t)D_{s'}(t') \rangle_{\mathrm{ss}}
\notag\\
&= \bar{D}_s(t)\,\bar{D}_{s'}(t')\,
\exp\left[-ss'\lambda^2\mathcal{W}^{>}(t,t')\right],
\label{eq:app_Css_closed}
\end{align}
with $C_{ss'}^{<}(t,t')\equiv\langle D_{s'}(t')D_s(t)\rangle_{\mathrm{ss}}$ given by the same expression with $\mathcal{W}^{>}\to\mathcal{W}^{<}$.
The connected propagator of Eq.~\eqref{eq:Dscr_ss_def} then follows as
\begin{align}
\mathscr{D}_{ss'}^{\gtrless}(t,t')
= -\mathrm{i}\bigl[C_{ss'}^{\gtrless}(t,t')-\bar{D}_s(t)\bar{D}_{s'}(t')\bigr].
\label{eq:app_Dscr_closed}
\end{align}

\paragraph*{Photon damping rate $\kappa$ via the quantum regression theorem.}
The Bogoliubov decomposition $\delta X_{\mathrm{lab}}(t)=f(t)b+f^*(t)b^\dagger$ [Eq.~\eqref{eq:deltaX_def}] reduces $\mathcal{W}(t,t')$ to two-time correlators of the Bogoliubov mode $b$ under the source Lindbladian $\mathcal{L}_{\mathrm{ph}}\rho=\kappa\mathcal{L}_{\mathrm{diss}}[b]\rho$ [Eq.~\eqref{eq:lindblad_bogoliubov_main}].
Taking expectation values of the Lindblad equation directly gives
\begin{align}
\partial_t \langle b\rangle = -\frac{\kappa}{2}\langle b\rangle,
\quad
\partial_t \langle b^\dagger b\rangle = -\kappa\langle b^\dagger b\rangle,
\label{eq:app_b_decay_rates}
\end{align}
the second of which confirms the population decay rate $\kappa$ of the engineered damping channel.
Equivalently, the Heisenberg adjoint acts as $\mathcal{L}_{\mathrm{ph}}^\dagger b=-(\kappa/2)b$ and $\mathcal{L}_{\mathrm{ph}}^\dagger b^\dagger=-(\kappa/2)b^\dagger$.
For a Markovian generator, the quantum regression theorem \cite{Lax1963, WallsMilburn2008} states that the same linear rates govern two-time correlators: if $t\geq t'$,
\begin{align}
\left\langle A(t)B(t')\right\rangle_{\mathrm{ss}}
= \Tr\Bigl[A\,\mathrm{e}^{\mathcal{L}_{\mathrm{ph}}(t-t')}(B\rho_{\mathrm{ph}}^{\mathrm{ss}})\Bigr],
\label{eq:app_qrt}
\end{align}
or, when a set of operators closes under one-time averages with a rate matrix $M_{ij}$,
\begin{align}
\partial_t \langle A_i(t)B(t') \rangle_{\mathrm{ss}}
= \sum_j M_{ij} \langle A_j(t)B(t') \rangle_{\mathrm{ss}}.
\label{eq:app_qrt_rates}
\end{align}
For the reverse ordering, $t<t'$, stationarity and the same theorem give the right-insertion form
\begin{align}
\left\langle A(t)B(t')\right\rangle_{\mathrm{ss}}
=
\Tr\Bigl[B\,\mathrm{e}^{\mathcal{L}_{\mathrm{ph}}(t'-t)}(\rho_{\mathrm{ph}}^{\mathrm{ss}}A)\Bigr].
\label{eq:app_qrt_reverse}
\end{align}
Here, $b(t)$ in the two-time correlators denotes the dissipative Heisenberg evolution generated by $\mathcal{L}_{\mathrm{ph}}^\dagger$; the carrier rotation remains explicit in the c-number coefficient $f(t)$.
At the vacuum steady state $\rho_{\mathrm{ph}}^{\mathrm{ss}}=|0_b\rangle\langle 0_b|$, the equal-time seeds are $\langle bb^\dagger\rangle_{\mathrm{ss}}=1$ and $\langle b^\dagger b\rangle_{\mathrm{ss}}=\langle b^2\rangle_{\mathrm{ss}}=0$.
Applying Eq.~\eqref{eq:app_qrt_rates} with $A_i=b$ (so $M=-\kappa/2$) for $t\geq t'$ therefore yields $\langle b(t)b^\dagger(t')\rangle_{\mathrm{ss}}=\mathrm{e}^{-\kappa(t-t')/2}$.
For $t<t'$, Eq.~\eqref{eq:app_qrt_reverse} gives
\begin{align}
\left\langle b(t)b^\dagger(t')\right\rangle_{\mathrm{ss}}
&=
\Tr\left[
b^\dagger
\mathrm{e}^{\mathcal{L}_{\mathrm{ph}}(t'-t)}
\left(\rho_{\mathrm{ph}}^{\mathrm{ss}}b\right)
\right]
=
\mathrm{e}^{-\kappa(t'-t)/2},
\label{eq:app_b_reverse_regression}
\end{align}
because $\rho_{\mathrm{ph}}^{\mathrm{ss}}b=|0_b\rangle\langle1_b|$ is an eigenoperator of $\mathcal{L}_{\mathrm{ph}}$ with eigenvalue $-\kappa/2$.
Combining the two time orderings gives
\begin{gather}
\langle b(t)b^\dagger(t')\rangle_{\mathrm{ss}} = \mathrm{e}^{-\kappa|t-t'|/2},
\label{eq:app_b_bdagger}
\\
\langle b^\dagger(t)b(t')\rangle_{\mathrm{ss}} = \langle b(t)b(t')\rangle_{\mathrm{ss}} = 0,
\label{eq:app_b_regression}
\end{gather}
where the remaining two correlators stay zero because their equal-time seeds vanish and the regression evolution only multiplies them by a damping factor.
Consequently,
\begin{gather}
\mathcal{W}^{>}(t,t') = f(t)f^*(t')\,\mathrm{e}^{-\kappa|t-t'|/2},
\label{eq:app_Wgreater_kappa}
\\
\mathcal{W}^{<}(t,t') =\mathcal{W}^{>}(t',t).
\label{eq:app_Wlesser_from_greater}
\end{gather}
The rate $\kappa$ entering Eq.~\eqref{eq:app_Wgreater_kappa} is the same photon damping rate that appears in the source Lindbladian $\mathcal{L}_{\mathrm{ph}}$ of the main formulation; the relative-time envelope of the two-time field covariance decays at the half-rate $\kappa/2$, while the underlying population decays at the full rate $\kappa$.

\paragraph*{Self-energy assembly and numerical procedure.}
With $\gamma_{+}(k)$ and $\gamma_{-}(k)$ from Eq.~\eqref{eq:hL_singleband}, the Born/SCBA self-energy of Eq.~\eqref{eq:dyson_schematic} is assembled as the vertex sum
\begin{align}
\varSigma_D^{\gtrless}(k;t,t') = \mathrm{i}\sum_{s,s'=\pm1}\gamma_s(k)\gamma_{s'}(k)\,G^{\gtrless}(k;t,t')\,\mathscr{D}_{ss'}^{\gtrless}(t,t'),
\label{eq:app_SigmaD_vertex_sum}
\end{align}
with the directed Peierls vertices $\gamma_{\pm}(k)$ of Eq.~\eqref{eq:hL_singleband}.
Numerically, $\mathscr{D}_{ss'}^{\gtrless}(t,t')$ and $G^{\gtrless}(t,t')$ are converted to Wigner coordinates $(t_{\mathrm{av}},\tau)$ using Eq.~\eqref{eq:wigner_conv}.
The pointwise product in Eq.~\eqref{eq:app_SigmaD_vertex_sum} becomes a convolution in Wigner harmonic and relative frequency,
\begin{align}
&\varSigma_{D,p}^{\gtrless}(k;\omega) \notag \\
&=
\mathrm{i}\sum_{s,s'=\pm1}\gamma_s(k)\gamma_{s'}(k)
\sum_q\int\frac{\mathrm{d}\nu}{2\pi}\,
G_q^{\gtrless}(k;\nu)
\mathscr{D}_{ss',p-q}^{\gtrless}(\omega-\nu).
\label{eq:app_SigmaD_wigner_convolution}
\end{align}
The resulting Wigner components are mapped to Sambe matrices with Eq.~\eqref{eq:wigner_to_sambe}.
The retarded component is reconstructed from the greater and lesser components by the standard time-domain relation
\begin{align}
\varSigma_D^{\mathrm{R}}(t,t') = \varTheta(t-t')\left[\varSigma_D^{>}(t,t')-\varSigma_D^{<}(t,t')\right].
\label{eq:app_SigmaR_theta}
\end{align}
For clarity, define the periodic mean-link Hamiltonian
\begin{align}
\bar h_k(t)
&\equiv
\sum_{s=\pm1}\gamma_s(k)\bar{D}_s(t),
\label{eq:app_mean_link_hamiltonian}
\end{align}
with Fourier harmonics $\bar h_{k,r}$.
Its Floquet--Sambe Hamiltonian is
\begin{align}
\bigl[H_{\mathrm{ML},\mathrm{F}}(k)\bigr]_{mn}
= \bar h_{k,m-n} - m\varOmega\,\delta_{mn}.
\label{eq:app_H_ML_F}
\end{align}
The associated mean-link Floquet Green's function $G_{\mathrm{ML}}$ is the bath-dressed solution of Eqs.~\eqref{eq:classical_sambe_matrix}--\eqref{eq:classical_lg_equation} with this drive and without the Born/SCBA resolvent regulator below; its spectrum $A_{\mathrm{ML}}^{\mathrm{R}}$ is the deterministic mean-link spectrum used in Sec.~\ref{sec:benchmark}.
Because $\langle\delta D_s(t)\rangle_{\mathrm{ss}}=0$, no separate first-order tadpole self-energy is included; the one-time mean link is already incorporated in $H_{\mathrm{ML},\mathrm{F}}(k)$.

The Born and SCBA Green's functions are obtained from the component Keldysh--Dyson equations
\begin{align}
G^{\mathrm{R}}(k,\omega)
&=
\Bigl[
\omega+\mathrm{i}\eta_{\mathrm{reg}}
-H_{\mathrm{ML},\mathrm{F}}(k)
\notag\\
&\quad
-\varSigma_{\mathrm{bath}}^{\mathrm{R}}(\omega)
-\varSigma_D^{\mathrm{R}}(k,\omega)
\Bigr]^{-1},
\label{eq:app_born_scba_GR}
\\
G^{\mathrm{A}}(k,\omega)
&=
\bigl[G^{\mathrm{R}}(k,\omega)\bigr]^\dagger,
\label{eq:app_born_scba_GA}
\\
G^{\gtrless}(k,\omega)
&=
G^{\mathrm{R}}(k,\omega)
\bigl[
\varSigma_{\mathrm{bath}}^{\gtrless}(\omega)
+\varSigma_D^{\gtrless}(k,\omega)
\bigr]
G^{\mathrm{A}}(k,\omega),
\label{eq:app_born_scba_Keldysh}
\end{align}
where all products are Sambe-matrix products.
Here, the bath self-energy is momentum independent, whereas $\varSigma_D$ depends on $k$ through the Peierls vertices.
The regulator $\eta_{\mathrm{reg}}=10^{-4}$ is introduced only in the numerical solution of the Born and SCBA equations.
Equivalently,
\begin{align}
(G^{\mathrm{R},-1})_{mn}
&=
(\omega+\mathrm{i}\eta_{\mathrm{reg}}+m\varOmega)\,\delta_{mn}
-\bar h_{k,m-n}
\notag\\
&\quad
-\varSigma_{\mathrm{bath},mn}^{\mathrm{R}}(\omega)
-\varSigma_{D,mn}^{\mathrm{R}}(k,\omega).
\label{eq:app_born_scba_GR_components}
\end{align}
The retarded fluctuation self-energy follows from Eq.~\eqref{eq:app_SigmaR_theta}.
For the Born input and the SCBA initialization, we use the regularized mean-link solution
\begin{align}
G_{\mathrm{ML,reg}}^{\mathrm{R}}(k,\omega)
&=
\Bigl[
\omega+\mathrm{i}\eta_{\mathrm{reg}}
-H_{\mathrm{ML},\mathrm{F}}(k)
-\varSigma_{\mathrm{bath}}^{\mathrm{R}}(\omega)
\Bigr]^{-1},
\label{eq:app_G_ML_reg}
\end{align}
together with the corresponding bath-only Keldysh components from Eq.~\eqref{eq:app_born_scba_Keldysh} at $\varSigma_D=0$.
This $G_{\mathrm{ML,reg}}$ differs from the unregularized $G_{\mathrm{ML}}$ of Sec.~\ref{sec:benchmark} only by $\eta_{\mathrm{reg}}$.

In the Born approximation, the electronic Green's function entering Eq.~\eqref{eq:app_SigmaD_vertex_sum} is fixed to $G^{\gtrless}\to G_{\mathrm{ML,reg}}^{\gtrless}$, and the resulting $\varSigma_D^{\mathrm{R},\gtrless}$ is evaluated once before solving Eqs.~\eqref{eq:app_born_scba_GR}--\eqref{eq:app_born_scba_Keldysh}.
In SCBA, the dressed $G^{\gtrless}$ generated together with $G^{\mathrm{R},A}$ by Eqs.~\eqref{eq:app_born_scba_GR}--\eqref{eq:app_born_scba_Keldysh} is reinserted into Eq.~\eqref{eq:app_SigmaD_vertex_sum} until self-consistency is reached.
The prescribed displacement correlator $\mathscr{D}_{ss'}$ is not updated.
The SCBA iteration is initialized with $G_{\mathrm{ML,reg}}$.
Let $\widetilde{G}_{n+1}^{x}$ denote the unmixed Dyson iterate for $x\in\{\mathrm{R},>,<\}$.
After computing this iterate, the retarded, lesser, and greater components are linearly mixed according to
\begin{align}
G_{n+1}^{x}
= (1-\eta_{\mathrm{mix}})G_n^{x} + \eta_{\mathrm{mix}}\widetilde{G}_{n+1}^{x},
\quad
x\in\{\mathrm{R},>,<\},
\label{eq:app_scba_mix}
\end{align}
with $\eta_{\mathrm{mix}}=0.02$, while $G^{\mathrm{A}}=(G^{\mathrm{R}})^\dagger$ is reconstructed from the mixed retarded component.
Convergence is assessed from the unmixed fixed-point residual,
\begin{align}
\epsilon_n
=
\max_{x\in\{\mathrm{R},>,<\}}
\frac{
\bigl\Vert\widetilde{G}_{n+1}^{x}-G_n^{x}\bigr\Vert
}{
\max\bigl(\Vert G_n^{x}\Vert,\,10^{-30}\bigr)
}
<10^{-6},
\label{eq:app_scba_tol}
\end{align}
The iteration is capped at $N_{\mathrm{iter}}^{\max}=10^{4}$.
All SCBA results reported in this work satisfy this convergence criterion before reaching $N_{\mathrm{iter}}^{\max}$.
Here, $\Vert\bullet\Vert$ denotes the Frobenius norm over the retained Sambe indices and frequency grid at fixed momentum.
The Born and SCBA calculations use a Sambe cutoff $M=10$ and retain Wigner harmonics $p=0,\pm 1, \ldots, \pm P$ with $P=2M=20$.
The external-frequency grid is $\omega\in[-12,12]$ with spacing $\Delta\omega=0.01$ ($N_\omega=2401$), covering a different interval from the shared-history window in Table~\ref{tab:benchmark_params}.
The convolution in Eq.~\eqref{eq:app_SigmaD_wigner_convolution} is evaluated on the same uniform grid $\nu_j=\omega_j$ with uniform weights $\Delta\nu/(2\pi)$ and $\Delta\nu=\Delta\omega$; shifted-frequency values outside the retained interval $[-12,12]$ are set to zero in the truncated convolution.
The Wigner transforms are discretized using $N_{t_{\mathrm{av}}}=128$ average-time points over one drive period and a relative-time grid with $\tau_{\max}=100$ and $N_\tau=8192$.

\section{Frequency-domain implementation of the six orderings}\label{sec:app_implementation}

This appendix collects the factorized frequency-domain building blocks that implement the shared-history lesser and greater components of Sec.~\ref{sec:bath_car}, complementing the algorithm overview of Sec.~\ref{sec:numerical_notes}.
The maps $\mathbf{A}_{\mathsf{L},\mathsf{R}}$ and $\mathbf{D}_{\mathsf{L},\mathsf{R}}$ are those of Eq.~\eqref{eq:sector_resolvent_chain}, and the sector resolvents $\mathbb{R}_{\mathsf{L}}$, $\mathbb{R}_{\mathsf{R}}$, and $\mathbb{R}_{\mathsf{LR}}$ are the instances of Eq.~\eqref{eq:branch_resolvents} for $q\in\{\mathsf{L},\mathsf{R},\mathsf{LR}\}$.
For every bath injection frequency, the charged one-leg injection vectors are
\begin{gather}
\Lket{u_{\mathsf{L}}(\nu)}
=
\mathbb{R}_{\mathsf{L}}(\nu)\mathbf{A}_{\mathsf{L}}\Lket{\rho_{\mathrm{F}}},
\label{eq:factorized_injection_L}
\\
\Lket{u_{\mathsf{R}}(\nu)}
=
\mathbb{R}_{\mathsf{R}}(-\nu)\mathbf{A}_{\mathsf{R}}\Lket{\rho_{\mathrm{F}}}.
\label{eq:factorized_injection_R}
\end{gather}
For every observation frequency $\omega$ of Eqs.~\eqref{eq:endpoint_charges} and \eqref{eq:keldysh_column_solve}, the terminal dual vectors are
\begin{gather}
\Lbra{\varLambda_{\mathsf{R}}(\omega)}
=
\Lbra{I_{\mathrm{F}}}\mathbf{D}_{\mathsf{R}}\mathbb{R}_{\mathsf{R}}(-\omega),
\label{eq:factorized_terminal_R}
\\
\Lbra{\varLambda_{\mathsf{L}}(\omega)}
=
\Lbra{I_{\mathrm{F}}}\mathbf{D}_{\mathsf{L}}\mathbb{R}_{\mathsf{L}}(\omega).
\label{eq:factorized_terminal_L}
\end{gather}
These injection and terminal objects are generated once and reused across components and bath frequencies.
The four overlap orderings are combined in the pairs $(3,6)$ and $(4,5)$ on the same bath-frequency grid before integration.
Define
\begin{gather}
\Lket{s_{\mathsf{LR}}^{\gtrless}}
=
\int_{-\infty}^{\infty}\frac{\mathrm{d}\nu}{2\pi}\,
\varSigma_{\mathrm{bath}}^{\gtrless}(\nu)
\left[
\mathbf{A}_{\mathsf{R}}\Lket{u_{\mathsf{L}}(\nu)}
+
\mathbf{A}_{\mathsf{L}}\Lket{u_{\mathsf{R}}(\nu)}
\right],
\label{eq:factorized_overlap_s}
\\
\Lket{v_{\mathsf{LR}}^{\gtrless}}
=
\mathbb{R}_{\mathsf{LR}}(0)\Lket{s_{\mathsf{LR}}^{\gtrless}}.
\label{eq:factorized_overlap_v}
\end{gather}
The overlap contribution is
\begin{align}
G_{k,\mathrm{ov}}^{\gtrless}(\omega)
&=
\Lbra{\varLambda_{\mathsf{R}}(\omega)}
\mathbf{D}_{\mathsf{L}}
\Lket{v_{\mathsf{LR}}^{\gtrless}}
+
\Lbra{\varLambda_{\mathsf{L}}(\omega)}
\mathbf{D}_{\mathsf{R}}
\Lket{v_{\mathsf{LR}}^{\gtrless}}.
\label{eq:factorized_overlap_green}
\end{align}
This pairing retains the cancellation of their leading large-frequency terms and avoids assigning separate quadrature errors to contributions that are only regular in their sum.

The two nonoverlap orderings contain the source-only resolvent.
Introduce
\begin{align}
\Lket{f_{\mathsf{L}}^{\gtrless}(\nu)}
&=
\varSigma_{\mathrm{bath}}^{\gtrless}(\nu)
\mathbf{D}_{\mathsf{L}}\Lket{u_{\mathsf{L}}(\nu)},
\\
\Lket{f_{\mathsf{R}}^{\gtrless}(\nu)}
&=
\varSigma_{\mathrm{bath}}^{\gtrless}(\nu)
\mathbf{D}_{\mathsf{R}}\Lket{u_{\mathsf{R}}(\nu)}.
\end{align}
The two source-only convolutions are
\begin{align}
\Lket{c_+^{\gtrless}(\omega)}
&=
\int_{-\infty}^{\infty}\frac{\mathrm{d}\nu}{2\pi}\,
\mathbb{R}_0(\nu-\omega)
\Lket{f_{\mathsf{L}}^{\gtrless}(\nu)},
\label{eq:factorized_nonoverlap_cplus}
\\
\Lket{c_-^{\gtrless}(\omega)}
&=
\int_{-\infty}^{\infty}\frac{\mathrm{d}\nu}{2\pi}\,
\mathbb{R}_0(\omega-\nu)
\Lket{f_{\mathsf{R}}^{\gtrless}(\nu)},
\label{eq:factorized_nonoverlap_convolution}
\end{align}
where $\mathbb{R}_0$ is the source-only resolvent of Eq.~\eqref{eq:replica_resolvent}.
Their readout is
\begin{align}
G_{k,\mathrm{no}}^{\gtrless}(\omega)
&=
\Lbra{\varLambda_{\mathsf{R}}(\omega)}
\mathbf{A}_{\mathsf{R}}
\Lket{c_+^{\gtrless}(\omega)}
+
\Lbra{\varLambda_{\mathsf{L}}(\omega)}
\mathbf{A}_{\mathsf{L}}
\Lket{c_-^{\gtrless}(\omega)}.
\label{eq:factorized_nonoverlap_green}
\end{align}
Its nondecaying part is evaluated from the full replica projector $P_{\mathrm{rep}}$.
Substituting the replica poles of Eq.~\eqref{eq:replica_resolvent} into Eqs.~\eqref{eq:factorized_nonoverlap_cplus}--\eqref{eq:factorized_nonoverlap_convolution} produces bath integrals of $\mathrm{i}/[\pm(\nu-\omega)+m\varOmega+\mathrm{i}0^+]$ against the projected sources $\Lket{f_{\mathsf{L,R}}^{\gtrless}(\nu)}$; each such pole contributes a delta and a principal-value term.
With $\nu_m=\omega\mp m\varOmega$ in the $c_\pm$ convolution and $F$ the corresponding projected bath-weighted source component, the principal-value integrals on the finite bath window $[a,b]$ are evaluated by the subtracted formula
\begin{align}
&\operatorname{PV}
\int_a^b\mathrm{d}\nu\,
\frac{F(\nu)}{\nu-\nu_m} \notag \\
&=
\int_a^b\mathrm{d}\nu\,
\frac{F(\nu)-F(\nu_m)}{\nu-\nu_m}
+
F(\nu_m)
\ln\left|
\frac{b-\nu_m}{\nu_m-a}
\right|.
\label{eq:subtracted_pv}
\end{align}
The subtraction removes sensitivity to whether a pole lies on a quadrature node without changing the continuum principal value.
The regular connected part is evaluated in the $Q_c$ sector, either through a frequency-domain shifted solve or through the equivalent time-domain memory integral.
The replica, overlap, residue, and connected contributions use one common bath-frequency grid, including the same endpoints and quadrature weights, because their leading discretization errors cancel only in the combined response.
Finite-$\eta$ resolvents provide an independent check of the $\eta\to0^+$ limit and do not define the pole prescription used here.

The Floquet harmonics $G^{<,>}_{k,m0}$ with $m\neq0$ of Eq.~\eqref{eq:m0_wigner_map} reuse these building blocks with a single left-leg Sambe shift.
Introduce the Sambe translation $(\mathbb{S}_r x)_a = x_{a-r}$ with vanishing entries outside the retained window, so that the shifted trace vectors of Eq.~\eqref{eq:replica_mode_projector} obey $\Lbra{I^{(m)}}=\Lbra{I_{\mathrm{F}}}\mathbb{S}_{-m}$.
The physical $G^{<,>}_{k,m0}$ is assembled by inserting $\mathbb{S}_{-m}$ at the left-leg deactivation boundary of each of the six orderings, while keeping the shared reduced-zone frequency assignment and the replica decomposition of Eqs.~\eqref{eq:replica_projector}--\eqref{eq:replica_resolvent}.
Equivalently, the shift can be implemented by moving the source column $\Lket{\rho_{\mathrm{F}}}\to \mathbb{S}_{-m}\Lket{\rho_{\mathrm{F}}}$ and the affected terminals to $\Lbra{I^{(m)}}$.
Extracting only the terminal $m$-block, without the shifted source column, does not reproduce the harmonic for $m\neq0$.
With $\mathbf{D}_{\mathsf{L}}^{(m)}=\mathbf{D}_{\mathsf{L}}\mathbb{S}_{-m}$ and the shifted left terminal $\Lbra{\varLambda_{\mathsf{L}}^{(m)}(\omega)}=\Lbra{I^{(m)}}\mathbf{D}_{\mathsf{L}}\mathbb{R}_{\mathsf{L}}(\omega)$, the overlap readout of Eq.~\eqref{eq:factorized_overlap_green} becomes
\begin{align}
G_{k,\mathrm{ov},m0}^{\gtrless}(\omega)
&=
\Lbra{\varLambda_{\mathsf{R}}(\omega)}
\mathbf{D}_{\mathsf{L}}^{(m)}
\Lket{v_{\mathsf{LR}}^{\gtrless}}
+
\Lbra{\varLambda_{\mathsf{L}}^{(m)}(\omega)}
\mathbf{D}_{\mathsf{R}}
\Lket{v_{\mathsf{LR}}^{\gtrless}},
\label{eq:factorized_overlap_m0}
\end{align}
and the nonoverlap readout of Eq.~\eqref{eq:factorized_nonoverlap_green} replaces $\mathbf{D}_{\mathsf{L}}\to\mathbf{D}_{\mathsf{L}}^{(m)}$ in $\Lket{f_{\mathsf{L}}^{\gtrless}(\nu)}$ for the left-deactivating ordering and $\Lbra{\varLambda_{\mathsf{L}}(\omega)}\to\Lbra{\varLambda_{\mathsf{L}}^{(m)}(\omega)}$ for the other.
The shared injection $u_{\mathsf{L,R}}$, overlap $v_{\mathsf{LR}}$, right terminal $\varLambda_{\mathsf{R}}$, replica resolvent $\mathbb{R}_0$, and bath grid are unchanged, so each harmonic adds only one shifted left terminal and one shifted nonoverlap column, and $m=0$ returns Eqs.~\eqref{eq:factorized_overlap_green} and \eqref{eq:factorized_nonoverlap_green}.
In the deterministic or source-decoupled limits, these harmonics reduce to the Floquet product $\sum_{l}G^{\mathrm{R}}_{ml}(\omega)\varSigma_{\mathrm{bath}}^{<}(\omega+l\varOmega)G^{\mathrm{A}}_{l0}(\omega)$ and, for a constant bath occupation $f_0$, to the identity $G^{<}_{m0}=-f_0\,(G^{\mathrm{R}}_{m0}-G^{\mathrm{A}}_{m0})$.
Outside those limits, the shared-history two-leg convolution is used instead, and the product formulas serve only as classical-limit checks.
The reduced-zone harmonics are identified without double counting by requiring $\varOmega/(2\Delta\varepsilon)\in\mathbb{N}$ for the frequency-grid spacing $\Delta\varepsilon$, so that $\omega+\tfrac{m}{2}\varOmega$ coincides with a discrete frequency point.

The direct lesser convolution contains a sharply varying constant wide-band contribution whose trapezoidal error can dominate occupied observables.
We reduce this discretization error with a CAR-based correction for the quadrature error.
The frequency-domain convolutions use the full kernels $\varSigma_{\mathrm{bath}}^{<,>}(\nu)$ of Eqs.~\eqref{eq:wbl_sigma_lesser_freq} and \eqref{eq:wbl_sigma_enlarged}, and therefore include the contact and noncontact contributions together.
The uncorrected numerical lesser component is
\begin{align}
G_{\mathrm{num}}^<(\omega)
&=
G_{k,\mathrm{ov}}^<(\omega)
+
G_{k,\mathrm{no}}^<(\omega).
\end{align}
Let $\Delta_{\mathrm{const}}$ denote the discrete response to the constant CAR kernel $\varSigma_{\mathrm{bath}}^>-\varSigma_{\mathrm{bath}}^<=-\mathrm{i}\varGamma$, evaluated with the same discretization as $G_{\mathrm{num}}^<$.
The corrected lesser component is
\begin{align}
G_{\mathrm{cv}}^<(\omega)
&=
G_{\mathrm{num}}^<(\omega)
+
f_{\mathrm{bath}}(\omega)
\,
\Delta_{\mathrm{const}}(\omega)
\nonumber\\
&\quad
-
f_{\mathrm{bath}}(\omega)
\left[
G^{\mathrm{R}}(\omega)-G^{\mathrm{A}}(\omega)
\right].
\label{eq:car_quadrature_correction}
\end{align}
The correction vanishes in the continuum limit and exactly restores $G^<=-f_0(G^{\mathrm{R}}-G^{\mathrm{A}})$ for a constant bath occupation $f_0$, up to the shifted-solve tolerance.
At the adopted spacing $\Delta\nu=0.0125$ on the most sensitive coherent $k=\pi/2$ probe, the maximum absolute correction is $\max|G_{\mathrm{cv}}^<-G_{\mathrm{num}}^<|=3.52\times10^{-4}$ ($1.07\times10^{-4}$ relative to $\max|G_{\mathrm{cv}}^<|$), while $\max|A_{\mathrm{cv}}^<-A_{\mathrm{num}}^<|=5.60\times10^{-5}$ and the occupation shift is $2.3\times10^{-5}$;
for constant $f_0$ the corrected residual falls to $4.4\times10^{-15}$.

\section{Numerical convergence}\label{sec:app_cutoff}

\begin{figure}[t]\centering
\includegraphics[scale=1]{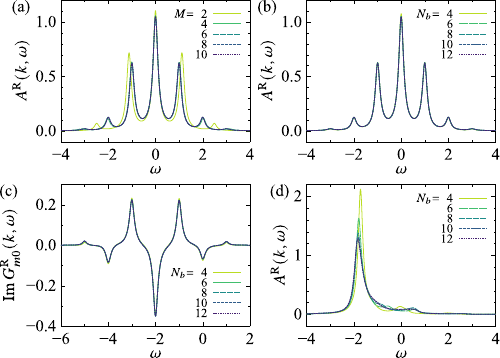}
\caption{Cutoff convergence of the shared-history spectra for $\kappa=0.1$, with $M_D=2M$ throughout.
Panels (a)--(c) use the coherent source with $A_{\mathrm{cl}}=1.1$ and $\lambda=1$ as in Fig.~\ref{fig:coherent_lam1} at $k=\pi/2$; panel (d) uses squeezed vacuum with $r=4$ and $\lambda=0.02$ as in Fig.~\ref{fig:squeezed_r4_kappa01} at $k=0$.
(a) period-averaged retarded spectrum $A^{\mathrm{R}}(k,\omega)$ under an $M$ sweep at fixed $N_b=8$;
(b) $A^{\mathrm{R}}(k,\omega)$ under an $N_b$ sweep at fixed $M=8$;
(c) harmonic-resolved $\im G^{\mathrm{R}}_{m0}(k,\omega)$ at $m=4$ under an $N_b$ sweep at fixed $M=8$;
(d) $A^{\mathrm{R}}(k,\omega)$ for squeezed vacuum under an $N_b$ sweep at fixed $M=8$.
}
\label{fig:convergence}
\end{figure}

Three cutoffs control the numerical accuracy of the shared-history calculations: the photon Fock dimension $N_b$, the Peierls harmonic cutoff $M_D$, and the Floquet harmonic cutoff $M$.
The primary finite photon regulator is the projected displacement operator $P_{N_b}D(\beta)P_{N_b}$ in Eq.~\eqref{eq:link_matrix_exp}, which is not exactly unitary at finite $N_b$.
Photon-cutoff convergence is monitored through the stability of the spectra and the Fock-space edge and tail weight after acting with $D_\ell(t)$ on representative low-lying photon states.
The Peierls harmonic cutoff is checked via the Fourier-harmonic edge weight near $|p|=M_D$.
On the Sambe side, convergence is assessed by increasing $M$, monitoring sideband edge weights of the computed Green's functions, and verifying stability of $A^{\mathrm{R}}, A^<, A^>$, and $f_{\mathrm{eff}}$.

Figure~\ref{fig:convergence} shows representative sweeps at the most stringent source damping used in the main figures, $\kappa=0.1$.
Panels (a)--(c) use the coherent-source parameters of Fig.~\ref{fig:coherent_lam1} ($A_{\mathrm{cl}}=1.1$, $\lambda=1$) at $k=\pi/2$, whereas panel (d) uses the squeezed-vacuum parameters of Fig.~\ref{fig:squeezed_r4_kappa01} ($r=4$, $\lambda=0.02$) at $k=0$.
For the coherent period-averaged spectrum $A^{\mathrm{R}}$, the Sambe sweep in panel (a) shows that $M=2$ misses the sideband structure entirely, whereas the difference from the $M=10$ result falls to about $6\times10^{-4}$ already at $M=8$; we therefore use $M=10$ with $M_D=2M=20$, the minimal Peierls cutoff covering all Sambe differences $|m-n|\le 2M$.
The photon sweep in panel (b) is milder: $N_b=4$ deviates by about $3\times10^{-2}$, dropping to $2\times10^{-3}$ at $N_b=8$ and $6\times10^{-4}$ at $N_b=10$.
Panel (c) uses the high sideband harmonic $\im G^{\mathrm{R}}_{m0}$ at $m=4$ as a more stringent photon-cutoff probe; the maximum difference from the $N_b=12$ curve decreases monotonically from about $10^{-2}$ at $N_b=4$ to $2\times10^{-4}$ at $N_b=10$.
The most sensitive case is squeezed vacuum at $k=0$ in panel (d), where the sharp band-edge peak requires the largest photon cutoff.
Successive $N_b$ overlays approach one another for $N_b\geq10$: the remaining pointwise difference between $N_b=10$ and $N_b=12$ reaches about $4\times10^{-2}$ near the peak, while the spectral sum and occupation change only by $\sim4\times10^{-6}$ and $\sim7\times10^{-4}$, respectively.
We therefore adopt $N_b=12$ for the reported calculations.
Because $\kappa=0.1$ is the smallest damping rate used in the main figures, these panels are conservative: calculations at larger $\kappa$ and for the lower-order components used elsewhere converge more rapidly than the curves shown here.

The occupied components require two additional numerical refinements.
The bath integral uses one common $\nu\in[-25,25]$ grid for the replica, overlap, residue, and connected contributions, with $\Delta\nu=0.0125$ and the subtracted principal value applied to every retained replica.
Comparisons of the reported $A^{\mathrm{R}}$ and CAR-corrected $A_{\mathrm{cv}}^{<}$ [from $G_{\mathrm{cv}}^{<}$ of Eq.~\eqref{eq:car_quadrature_correction}] with staggered and twice-finer grids give differences below $10^{-11}$.
At fixed coarse spacing, the finite-window residual is essentially flat for $\nu_{\max}$ from $8$ to $32$, so the adopted window $\nu\in[-25,25]$ is not the limiting error source relative to $\Delta\nu$.
The connected memory integral uses the plain trapezoidal rule at $\Delta\tau=0.00625$ and $\epsilon_{\mathrm{mem}}=10^{-10}$, with $\tau_{\max}=-2\ln(\epsilon_{\mathrm{mem}})/\kappa$.
Doubling the memory window from $-\ln(\epsilon_{\mathrm{mem}})/\kappa$ to this amplitude-rate choice changes $A^<$ by at most $5\times10^{-12}$ on the $\kappa=0.1$ probes (coherent and squeezed vacuum at $k=0$ and $\pi/2$), while $A^{\mathrm{R}}$ is unchanged.
The Peierls time grid uses $N_t=64$, which places the Nyquist harmonic above $M_D=20$; refining to $N_t=128$ and $256$ changes $A^{\mathrm{R}}$ and $A^<$ by at most $\sim10^{-11}$.
These bath-frequency, memory, and time-grid cutoffs are refined independently of $N_b$, $M$, and $M_D$.

As a consistency check of the independently assembled occupied components $G_{\mathrm{num}}^{<,>}$ prior to CAR reconstruction of the reported components, define $\Delta_{\mathrm{sub}} = G_{\mathrm{num}}^{>} - G_{\mathrm{num}}^{<}$, and compare it with the discrete response $\Delta_{\mathrm{const}}$ to the constant wide-band kernel $-\mathrm{i}\varGamma$, evaluated with the same six-ordering quadrature.
At the adopted cutoffs $M=10$, $N_b=12$, $M_D=20$, and $\Delta\nu=0.0125$, we obtain $\Vert\Delta_{\mathrm{sub}}-\Delta_{\mathrm{const}}\Vert_{\infty}=6.8\times10^{-10}$ for the coherent source at $k=\pi/2$, and $1.8\times10^{-10}$ for squeezed vacuum with $r=4$, $\kappa=0.1$, and $k=0$.
Here, $\Vert\bullet\Vert_{\infty}$ denotes the discrete maximum norm over the retained observation-frequency grid at fixed momentum, for the period-averaged ($m=0$) components.
The corresponding residuals relative to $G^{\mathrm{R}}-G^{\mathrm{A}}$, $\Vert\Delta_{\mathrm{sub}}-(G^{\mathrm{R}}-G^{\mathrm{A}})\Vert_{\infty}$, are $5.8\times10^{-5}$ and $6.5\times10^{-6}$, respectively, and decrease systematically under bath-frequency-grid refinement.
These residuals measure the absolute mismatch of the uncorrected $G_{\mathrm{num}}$ assembly to $G^{\mathrm{R}}-G^{\mathrm{A}}$.
Because $\Delta_{\mathrm{sub}}$ tracks $\Delta_{\mathrm{const}}$ far more closely than $G^{\mathrm{R}}-G^{\mathrm{A}}$, the remaining discrepancy of the uncorrected assembly is controlled by the finite-grid representation of the wide-band contact contribution rather than by the relative lesser/greater channel assembly.
The correction $G_{\mathrm{cv}}^{<}-G_{\mathrm{num}}^{<}$ in Eq.~\eqref{eq:car_quadrature_correction} absorbs this contact quadrature error and vanishes in the continuum limit.

As an independent assembly check in a solvable limit, a deterministic finite-amplitude c-number Peierls calculation shows that the directly assembled two-leg $G^<$ agrees with the standard Floquet expression $G^{\mathrm{R}}\varSigma_{\mathrm{bath}}^{<}G^{\mathrm{A}}$ of Eq.~\eqref{eq:classical_lesser_equation} within a relative difference of $6.1\times10^{-4}$ in the $\Vert\bullet\Vert_{\infty}$ norm.

\begin{table}[t]
\centering
\small
\caption{Full-Brillouin-zone numerical checks of the shared-history occupied components for the six parameter sets spanning the source families of Sec.~\ref{sec:benchmark}.
The extrema are taken over 1536 momentum-resolved spectra, each sampled at 1601 frequencies in $\omega\in[-8,8]$.
The $A^>$ values in this table use the CAR-reconstructed component of Eq.~\eqref{eq:greater_from_car} employed for the reported spectra; the independent $G_{\mathrm{num}}^{<,>}$ consistency check prior to that reconstruction is described above.}
\label{tab:keldysh_full_bz_checks}
\begin{tabular*}{\columnwidth}{@{\extracolsep{\fill}}ll@{}}
\hline\hline
Quantity & Observed value \\
\hline
$\min A^<(k,\omega)$ & $6.444\times10^{-13}$ \\
$\min A^>(k,\omega)$ & $6.831\times10^{-13}$ \\
$n_k$ range & $[0.020459,\,0.971203]$ \\
$f_{\mathrm{eff}}(k,\omega)$ range & $[1.296\times10^{-9},\,0.999999999]$ \\
$\max|\mathcal{S}_k-1|$ & $8.556\times10^{-3}$ \\
$\max|\Tr_{\mathrm{ph}} W_k-1|$ & $4.383\times10^{-10}$ \\
\hline\hline
\end{tabular*}
\end{table}

We verify over the full Brillouin zone, for the six parameter sets spanning the source families of Sec.~\ref{sec:benchmark}, the positivity of $A^{<,>}$, the fermionic bounds on $n_k$ and $f_{\mathrm{eff}}$, and the single-band retarded sum rule.
All frequency integrals quoted here are evaluated over the window $\omega\in[-8,8]$ of Table~\ref{tab:benchmark_params}.
Both $A^<(k,\omega)$ and $A^>(k,\omega)$ remain nonnegative at every sampled point, and the integrated occupations and effective spectral occupancies remain within their fermionic bounds.
The Gaussian-probe reconstructions of $A^{<}_{\mathrm{probe}}$ also remain nonnegative over the sampled probe times, as required for a windowed photoemission intensity.
For the single-band retarded sum
\begin{align}
\mathcal{S}_k
&\equiv
\int_{-8}^{8}\mathrm{d}\omega\,A^{\mathrm{R}}(k,\omega),
\label{eq:app_retarded_sum}
\end{align}
the maximum deviation from unity is $8.556\times10^{-3}$.
Table~\ref{tab:keldysh_full_bz_checks} summarizes the envelope over these calculations.

\bibliography{reference,reference_unpublished}

\end{document}